# Why Money Trickles Up – Wealth & Income Distributions

## Geoff Willis

gwillis@econodynamics.org

## 0.0 Abstract

This paper combines ideas from classical economics and modern finance with the general Lotka-Volterra models of Levy & Solomon to provide straightforward explanations of wealth and income distributions.

Using a simple and realistic economic formulation, the distributions of both wealth and income are fully explained. Both the power tail and the log-normal like body are fully captured. It is of note that the full distribution, including the power law tail, is created via the use of absolutely identical agents.

It is further demonstrated that a simple scheme of compulsory saving could eliminate poverty at little cost to the taxpayer.

## 0.1 Contents

## 0.2 Introduction

This paper is a condensed extract from the full paper 'Why Money Trickles Up' which is available at econodynamics.org. This paper is formed from the first section of the full paper, but has been edited down significantly.

This paper introduces the basic mathematical model, a form of the General Lotka Volterra (GLV) model of Levy & Solomon and shows its application to the field of wealth and income distributions. The models use basic economic variables to give straightforward explanations of the distributions of wealth and income in human societies. Usefully, the models also provide simple effective methods for eliminating poverty without using tax and welfare.

The agents in the initial models were identical, and painfully simple in their behaviour. They worked for money, saved some of their money, spent some of their money, and received interest on the money accumulated in their bank accounts.

Because of this the agents had no utility or behavioural functions of the sort commonly used in agent-based economic modelling. As such the models had no initial underlying references to neoclassical economics, or for that matter behavioural economics. There simply was no need for neoclassicism or behaviouralism.

As the modelling progressed, somewhat to my surprise, and, in fact to my embarrassment, it became clear that the models were modelling the economics of the classical economists; the economics of Smith, Ricardo, Marx, von Neumann (unmodified) and Sraffa. In this model wealth is implicitly conserved in exchange, but created in production and destroyed in consumption. Despite the rejection of neoclassicism, the models work, classical economics works.

Where the classical economists were undoubtedly wrong was in their belief in the labour theory of value. They were however absolutely correct in the belief that value was intrinsic, and embodied in the goods bought, sold and stored as units of wealth. Once intrinsic wealth, and so the conservation of wealth is recast and accepted, building economic models becomes surprisingly easy.

I would like to note that Ian Wright, Makoto Nirei & Wataru Souma have produced work on similar lines to my own, the parallels between their work and my own is discussed in section 12. Also, not a word of this paper would have been written without the work of Levy & Solomon and their GLV models. Manipulation of the GLV is beyond my mathematical ability. Although Levy & Solomon's economic explanations are naïve, their gut feeling of the applicability of the GLV to economics in particular, and complex systems in general, was correct. I believe their work is of profound general importance.

## 0.3 Structure of the Paper

Section 1.1 gives a brief review of empirical information known about wealth and income distributions while section 1.2 gives background information on the Lotka-Volterra and General Lotka-Volterra models. Sections 1.3 to 1.5 gives details of the models, their outputs and a discussion of these outputs.

Section 1.6 discusses the effects that changing the ratio of waged income to earnings from capital has on wealth and income distributions.

Sections 1.7 discusses an effective, low-cost option for modifying wealth and income distributions and so eliminating poverty.

Section 12 gives a history of the gestation of this paper and an opportunity to thank those that have assisted in its formation. While sections 15 and 16 give the references and figures respectively.

## 1.1 Wealth & Income Data – Empirical Information

Within theoretical economics, the study of income and wealth distributions is something of a backwater. Neo-classical economics starts from given exogenous distributions of wealth and then looks at the ensuing exchange processes. Utility theory assumes that entrepreneurs and labourers are fairly rewarded for their efforts and risk appetite. The search for deeper endogenous explanations within mainstream economics has been minimal. This is puzzling, because it has been clear for a century that income distributions show very fixed uniformities.

Vilfredo Pareto first showed in 1896 that income distributions followed the power law distribution that now bears his name [Pareto 1896].

Pareto studied income in Britain, Prussia, Saxony, Ireland, Italy and Peru. At the time of his study Britain and Prussia were strongly industrialised countries, while Ireland, Italy and Peru were still agricultural producers. Despite the differences between these economies, Pareto discovered that the income of wealthy individuals varied as a power law in all cases.

Extensive research since has shown that this relationship is universal across all countries, and that not only is a power law present for high income individuals, but the gradient of the power law is similar in all the different countries.

Typical graphs of income distribution are shown below. This is data for 2002 from the UK, and is an unusually good data set [ONS 2003].

Figure 1.1.1 here

Figure 1.1.1 above shows the probability density function. As can be seen this shape has a large bulge towards the left-hand side, with a peak at about £300 per week. To the right hand side there is a long tail showing smaller and smaller numbers of people with higher and higher earnings.

Also included is a log-normal distribution fitted to the curve, on these scales the log-normal appears to give a very good fit to the data. However there are problems with this.

Figure 1.1.2 here

Figure 1.1.2 above shows the same data, on a log-linear scale. Although the log-normal gives a very good fit for the first two thirds of the graph, somewhere around a weekly wage level of £900 the data points move off higher than the log-normal fit. The log-normal fit cannot describe the income of high-earners well.

Figure 1.1.3 here

Figure 1.1.3 above shows the same data as a cumulative density function (cdf). In figure 1.1.3 about 10% of people, a proportion of 0.1, earn more than £755 per week.

It can be seen that the curve has a curved section on the left-hand side, and a straight line power tail section on the right-hand side. This section of the data obeys the power-law described by Pareto 100 years ago.

The work of Pareto gives a remarkable result. An industrial manufacturing society and an agrarian society have very different economic systems and societal structures. Intuitively it seems reasonable to assume that income would be distributed differently in such different societies.

The other big big unexpected conclusion from the data of Pareto and others is the existence of the power tail itself. Traditional economics holds that individuals are fairly rewarded for their abilities, a power tail distribution does not fit these assumptions.

Human abilities are usually distributed normally, or sometimes log-normally. The earning ability of an individual human being is made up of the combination of many different personal skills.

Logically, following the central limit theorem, it would be reasonable to expect that the distribution of income would be a normal or log-normal distribution. A power law distribution however is very much more skewed than even a log-normal distribution, so it is not obvious why individual skills should be overcompensated with a power law distribution.

While the income earned by the people in the power tail of income distribution may account for approximately 50% of total earnings, the Pareto distribution actually only applies to the top 10%-20% of earners. The other 80%-90% of middle class and poorer people are accounted for by a different 'body' of the distribution.

Going back to the linear-linear graph in figure 1.1.1 it can be seen that, between incomes of £100 and £900 per week, there is a characteristic bulge or hump of individuals, with a skew in the hump towards the right hand side.

In the days since Pareto the distribution of income for the main 80%-90% of individuals in this bulge has also been investigated in detail.

The distribution of income for this main group of individuals shows the characteristic skewed humped shape similar to that of the log-normal distribution, though many other distributions have been proposed.

These include the gamma, Weibull, beta, Singh-Maddala, and Dagum. The last two both being members of the Dagum family of distributions. Bandourian, McDonald & Turley [Bandourain et al 2002] give an extensive overview of all the above distributions, as well as other variations of the general beta class of distributions. They carry out a review of which of these distributions give best fits to the extensive data in the Luxembourg Income Study. In all they analyse the fit of eleven probability distributions to twenty-three different countries. They conclude that the Weibull, Dagum and general-beta2 distributions are the best fits to the data depending on the number of parameters used.

For more information, readers are referred to 'Statistical Size Distributions in Economics and Actuarial Sciences' [Kleiber & Kotz 2003] for a more general overview of probability distributions in economics, and also to Atkinson and Bourguignon [Atkinson & Bourguignon 2000] for a very detailed discussion of income data and theory in general.

The author has analysed a particularly good set of income data from the UK tax system, one example is shown in figures 1.1.1-3 above. This data suggests that a Maxwell-Boltzmann distribution also provides a very good fit to the main body of the income data that is equal to that of the log-normal distribution [Willis & Mimkes 2005].

The reasons for the split between the income earned by the top 10% and the main body 90% has been studied in more detail by Clementi and Gallegati [Clementi & Gallegati 2005a] using data from the US, UK, Germany and Italy. This shows strong economic regularities in the data. In general it appears that the income gained by individuals in the power tail comes primarily from income gained from capital such as interest payments, dividends, rent or ownership of small businesses. Meanwhile the income for the 90% of people in the main body of the distribution is primarily derived from wages. These conclusions are important, and will be returned to in the models below.

This view is supported, though only by suggestion, by one intriguing high quality income data set. This data set comes from the United States and is from a 1992 survey giving proportions of workers earning particular wages in manufacturing and service industries.

The ultimate source of the data is the US Department of Labor; Bureau of Statistics, and so the provenance is believed to be of the good quality. Unfortunately, enquiries by the author has failed to reveal the details of the data, such as sample size and collection methodology.

The data was collected to give a comparison of the relative quality of employment in the manufacturing and service sectors. Although the sample size for the data is not known, the smoothness of the curves produced suggest that the samples were large, and that the data is of good statistical quality. The data for services is shown in figures 1.1.4 & 1.1.5 below, the data for manufacturing is near identical.

Figure 1.1.4 here

Figure 1.1.5 here

Like the UK data, there appears to be a clear linear section in the central portion of the data on a log-linear scale in figure 1.1.5, indicating an exponential section in the raw data. Again this data can be fitted equally well with a log-normal or a Maxwell-Boltzmann distribution.

What is more interesting is that, beyond this section, the data heads rapidly lower on the logarithmic scale. This means it is heading rapidly to zero on the raw data graph. With these two distributions there is no sign whatsoever of the 'power tail' that is normally found in income distributions.

It is the belief of the author that the methodology for this US survey restricted the data to 'earned' or 'waged' income, as the interest in the project was in looking at pay in services versus manufacturing industry. It is believed income from assets and investments was not included as this would have been irrelevant to the investigation.

This US data set has been included for a further reason, a reason that is subtle; but in the belief of the author, important.

Looking back at figure 1.1.1 for the UK income data, there is a very clear offset from zero along the income axis. That is the curve does not start to rise from the income axis until a value of roughly £100 weekly wage.

The US data shows an exactly similar offset, with income not rising until a weekly wage of $100.

This is important, as the various curves discussed above (log-normal, gamma, Weibull, beta, Singh-Maddala, Dagum, Maxwell-Boltzmann, etc) all normally start at the origin of the axis, point (0,0) with the curve rising immediately from this point.

While it is straightforward enough to put an offset in, this is not normally necessary when looking at natural phenomena.

Following the work of Pareto, the details of income and wealth distributions have rarely been studied in mainstream theoretical economics. Despite the lack of interest within economics, this area has had a profound attraction to those outside the economics profession for many years, a review of this history is provided by Gabaix [Gabaix 2009].

In recent years, the study of income distributions has gone through a small renaissance with new interest in the field shown by physicists with an interest in economics, and has become a significant element of the body of research known as 'econophysics'.

Notable papers have been written in this field by Bouchaud & Mezard, Nirei & Souma, Dragulescu & Yakovenko, Chatterjee & Chakrabarti, Slanina, Sinha and many, many, others [Bouchaud & Mezard 2000, Dragulescu & Yakovenko 2001, Nirei & Souma 2007, Souma 2001, Slanina 2004, Sinha 2005].

The majority of these papers follow similar approaches; inherited either from the work of Gibrat, or from gas models in physics. Almost all the above models deal with basic exchange processes, with some sort of asymmetry introduced to produce a power tail. Chatterjee et al 2007, Chatterjee & Chakrabarti 2007 and Sinha 2005 give good reviews of this modelling approach.

The approaches above have been the subject of some criticism, even by economists who are otherwise sympathetic to a stochastic approach to economics, but who are concerned that a pure exchange process is not appropriate for modelling modern economies [Gallegati et al 2006].

An alternative approach to stochastic modelling has been taken by Moshe Levy, Sorin Solomon, and others [Levy & Solomon 1996].

They have produced work based on the 'General Lotka-Volterra' model. Unsurprisingly, this is a generalised framework of the 'predator-prey' models independently developed for the analysis of population dynamics in biology by two mathematicians/physicists Alfred Lotka and Vito Volterra.

A full discussion of the origin and mathematics of GLV distributions is given below in section 1.2.

These distributions are interesting for a number of reasons; these include the following:

- the fundamental shape of the GLV curve
- the quality of the fit to actual data
- the appropriateness of the GLV distribution as an economic model

Figure 1.1.6 here

Figure 1.1.7 here

With regard to the fundamental shape of the GLV curve, figures 1.1.6 and 1.1.7 above show plots of the UK income data against the GLV on a linear-linear and log-log plot.

The formula for this distribution is given by:

$$P(w) = K(e^{-(\alpha-1)/(w/L)})/((w/L)^{(1+\alpha)}) \qquad (1.1a)$$

and it has three parameters; K is a general scaling parameter, L is a normalising constant for w, and $\alpha$ relates to the slope of the power tail of the distribution.

It should firstly be noted that the GLV has both a power tail and a 'log-normal'-like main body. That is to say it can model both the main population and the high-end earners at the same time. This is a very significant advantage over other proposed distributions.

The second and more subtle point to note is that the GLV has a 'natural' offset from zero. It is in the nature of the GLV that the rise from zero probability on the y-axis starts at a non-zero value on the x-axis, this is discussed further in section 1.2 Below.

Finally the detailed fit of the GLV appears to be equivalent or better than the log-normal distribution.

| **Figure 1.1.8** | Reduced Chi Squared | |
|---|---|---|
| | Full Data Set | Reduced Data Set |
| Boltzmann Fit | 3.27 | 1.94 |
| Log Normal Fit | 2.12 | 3.02 |
| GLV Fit | 1.21 | 1.83 |

Figure 1.1.8 above gives results from a basic statistical analysis using the GLV, log-normal and Maxwell-Boltzmann distributions. (The values in the table are the reduced chi-squared values, using an assumed standard measurement error of 100. The actual measurement error is not known, so the values above are not absolute, however, changing the measurement value will change the values in the table by equal proportions, so the relative sizes of the values in the table will stay the same.)

It can be seen from the figures in the first column that the GLV, with the lowest value of chi-squared, gives the best fit. In itself this is not altogether surprising, as it is known that the log-normal and the Maxwell-Boltzmann have exponential tails, and so are not able to fit power tails.

More remarkably, the figures in the second column show the same analysis carried out using a truncated data set with an upper limit of £800 per week. This limit was taken to deliberately exclude the data from the power tail. Again it can be seen that the GLV still just gives the best fit to the data. This in itself suggests that the GLV should be preferred to the log-normal or the Maxwell-Boltzmann distributions.

It is also of note that in parallel to the work of Solomon et al, Slanina has also proposed an exchange model that produces the same output distribution as the GLV [Slanina 2004].

Unfortunately the modelling approaches of Solomon et al, and Slanina use economic models that are not wholly convincing, and as such have significant conceptual shortcomings.

It is the belief of the author that an alternative economic analysis, using more appropriate analogies allows a much more effective use of GLV distributions in an intuitive and simple economic formulation. This is the third main reason for preferring the GLV distribution, and forms the key content of the initial sections of this paper. As previously noted Souma & Nirei have also pursued research in this direction.

For those who want more background on the formation of power laws, log-normal laws and related processes, there are three very good background papers by Newman [Newman 2005], Mitzenmacher [Mitzenmacher 2004] and Simkin & Roychowdhury [Simkin & Roychowdhury 2006].

The papers by Newman and Mitzenmacher give very good overviews of what make power law and log-normal normal distributions without being mathematically complex.

One basic point from the papers is that there are many different ways of producing power law distributions, but the majority fall into three main classes.

The first class gives a power law distribution as a function of two exponential distributions; of two growth processes.

The second class gives power law distributions as an outcome of multiplicative models. This is the route that Levy and Solomon have followed in their work, and forms the basis for the GLV distribution discussed in detail in the next section.

The third class for producing power laws uses concepts of 'self-organised criticality' or 'SOC'.

A second basic point, discussed in Mitzenmacher, is that the difference between a log-normal distribution and a power law distribution is primarily dependent on the lower barrier of the distribution, if the lower barrier is at zero, then you get a log-normal distribution, if the barrier is above zero, then the distribution gives a power tail. A non-zero barrier, provided by wage income, is an essential part of the GLV model discussed in section 1.2 below.

The paper of Simkin and Roychowdhury is illuminating and entertaining. It shows that the same basic mechanisms for producing power laws, and branching processes in general, have been rediscovered dozens of times, and that most power law / branching processes are in fact analogous. As an example, the models of Levy & Solomon follow processes previously described by Champernowne in economics, and ultimately by Yule and Simon almost a century ago. This is not to devalue the work of Solomon and Levy; their approach allows for dynamic equilibrium formation, this includes an element missing from most branching models that in my opinion makes the Solomon and Levy model much more powerful as a general model. This is returned to in section 1.2 below.

Finally it is important to note the difference between income and wealth.

Income data is relatively easy to collect from income tax returns. Pareto's original work and almost all subsequent analysis of data is based on that from income data.

Wealth data of any quality is very difficult to find. Where this data has been collected it almost exclusively pertains to the richest portion of society, and suggests that wealth is also distributed as a power law for these people.

I am not aware of any data of sufficient quality to give any conclusions about the distribution of wealth amongst the bottom 90% of individuals.

This has led to some very unfortunate consequences within the econophysics community.

Without exception all the exchange models by all the various authors above, including those of Solomon and Slanina, are wealth exchange modelled. I have not yet seen a model where income (trivially the time derivative of wealth) is modelled.

Despite this, the output distributions from these wealth models are often judged to be successful when they map well onto data derived from income studies. An explanation for why wealth models can give outputs that can then define income data successfully is given in section 1.4.4 below.

Before moving on to the modelling of income and wealth distributions, I would first like to discuss the derivation and mechanics of the Lotka-Volterra distribution and the GLV distribution in more detail.

## 1.2 Lotka-Volterra and General Lotka-Volterra Systems

### 1.2.1 Lotka-Volterra systems

Lotka-Volterra systems were independently discovered by Alfred Lotka [Lotka 1925] and Vito Volterra [Volterra 1926] and are used to describe the dynamics of populations in ecological systems. Ultimately this dynamic approach goes back directly to the economic growth equations of Malthus and Sismondi.

A basic Lotka-Volterra system consists of a population of prey (say rabbits) whose size is given by x, and a population of predators (say foxes) given by y.

Not explicitly given in this simple case, it is further assumed that there is a steady supply of food (eg. grass) for the prey.

When no predators are present this means that the population of the rabbits is governed by:

$$\frac{dx}{dt} = ax \qquad (1.2.1\,a)$$

where a is the population growth rate.

Left to their own business, this would give exponential, Malthusian growth in the population of the rabbits.

In the absence of any rabbits to eat, it is assumed that there is a natural death rate of the foxes:

$$\frac{dy}{dt} = -cx \qquad (1.2.1\,b)$$

where c is the population die-off rate, and the negative sign indicates a decline in the population. This would give an exponential fall in the fox population.

When the foxes encounter the rabbits, two further effects are introduced, firstly the rate at which rabbits are killed is proportional to the number of rabbits and the number of foxes (ie the chance of foxes encountering rabbits), so:

$$\frac{dx}{dt} = -\alpha\,x\,y \qquad (1.2.1\,c)$$

where $\alpha$ is a constant, and the –ve sign indicates that such encounters are not good for the rabbits. However these interactions are good for the foxes, giving:

$$\frac{dy}{dt} = \gamma\,x\,y \qquad (1.2.1d)$$

Where γ is again a fixed constant.

Taken together, the results above give a pair of differential equations:

$$\begin{aligned}\frac{dx}{dt} &= ax - \alpha x y \\ &= x(a - \alpha y) \quad (1.2.1\,e)\end{aligned}$$

for the rabbits, and:

$$\begin{aligned}\frac{dy}{dt} &= \gamma x y - cy \\ &= y(\gamma x - c) \\ &= y(-c + \gamma x) \quad (1.2.1\,f)\end{aligned}$$

for the foxes.

The most important point about this pair of equations is that x depends on y, while at the same time, y depends on x. The dependency goes in both directions, this make things fun.

While it is possible for these equations to have a single stable solution, this is often not the case. Commonly the populations of both rabbits and foxes fluctuates wildly. An example is given in figure 1.2.1.1 below for lynx preying on arctic hares [BBC]:

Figure 1.2.1.1 here

The data for the graph above comes from long-term records of pelts collected by the Hudson Bay Company. The graph shows very closely the recurrent booms and busts in population of the two types of animals. In the short term the population and total biomass of both lynx and hares can increase or decrease substantially. The population of lynx can be large or small in proportion to that of the hares. The populations of both are highly unstable.

A subtlety to note is that the population of the lynx follows, 'lags', the population of the hares. It is also worth considering, even at this early stage, the behaviour, or indeed the 'behaviouralism' of the lynx in particular.

Following a previous collapse, the population of hares can expand rapidly as there are very few lynx to hunt them.

As the population of hares increases rapidly, the lynx behave 'rationally' (at least given the absence of long-term, liquidly tradable, hare futures) in both eating lots of hares, and also giving birth to lots of new lynx to feed on the excess of hares.

Eventually, of course there are too many lynx for the population of hares, and ultimately there are too many lynx and hares for the underlying amount of grass available.

At the peaks of hare and lynx populations there is simply too much biomass wandering around for the land to support.

Despite the substantial fluctuations seen in figure 1.2.1.1 above, the populations of both lynx and hares show stable fluctuations around long term averages; roughly 40,000 or so for the hares and 20,000 or so for the lynx, though note that the populations pass through these average values very quickly.

In fact the values of the two populations are confined to a band of possible values. The population can move round in a limited set of possible options, this is shown for example in the two figures from simulations below.

Figure 1.2.1.2 here

Note also the figure 1.2.1.2 shows the same leads and lags in predator and prey populations as the real data. The populations of wolves and rabbits can be displayed on one graph, this then produces the phase diagram in figure 1.2.1.3 below showing how the population of wolves and rabbits vary with each other, and how they are constrained to a particular set of paths.

Figure 1.2.1.3 here

These diagrams are taken from the website of Kumar, [Kumar 2006], which gives a very good brief introduction to the maths and modelling of Lotka-Volterra systems.

It can be seen that the simulated population of wolves and rabbits wanders continuously around average values of approximately seventeen rabbits and six wolves.

In contrast, figures 1.2.1.4 & 5 below show the same system with minor changes to the rates of growth. In this model the oscillations slowly die down to stable long-term values. Another alternative is that the oscillations can grow in size unstably and explode to infinity.

Figure 1.2.1.4 here

Figure 1.2.1.5 here

One of the important things to note about non-linear dynamic systems such as these is that relatively minor changes in parameters can result in dramatic differences in system behaviour.

Note that you don't need both predators and prey, a solitary animal population that grows too quickly can also suffer from population booms and crashes. An example is that of Soay sheep on the island of Soay (in this case the grass can be considered to be the prey, though a better solution would be to use the logistic equation or a similar carrying capacity based approach).

### 1.2.2 General Lotka-Volterra (GLV) systems

As the name implies, the General Lotka-Volterra system (GLV) is a generalisation of the Lotka-Volterra model to a system with multiple predators and prey. This can be represented as:

$$\frac{dx_i}{dt} = x_i r_i + \sum_{j=1}^{N} a_{i,j} x_i x_j \qquad (1.2.2\,a)$$

$$= x_i (r_i + \sum_{j=1}^{N} a_{i,j} x_j) \qquad (1.2.2\,b)$$

here, $dx_i/dt$ is the overall rate of change for the i-th particular species, out of a total of N species. This is made up of two terms.

The first term is the natural growth (or death) rate, $r_i$, for the species, where $x_i$ is the population of species i. This rate $r_i$ is equivalent to the growth rate 'a' in equation (1.2.1e) or the death rate '-c' in equation (1.2.1f).

The second term gives the sum of all the interactions with the j number of other species. Here $a_{i,j}$ is the interaction rate defining the relationship between species i and j.

$a_{i,j}$ is negative if species j is a predator feeding on species i, positive if species i is a predator feeding on species j, or can be of either sign for a heterotroph. $a_{i,j}$ is equivalent to the $\alpha$ of equation (1.2.1e) or the $\gamma$ of equation (1.2.1f).

Hopefully it is clear that equations (1.2.2a) and (1.2.2b) are generalisations of equations (1.2.1e) and (1.2.1f) for many interacting species.

For each species in the system, potentially N-1 interaction rates $a_{i,j}$ are needed, while N! separate differential equations are needed to describe the whole system. This makes direct solution of the equations for the system somewhat problematic.

Fortunately in many systems it is possible to make simplifying assumptions. As an example Solomon [Solomon 2000] proposes the following difference equation as a possible explanation for the power law distribution of city population sizes. This equation describes changes in the distribution in terms of discrete time-steps from time t to time t+1:

$$w_{i,t+1} = \lambda_t w_{i,t} + a_t \bar{w}_t - c_t \bar{w}_t w_{i,t} \qquad (1.2.2\,c)$$

The terms on the right hand side, in say the year 2003, the year t, add up to give the population w of city i in the year 2004 on the left hand side, which is at time t+1.

Such equations are typically used in simulations, one after the other, to give a model of how populations change. Sometimes, though often not, clever mathematicians can derive output population distributions from the underlying difference equations.

In equation (1.2.2c), $\lambda$ is the natural growth rate of the population w of city i, but is assumed that $\lambda$ is the same for each city.

$a_t$ is the arrival rate of population from other cities, which is multiplied by the average population $\bar{w}$ of all the cities.

The final term gives the rate of population leaving each city, which is due to the probability $c_t$ of an individual meeting a partner from another city. This is given by multiplying the average population $\bar{w}$ with the population of city i.

Leaving aside the detail of the model, important generalisations have been made to produce a more tractable model.

In this case $\lambda$, a and c are universal rates, applicable to all members of the system.

$\lambda$ and a both give 'positive autocatalytic' (positive feedback) terms which increase the population w of each city. While the negative value of c ensures that the population of each city has an element of decrease.

In the absence of the negative feedback term, the populations of the cities can increase indefinitely to infinity without reaching a stable solution.

In the absence of the positive autocatalytic growth of the $\lambda$ in the first term on right hand side, the second and third terms will cause all of the population to end up in a single city.

Normally one or more variables are assumed to be stochastic; that is they can vary randomly. In Solomon's example above, all three of $\lambda$, a and c are assumed to be stochastic. This stochasticity need not be large; it can be small fluctuations around a long-term mean, but it ensures that a locally stable solution is not reached, and that the system evolves into a single long term equilibrium solution.

While the above may seem complex, it is argued in section 7.3 of the full paper that this model can be seen as a very general model across many different real world complex systems.

It is possible to show (though not by me) that the above system can give a stable resultant probability distribution function of populations over the various cities of the form:

$$P(w) = (e^{-(\alpha-1)/w})/(w^{(1+\alpha)}) \qquad (1.2.2\,d)$$

Which is the general form of the GLV distribution. Or more specifically:

$$P(w) = K(e^{-(\alpha-1)/(w/L)})/((w/L)^{(1+\alpha)}) \qquad (1.2.2\,e)$$

As has been shown above in section 1.1 this formula gives a very good fit to income data.

As well as giving a good fit, the GLV can also fit both the main body and power tail of income distributions. The GLV also has a natural offset from zero.

However the main reason for using the GLV is that it naturally describes complex dynamic flow systems that have reached a maximum entropy production equilibrium. Economics is such a complex dynamic flow system, and it will be seen that the straightforward models described below model real economic outcomes surprisingly well.

Solomon further proposes a similar model as an explanation for income distribution:

$$w_{i,t+1} = \lambda_t w_{i,t} + a_t \bar{w}_t - c_t \bar{w}_t w_{i,t} \qquad (1.2.2\,\text{f})$$

In this case $\lambda$ is proposed to be positive gains by individuals with origins on the stock market, 'a' is assumed to represent wealth received in the form of *'subsidies, services and social benefits'*, while 'c' is assumed to represent competition for scarce resources, or *'external limiting factors (finite amount of resources and money in the economy, technological inventions, wars, disasters, etc.) as well as internal market effects (competition between investors, adverse influence of bids on prices such as when large investors sell assets to realize their value and prices fall as a result'.*

While it is the author's belief that a form of the GLV is appropriate for modelling wealth and income distributions, it is believed that the above economic mechanisms are not realistic. In the next section an economic model is proposed that I believe much more closely represents real life economic mechanisms.

## 1.3 Wealth & Income Models - Modelling

Figure 1.3.1 here

Figure 1.3.1 above shows a simple macroeconomic model of an economy. This model is taken from figure 1 of chapter 2 of 'Principles of Economics', by Gregory Mankiw [Mankiw 2004].

Figure 1.3.2 below shows a modified version of the diagram. The two 'markets' between the firms and households have been removed, investment and saving streams have been added, as well as the standard economics symbols for the various flows.

Figure 1.3.2 here

All standard economics textbooks use similar diagrams to figures 1.3.1 and 1.32 for macroeconomic flows; I have chosen that of Mankiw as his is one of the most widely used.

Flows of goods and services are shown in the black lines. The lighter broken lines show the flows of money. Note that Mankiw shows households owning 'factors of production' such as land and capital, which the households are then shown as selling to firms. This is indicated as a flow of land and capital (along with labour) from households to firms. We will return to this particular 'flow' later.

Note also that the total system shows a contained circularity of flow, with balances between supply and demand of goods and services.

In this circular flow model economic textbooks assume some basic equalities:

$$C = G \qquad (1.3a)$$

$$C = Y \qquad (1.3b)$$

Equation (1.3b) state that the total income gained from firms adding value is equal to the total consumption of goods and services.

[Nb. In writing this paper I have attempted to use standard notation from economics wherever possible. This occasionally results in confusion. It should be noted that the capital letter Y is used as standard in (macro) economics for income, while small y is used as standard in (micro) economics for outputs from companies. This is not normally a problem, as the two are rarely discussed at the same time in standard economic models. In the discussions of income that follows y is not actually necessary for the analysis, and Y invariably refers to income in the equations of the mathematical model and is normally subscripted.]

Figure 1.3.3 here

In figure 1.3.3 above I have modified this standard model to reflect what I believe is something closer to reality.

Firstly in this model households have been changed to individuals, this is simply to bring the model more in line with the standard analysis of statistical physics and agent based, modelling techniques. This amounts to little more than pickiness. This distinction can be made irrelevant by simply assuming that all households consist of a single individual.

Much more importantly, the flow pattern has been changed and the circularity has been disturbed.

In the real world most goods and services are consumed in a relatively short period of time. To show this, Consumption C, has been changed to represent the actual consumption of goods. This is a real flow of goods, and represents a destruction of value. Note that this is a change from the standard use of C in economics textbooks.

That which was previously shown as consumption is now shown as 'y' the material output of goods and services, which are provided to consumers from the firms operating in the economy.

The money paid for these goods and services is shown by My.

As can be seen in figure 1.3.3 above, the income stream Y has been split into two components, one, e is the earnings; the income earned from employment as wages and salaries, in return for the labour supplied.

$\pi$ is the 'profit' and represents the payments made by firms to the owners of capital, this can be in the form of dividends on shares, coupons on bonds, interest on loans, rent on land or other property, etc.

The flow of capital has been shown as a dotted line. This is because capital doesn't flow. Householders do not hold stocks of blast furnaces in their backyards in the hope of selling them to firms in exchange for profit or interest on their investments.

Capital, such as machine tools and blast furnaces, is normally bought in by firms from other firms, sometimes using money provided by households, but mostly by retained earnings.

In fact all the various models that follow in this paper ignore both investment I, and saving S. In the income models it is always assumed that the overall economy is in a steady state and so, firstly, that all funds required for wear & repair are taken from internal flows. More importantly, in later models; both for companies and macroeconomic modelling, it is also assumed that all new capital is produced from retained earnings within companies.

For many economists this will be seen as a serious flaw. Since at least the time of Keynes, investment and saving have been at the heart of macroeconomic modelling, and this is true of neo-classical and other heterodox modelling, not just that in the Keynesian tradition. However in the real world:

*"Most corporations, in fact, do not finance their investment expenditure by borrowing from banks."* [Miles & Scott 2002, 14.2]

As examples, Miles & Scott give the following table for proportions of investment financing for four countries averaged over the years 1970-1994.

Figure 1.3.4 here
[Miles & Scott 2002 / Corbett & Jenkinson 1997]

As can be seen the maximum possible proportion of external financing (the IS of economists) is 36.8% for Japan. For the UK it doesn't even reach 20%. This financing is small to negligible in importance. In the real world most financing is taken from cash flow.

Going back to capital; real capital, in the form of land, machine tools, computers, buildings, etc will be represented in the diagram as fixed stocks of real capital K, held by the companies.

All of this real capital is assumed to be owned by households, in the form of paper assets, W, representing claims on the real assets in the form of stocks or shares. In the following discussions bonds and other more complex assets will be ignored, and it will be assumed that all the wealth of K is owned in the form of shares (stocks) in the various firms.

This paper wealth will be represented as W in total, or $w_i$ for each of i individuals.

For the income models in the first part of this paper it will further be assumed that the paper wealth of the households accurately represents the actual physical capital owned by the companies, so:

$$\text{total } W \ = \ \text{total } K \qquad (1.3c) \qquad \text{or:}$$

$$\sum w_i \ = \ W \ = \ K \qquad (1.3d)$$

the total real capital invested in the firms is equal to the total value of financial assets held by individuals.

The dotted line in the figure 1.3.3 indicates the assumed one to one link between the financial assets W and the real assets K. It is dotted to show that it is not a flow, it simply indicates ownership.

This mapping of real and financial assets assumes that the financial assets are 'fairly' priced, and can be easily bought and sold in highly liquid markets.

In the models below it is assumed that there is a steady state, so the totals of W and K are both constant. This means that the model has no growth, and simply continues at a steady equilibrium of production and consumption. There is no change in population, no change in technology, no change in the efficiencies of the firms. The example of Japan over the last two decades has shown that economies can continue to function in a normal manner with extended periods of negligible growth. For a modern economy the difference between the creation and the destruction is economic growth of the GDP, and at 2%-4% or so per annum is pretty close to being stable.

It is important to note that the capital discussed here is only the capital vested in productive companies. Other personal capital is excluded, the most important of these is housing. I have ignored the role of housing in these early models, though clearly this is a major simplification. This is discussed further in section 1.9.1 of the full paper. For the moment all wealth held is assumed to be financial assets. All other personal assets such as housing, cars, jewellery, etc are ignored.

There are some other important base assumptions of the model. These are discussed briefly below:

The economy is isolated; there are no imports or exports.

There is no government sector, so no taxation or welfare payments, government spending, etc.

There is no unemployment; all individuals are employed, with a given wage, either from a uniform distribution or a normal distribution depending on the model.

Labour and capital are assumed to be complementary inputs and are not interchangeable at least in the short term. It turns out, much later, that this assumption is not only true, but of profound importance, this is discussed at some length in the full paper.

The role of money is ignored in these models, for the sake of argument, it can be assumed that payments are made in the form of units δW of the paper assets held by the individuals, say in units of DJI or FTSE all share trackers.

Finally there is no debt included in the income models.

Figure 1.3.5 below shows some of the assumptions above, it also adds in some more flows to help bring the model closer to the real world.

Figure 1.3.5 here

There are two main reasons for changing the diagram in this manner. One reason is to bring the diagram into line with the ideas of the classical economists such as Smith, Ricardo, Marx and Sraffa. The second is to help the model comply with some of the more basic laws of physics.

Starting with the classical economics. It has previously been defined that consumption by the individuals means the destruction of value in the form of using up resources. This consumption could be food eaten in a few days, clothes which wear out in a few months or cars and furniture that take years to wear out, but which ultimately need to be replaced periodically. The consumption can also be services such as meals in restaurants, going to see films, receiving haircuts, going on holiday, etc. All value destruction is assumed to take place within households as consumption.
In physics terms, this destructive process is characterised as a local increase in entropy.

To balance this destruction, it is assumed that all value is created in the processes of production, and that all this value is created within firms.

I am going to follow in Schrödinger's footsteps and describe this increase in value as the creation of something called 'negentropy'. For physicists a better term might be 'humanly useful free energy'. Non-physicists who require a more detailed explanation can consult part B of the full paper. For the moment the important thing to grasp is that negentropy is equivalent to economic value, the more negentropy something has, the more you are willing to pay for it.

Although the discussions in these models use production of manufactured goods as an easily understandable example; it should be noted that 'production' is any process that adds value, and produces higher value inputs than the outputs. So agriculture, mining, power generation, as well as distribution, retail, personal and financial services are all forms of production.

Indeed, almost any process that is done within a company is production. That is why companies exist, so that the value added is kept securely within the company.

In general, exchange processes don't create value, they are simply a means for swapping goods from different points along the supply chain leading up to the final point of consumption. Exchanges are simply a result of the division of labour between different companies or individuals who have particular sets of skills and abilities.

The model in figure 1.3.5 above essentially goes back to the ideas of the classical economists; of Smith, Ricardo, Marx, Sraffa and others. It assumes that goods and services have meaningful, long term, intrinsic values, and that long-term prices reflect these values. Short-term prices may move away from these values, primarily to allow generation of new capital.

This paper explicitly rejects the marginalist view that value is exogenously set by the requirements and beliefs of individuals, and that exchange between such individuals creates value.

Figure 1.3.6 here

Figure 1.3.6 above figure demonstrates these assumptions for a more complex model of linear flows of value added.

In figure 1.3.6, all the horizontal flows (flows through the side walls) are direct exchanges of actual goods for monetary tokens. Assuming a free market with fair pricing, and that the currency is a meaningful store of value, then all the horizontal exchange flows have zero net value.

$$x_1 + Mx_1 = 0 \quad \text{or:}$$

$$x_1 = -Mx_1, \quad x_2 = -Mx_2, \quad \ldots \quad x_k = -Mx_k, \quad \text{etc}$$

Vertical flows, through the top and bottom of the boxes, involve changes; increases or decrease in negentropy.

In economic terms this is stated as value being added or wealth being created. In figure 1.3.6 above the values of the final output y and the series of inputs x are related by:

$$y > x_3 > x_2 > x_1 \quad \text{and clearly}$$

$$My > Mx_3 > Mx_2 > Mx_1$$

The differences between these values represents the wealth created by the employees and capital of the firm acting on the inputs to create the outputs. The employees are rewarded for this wealth creation via their wage earnings, while the owners of the capital are rewarded with returns on their capital.

Figure 1.3.7 here

Figure 1.3.7 above gives another layout that shows that the whole system doesn't have to be linear, but that the same assumptions regarding adding value still hold.

Finally to satisfy the physicists reading; waste streams are included so that the $2^{nd}$ law is not violated. The total entropy created by the waste streams from the firms, principally low grade heat, is greater than the negentropy created in the products of the firms.

Essentially figures 1.3.5 to 1.3.7 bring together the economic and physical diagrams discussed in Ayres & Nair [Ayres & Nair 1984]; so that the circulation of wealth and money complies with the laws of physics as well as the laws of finance. The discussions of Ayres & Nair clearly have strong antecedents in the theories of Georgescu-Roegen [Georgescu-Roegen 1971].

Figure 1.3.5 here

So, going back to figure 1.3.5, we are now at a point where we can move into the detail of the mathematical model.

Firstly we will assume that $x = Mx$ and that both are irrelevant to the rest of the debate.

We will also assume that $L = e$, ie that labour is fairly rewarded for the value of its input. In later sections this is discussed in more depth.

Next we will assume $y = My$, ie that 'fair' prices are being paid for the goods sold to the consumers. We will eventually relax this assumption in later models.
In this model it will further be assumed that:

$$\text{total } C = \text{total } Y = \text{total } My$$

at steady state equilibrium.

It will be seen later that this is actually a natural outcome of the model, and doesn't need to be forced. Note that although the totals of C and Y are the same, they may not be the same for individuals. Some individuals may consume less than they earn, or vice versa.

In these earlier models, we are not interested in the detail of the firms so we are going to ignore the difference between the capital K and it's financial equivalent W.
We will assume that total $K$ = total $W$, and so assume that companies are fairly and accurately priced in the financial markets. These assumptions will be relaxed later, again with interesting consequences.

The paper wealth W will be split between N individuals, so from individual $i = 1$ to individual $i = N$.

Going back to figure 1.3.5 and equation (1.3d) above; although the total capital and wealth is fixed, individual wealth is allowed to vary, so:

$$\sum w_{i,t} = \sum w_{i,t+1} = W = K = \text{constant} \qquad (1.3e)$$

Where $w_i$ is the wealth of individual i.

Looking at a single individual in the box on the right of figure 1.3.5, in one time unit, from t to t+1, the change in wealth is given by the following equation:

$$w_{i,t+1} = w_{i,t} + y_{i,t} - My_{i,t} + e_{i,t} + \pi_{i,t} - C_{i,t} - \text{labour}_{i,t} - \text{capital}_{i,t} \qquad (1.3f)$$

This equation states that the wealth for a single individual at time t+1, on the left hand side, is equal to the wealth at time t, plus the contributions of the seven arrows going into or out of the box on the right hand side of figure 1.3.5.

However equation (1.3f) is not meaningful as it is trying to add apples and oranges. The items y, C, labour and capital are real things, while w, My, e and $\pi$ are all financial quantities. Adding the non-financial things is not appropriate, however all the financial flows must ultimately add up.

So looking then at the financial flows, we have the following equation:

$$w_{i,t+1} = w_{i,t} - My_{i,t} + e_{i,t} + \pi_{i,t} \qquad (1.3g)$$

This now counts things that are the same. As stated above, although the totals of My = Y = C some individuals can consume less than y, and so accumulate more wealth W, others can consume more than Y and so reduce their total W.

To make this process clearer, I am going to use – $C_{i,t}$ in place of – $My_{i,t}$ in equation (1.3g).

In this case $C_{i,t}$ is now a monetary unit, and effectively reverts to standard economics usage. To keep the units correct, it is assumed that in practice heavy consumers exchange part of their wealth W with some heavy savers, in return for some of the savers real goods y.

Substituting and rearranging, this then leaves us with the following equation:

$$w_{i,t+1} = w_{i,t} + e_{i,t} + \pi_{i,t} - C_{i,t} \qquad (1.3h)$$

This then is the difference equation for a single agent in this model.

In a single iteration, the paper wealth w of an individual i increases by the wages earned e plus the profits received $\pi$. The individual's paper wealth also reduces by the amount spent on consumption C.

We now need to investigate the mechanics of this in more detail. Looking at the second, third and fourth terms on the right hand side of (1.3h) in order, we start with earned income; e.

In the first model, Model 1; it is assumed that all agents are identical, and unchanging in their abilities in time, so:

$$e_{i,t} = e = \text{constant}; \quad (1.3i) \quad \text{for all i agents.}$$

The assumption above effectively assumes that the economy as a whole is in dynamic equilibrium (the difference between static and dynamic equilibria is discussed at length in section 6 of the full paper), there is no technological advancement, no education of employees, etc. It assumes that all individuals have exactly the same level of skills and are capable of producing the exact same level of useful output as one another; and that this is unchanging through time.

We move next to $\pi$, the income from returns. We assume that the economy consists of various companies all with identical risk ratings, all giving a uniform constant return; r on the investments owned, as paper assets, by the various individuals. Here r represents profits, dividends, rents, interest payments, etc to prevent confusion with other variables, r will normally be referred to as the profit rate.

This gives:

$$\pi_{i,t} = w_{i,t} r \quad (1.3j) \quad \text{for each of the i agents.}$$

Given r as constant, then:

$$\sum \pi_i = r \sum w_i \quad (1.3k) \quad \text{so:}$$

$$r = \frac{\sum \pi_i}{\sum w_i} \quad \text{and}$$

$$r = \frac{\sum \pi_i / N}{\sum w_i / N} \quad \text{giving:}$$

$$r = \frac{\bar{\pi}}{\bar{w}} \quad (1.3l)$$

where $\bar{\pi}$ and $\bar{w}$ are the average values of $\pi$ and W respectively.
Note that r, $\bar{w}$ and $\bar{\pi}$ are all fixed constants as a consequence of the definitions.

So for an individual:

$$\pi_{i,t} = w_{i,t} \frac{\bar{\pi}}{\bar{w}} \quad (1.3m)$$

For the final term consumption; C is assumed to be a simple linear function of wealth. As wealth increases, consumption increases proportionally according to a fixed rate $\Omega$ (a suggested proof that this might be reasonable a assumption is given in Burgstaller [Burgstaller 1994], the constancy of $\Omega$ is discussed in depth in section 4.5).

So:

$$C_{i,t} = w_{i,t}\Omega \qquad (1.3n)$$

This final assumption gives the conceptual reason for using C rather than My for this final term.

Clearly a linear consumption function is not realistic, and a concave consumption function would reasonably be expected, with the rate of consumption declining as wealth increased. For most of the modelling, this simple consumption function is sufficient to demonstrate the required results, this is examined further in section 1.9.1 in the full paper.

In model 1A, $\Omega$ is made to be stochastic, with a base value of 30% multiplied by a sample from a normal distribution which has a variance of 30% of this base value.

Consumption is chosen as the stochastic element, as being realistic in a real economy. While earnings are usually maximised and fixed as salaries, choosing to save or spend is voluntary. It should be noted that all agents remain fully identical. While the proportion consumed by each agent changes in the model in each iteration, on average each agent spends exactly 30% of its wealth. This is critically important, in model 1A all the agents are identical and have the same long-term average saving propensity, as well as earning ability.

Taken together and substituting into (1.3h) this gives the difference equation for each agent as follows:

$$w_{i,t+1} = w_{i,t} + e + w_{i,t}\frac{\overline{\pi}}{\overline{w}} - w_{i,t}\Omega \qquad \text{or simply:}$$

$$w_{i,t+1} = w_{i,t} + e + w_{i,t}r - w_{i,t}\Omega \qquad (1.3o)$$

Equation (1.3o) is the base equation for all the income models. Note that equation (1.3o) is for a single individual in the model.

Although this is a little different to the standard GLV equations quoted in section 1.2 above, it shares the same basic functions.

Firstly it is worth noting how simple this equation is. Here w is the only variable. e, r and $\Omega$ are all constants of one form or another, depending on the modelling used.

In future models e, r and $\Omega$ may be different constants for different individuals. However, in this first model, e and r are constant, and the same for all individuals.

$\Omega$ is slightly different. It is the same for all individuals, and is constant over the long term, but varies over the short term due to stochastic variation.

The second term on the RHS, the earned income e, provides a constant input that prevents individual values of wealth collapsing to zero. Note that this is additive, where in the models of Levy & Solomon in section 1.2 above this term was multiplicative.

The third term on the RHS is a multiplicative term and gives a positive feedback loop. The fourth term is also multiplicative and gives negative feedback.

In all the income models studied, the total income Y per time unit was fixed, and unless otherwise specified, the earned income was fixed equal to the returns income. So:

$$Y = \sum e_i + \sum \pi_i = \text{constant}, \quad \text{always} \quad (1.3p) \quad \text{and}$$

$$\sum e_i = \sum \pi_i = \frac{Y}{2} \quad \text{usually} \quad (1.3q)$$

So unless otherwise specified, the total returns to labour are equal to the total returns to capital. This last relationship; that total payments in salaries and total profits are similar in size is not outlandish. Depending on the level of development of an economy, the share of labour earnings out of total income can vary typically between 0.75 and 0.5.

Although the value appears to vary cyclically, in developed economies the value tends to be very stable in the region of 0.65 to 0.75. This was first noted by a statistician, Arthur Bowley a century ago, and is known as Bowley's Law, and represents as close to a constant as has ever been found in economics, figure 1.3.8 below gives an example for the USA. In developing economies, with pools of reserve subsistence labour, values can vary more substantially. Young gives a good discussion of the national income shares in the US, noting that the overall share is constant even though sector shares show long-term changes [Young 2010]. Gollin gives a very thorough survey of income shares in more than forty countries [Gollin 2002].

Figure 1.3.8 here
[St Louis Fed 2004]

We will come back to Bowley's Law in some depth in sections 1.6 and 4.5-4.8 as it turns out that Bowley's law is of some importance. Because of this importance, it is useful to define some ratios. We already have:

$$\text{Profit rate} \quad r = \frac{\sum \pi}{\sum w} \quad (1.3r)$$

Where profit can refer to any income from paper assets such as dividends, rent, coupons on bonds, interest, etc.

To this we will add:

$$\text{Income rate} \quad \Gamma = \frac{\sum Y}{\sum w} \quad (1.3\text{s})$$

which is the total earnings over the total capital. Here total earnings is all the income from wages and all the income from financial assets added together.

To these we add the following:

$$\text{Bowley ratio} \quad \beta = \frac{\sum e}{\sum Y} \quad (1.3\text{t})$$

$$\text{Profit ratio} \quad \rho = \frac{\sum \pi}{\sum Y} \quad (1.3\text{u})$$

These two define the wages and profit respectively as proportions of the total income. Following from the above, the following are trivial:

$$\beta + \rho = 1 \quad (1.3\text{v})$$

$$\text{Profit ratio} \quad \rho = \frac{r}{\Gamma} \quad (1.3\text{w})$$

Finally, in most of the following models, unless otherwise stated $\beta = \rho = 0.5$

Going back to equation (1.3o), at equilibrium, total income is equal to total consumption, so:

$$\sum w_{i,t+1} = \sum w_{i,t} \quad \text{so:}$$

$$\sum Y_{i,t+1} = \Omega \sum w_{i,t}$$

where $\sum Y_i$ is the total income from earnings and profit.

$$\bar{w} = \frac{\bar{Y}}{\Omega} \quad (1.3\text{x})$$

so the average wealth is defined by the average total income and the consumption rate.

There is an important subtlety in the discussion immediately above. In the original textbook economic model the total income and consumption are made equal by definition. In the models in this paper, income is fixed, but consumption varies with wealth. The negative feedback of the final consumption term ensures that total wealth varies automatically to a point where consumption adjusts so that it becomes equal to the income.

This automatically brings the model into equilibrium. If income is greater than consumption, then wealth, and so consumption, will increase until C=Y.

If income is less than consumption, the consumption will decrease wealth, and so consumption, until again, C=Y.

### 1.4 Wealth & Income Modelling - Results

#### 1.4.1   Model 1A   Identical Waged Income, Stochastic on Consumption

In the first model, Model 1A, the model starts with each agent having an identical wealth.

The distribution of earning power, that is the wages received e, is completely uniform. Each agent is identical and earns exactly 100 units of wealth per iteration.

The split between earnings to labour and earnings to capital are fifty-fifty, ie half to each.

The consumption of each agent is also identical, at an average of 30% of wealth. So 70% of wealth is conserved by the agent on average through the running of the model.

However the consumption of the agents is stochastic, selected from a normal range so that almost all the agents have a consumption rate between zero and 60% on each iteration.

So although the consumption of each agent is identical on average, consumption varies randomly from iteration to iteration. So an agent can consume a large amount on one iteration, followed by a small amount of consumption on the next iteration.

It is restated, in the very strongest terms, that all these agents are identical and indistinguishable.

The models were run for 10,000 iterations, the final results were checked against the half-way results, and this confirmed that the model quickly settled down to a stable distribution.

The results in figure 1.4.1.1 show the probability density function, showing the number of agents that ended up in each wealth band. This is a linear-linear plot. Also shown is the fit for the GLV function.

Figure 1.4.1.1 here

It can be seen that the data has the characteristic shape of real world wealth and income distributions, with a large body at low wealth levels, and a long declining tail of people with high levels of wealth.

As expected, the GLV distribution gives a very good fit to the modelling data.

Figure 1.4.1.2 shows the cumulative distribution for wealth for each of the agents in the model on a log-log plot. The x-axis gives the amount of wealth held by the agent, the y-axis gives the rank of the agents with number 1 being the richest and number 10,000 Being the poorest.

So the poorest agent is at the top left of the graph, while the richest is at the bottom right.

Figure 1.4.1.3 shows the top end of the cumulative distribution. It can be seen from figure 1.4.1.3 that there is a very substantial straight-line section to the graph for wealth levels above 1000 units. It can also be seen that this section gives a very good fit to a power law, approximately 15% of the total population follow the power law.

Figures 1.4.1.2 here

Figures 1.4.1.3 here

The earnings distribution for this model is uniform, so the Gini coefficient for the earnings is strictly zero.

The Gini coefficient for wealth however is 0.11. In this wealth distribution, the wealth of the top 10% is 1.9 times the wealth of the bottom 10%. The wealthiest individual has slightly more than four times the wealth of the poorest individual.

So the workings of a basic capitalist system have created an unequal wealth distribution out of an absolutely equal society.

This model, gives probably the most important result in this paper.

A group of absolutely identical agents, acting in absolutely identical manners, when operating under the standard capitalist system, of interest paid on wealth owned, end up owning dramatically different amounts of wealth.

The amount of wealth owned is a simple result of statistical mechanics; this is the power of entropy. The fundamental driver forming this distribution of wealth is not related to ability or utility in any way whatsoever.

In the first model, the random nature of changes in consumption / saving ensure that agents are very mobile within the distribution; individual agents can go from rags to riches to rags very quickly.

As a consequence, income changes are very rapid as they depend on the amount of wealth owned. So individual incomes are not stable. For this reason the distribution for income is not shown for model 1A.

### 1.4.2 Model 1B     Distribution on Waged Income, Identical Consumption, Non-stochastic

In model 1B, the characteristics of the agents are changed slightly.

Firstly, the agents are assumed to have different skills and abilities, and so different levels of waged income (it is also assumed the are being fairly rewarded for their work).

It is still assumed that all agents has an average earning power of 100, and the total split of earnings to capital is still 50%-50%.

However, prior to starting the model, each agent is allotted an earnings ability according to a normal distribution so earning ability varies between extremes of about 25 units and 175 units.

The worker retains exactly the same working ability throughout the model.

Meanwhile the saving propensity in this model is simplified. Throughout the running of the model, each agent consumes exactly 20% of its wealth. There is no longer a stochastic element for the saving, and all agents are identical when it comes to their saving propensity.

It should be noted that, although there is a random distribution of earning abilities prior to running the model, because this distribution is fixed and constant throughout the simulation, the model itself is entirely deterministic. This is not a stochastic model.

It turns out this model is in fact very dull. With equal savings rates the output distributions for wealth and income are exactly identical in shape to the input earnings distribution. All three distributions have exactly the same Gini coefficient.

### 1.4.3 Model 1C    Identical Waged Income, Distribution on Consumption, Non-stochastic

In model 1C, the characteristics of the agents are reversed to those in model 1B.

As with model 1A, the agents are assumed to have absolutely identical skills and abilities, and so identical levels of waged income.

It is again assumed that each agent has an earning power of exactly 100, and the total split of earnings to capital is still 50%-50%.

However, prior to starting the model, each agent is allotted a consumption propensity according to a normal distribution so average consumption rates are 20%, but vary between extreme values of 12% and 28%, while 95% of values fall between 16% and 24%. This is a much narrower range of consumption rates than model 1A with rates only varying plus or minus 20% from the normal rate for the vast majority of people. The big difference to model 1A is that each worker retains exactly the same saving propensity throughout the model, from beginning to end.

Again it should be noted that, although there is a random distribution of saving propensity prior to running the model, because this distribution is fixed and constant throughout the simulation, the model itself is entirely deterministic. This is not a stochastic model.

Figures 1.4.3.1 here

Figures 1.4.3.2 here

Figure 1.4.3.1 and 1.4.3.2 show the distributions of the wealth data. Figure 1.4.3.1 is the probability density function in linear-linear space while figure 1.4.3.2 is the cumulative density function in log-log space.

Again it can be seen that the GLV distribution fits the whole distribution, and that the tail of the distribution gives a straight line, a power law.

The fit to the GLV distribution is now less good, especially when compared with figure 1.4.1.1 for model 1A. This is because model 1C is not a 'true' GLV distribution. In the original GLV model described in sections 1.2 and 1.3, and modelled in model 1A, the consumption function was stochastic, and balanced out to a long-term average value. All the agents were truly identical. In model 1C the distribution of consumption is fixed at the outset and held through the model, the agents are no longer identical. As a result the underlying consumption distribution can influence the shape of the output GLV distribution. This is explored in more detail in sections 1.4.4 and 1.9.1.

In this model, because the consumption ratios are fixed and constant throughout, the hierarchy of wealth is strictly defined. The model comes to an equilibrium very quickly, and after that wealth, and so income, remain fixed for the remainder of the duration of the modelling run.

This allows a meaningful sample of income to be taken from the last part of the modelling run.

Figures 1.4.3.3 and 1.4.3.4 below show the pdf and cdf for the income earned by the agents in model 1C.

Figures 1.4.3.3 here

Figures 1.4.3.4 here

Figure 1.4.3.4 shows a very clear power law distribution for high earning agents. However figure 1.4.3.3 shows that a fit of the GLV distribution to this model distribution for income is very poor. This income distribution does not match the real life income distributions seen in section 1.1 above. There is a very good reason for this. This is most easily explained by going on to model 1D.

Not withstanding this, it is worth looking at some of the outputs of the model, compared to the inputs. The inputs are exactly equal earning ability; so a Gini index of zero, and a consumption propensity that varied between 0.16 and 0.24 for 95% of the population – hardly a big spread.

The outputs are a Gini index of 0.06 for income and 0.12 for wealth. The top 10% of the population have double the wealth of the bottom 10%, and the richest individual has more than six times the wealth of the poorest individual.

As with model 1A, near equality of inputs results in gross wealth differences on outputs.

### 1.4.4 Model 1D    Distribution on Consumption and Waged Income, Non-stochastic

In model 1D the distribution of wages is a normal distribution as in model 1B, however the distribution is narrower than that for model 1B. The average wage is 100 and the extremes are 62 and 137. 95% of wages are between 80 and 120. The Gini coefficient for earnings is 0.056 and the earnings of the top 10% is 1.43 times the earnings of the bottom 10%.

The distribution of consumption is exactly as model 1C.

Importantly the distributions of wages and consumption propensity are independent of each other. Some agents are high earners and big savers, some are high earners and big spenders, similarly, low earners can be savers or spenders.

As in models 1B & 1C, the earning and consumption abilities are fixed at the beginning of the model run and stay the same throughout. Again the model is deterministic, not stochastic.

Figures 1.4.4.1 here

Figures 1.4.4.2 here

Figures 1.4.4.1 and 1.4.4.2 show the distributions of the wealth data. Figure 1.4.4.1 is the probability density function in linear-linear space while figure 1.4.4.2 is the cumulative density function in log-log space.

Again it can be seen that the GLV distribution fits the whole distribution, and that the tail of the distribution gives a power law section. Again, as with model 1C, there are small variations from the GLV due to the influence of the input distributions.

In this model the hierarchy of wealth is strictly defined. The model comes to an equilibrium very quickly, and after that wealth, and so income, remain fixed for the remainder of the duration of the modelling run.

Figure 1.4.4.3 and 1.4.4.4 below show the pdf and cdf for the income earned by the agents in model 1D.

Figures 1.4.4.3 here

Figures 1.4.4.4 here

It can be seen that the GLV distribution gives a good fit to the curve, much better than that for model 1C. On the face of it the curve for income distribution appears to be a GLV and the power law tail is also evident.

However these assumptions are not quite correct.

The power law tail is a direct consequence of the income earned from capital. For the individuals who are in the power tail the amount of income earned from capital is much higher than that earned from their own labour, and the capital income dominates the earned income. So the power tail for income is directly proportional to the power tail for capital.

In the main body, things are slightly different. This is not in fact a GLV distribution. The income distribution is actually a superposition of two underlying distributions.

The first element of the income distribution is the investment income. This is proportional to the wealth owned. The wealth owned is a GLV distribution; as found above, so the distribution of investment income is also a GLV distribution.

The second element of income distribution is just the original distribution of earned income. This input was defined in the building of the model as a normal distribution. By definition the graph is a sum of the two components of Y, that is e for wage earnings, and $\pi$ for payments from investments. The full distribution of income is the sum of these two components.

This then explains why the income graph in model 1C fitted reality so badly. In model 1C the underlying earnings distribution was a flat, uniform distribution. This is highly unrealistic, so reality shows a different distribution.

In fact model 1D would have been better modelled with a log-normal distribution as figure 1.1.4.

Finally this model represents a more realistic view of the real world, with variations in both earning ability and consumption propensity. It is again worth looking at the outcomes for different individuals. Earnings ability varies by only plus or minus 20% for 95% of individuals in this model. Similarly consumption propensity only varies by plus or minus 20% for 95% of people.

Despite this the top ten percent of individuals earn more than twice as much as the poorest 10% and the most wealthy individual has 11 times the wealth of the poorest. The outputs give a Gini index of 0.082 for income and 0.131 for wealth.

## 1.5 Wealth & Income Modelling - Discussion

To start a discussion of the results above, it is worth firstly looking back at figure 1.4.4.2 above. There is a changeover between two groups in this distribution. The bottom 9000 individuals, from 1000 to 10,000 (the top quarter of the graph) are included in the main, curved, body of the distribution. The top 1000 individuals are included in the straight-line power tail. In this, very simple, model class segregation emerges endogenously.

The distribution has a 'middle class' which includes middle income and poor people; 90% of the population. This group of individuals are largely dependent on earnings for their income. Above this there is an 'upper class' who gain the majority of their income from their ownership of financial assets.

As discussed in 1.4.1 above, the rewards for this group are disproportionate to their earnings abilities, this is most obvious in model 1A where earnings abilities are identical.

In economic terms this is a very straightforward 'wealth condensation model'. The reason for this wealth condensation is due to the unique properties of capital. In the absence of slavery, labour is owned by the labourer. Even with substantial differences in skill levels, assuming approximately fair market rewards for labour, there is a limit to how much any single person can earn. In practice only a very limited number of people with special sporting, musical, acting or other artistic talents can directly earn wages many times the average wage, and in fact, such people can be seen as 'owning' monopolistic personal capital in their unique skills.

Capital however is different.
Crucially, capital can be owned in unlimited amounts.

And with capital, the more that is owned, the more that is earned. The more that is earned, then the more that can be owned. So allowing more earning, and then more ownership.

Indeed, in the absence of the labour term providing new wealth each cycle, the ownership of all capital would inevitably go to just one individual.

In the various income models above, the new wealth input at the bottom (due solely to earnings not capital) prevents the condensation of all wealth to one individual, and results in a spread of wealth from top to bottom. But this still results in a distribution with a large bias giving most of the wealth to a minority of individuals.

Going back to the Lotka-Volterra and GLV models discussed in section 1.2, it is better to abandon the predator-prey model of foxes killing rabbits, and instead think in terms of a 'grazing' model where the 'predators' are sheep and the 'prey' is grass. In this model the prey is not killed outright, but is grazed on, with a small proportion of its biomass being removed.

The wealth condensation process can then be thought of in terms of a complex multi-tier grazing model, a little analogous to the tithing model in medieval Europe.

In a simple tithing system, the peasants don't own the land, but are tied to the land-owners. They are allowed to work the land and keep a proportion of the crops grown. However they are obliged to pay a portion of the tithes to the lord of the manor, and also some to the church. The tithes form the rent payable for being allowed to use the land. The lord of the manor may be obliged to pay taxes to the local noble. The noble will be obliged to pay taxes to the king. As national institutions the church and king can gain substantial wealth, even with a relatively low tax, as they can tax a lot more people.

In a modern capitalist system things are similar but the payments are now disintermediated. People supply their labour to employers, and receive payments in wages as compensation. Payments to capital are returned in the form of interest on the owners of the capital. The more capital you have, the more return you get. The more capital you have, the bigger grazer you are in a near infinite hierarchy of grazers. The higher up you get the grazers get bigger but fewer.

So, to take an example, Rupert Murdoch is a fairly high level grazer as he owns many national newspapers and television stations, so many people make use of his business, and reward him with a small percentage of profit.

At the time of writing, Bill Gates is the apex grazer, because even Rupert Murdoch's companies use lots of computers with Windows software.

The more capital you have got, the more grazing you get to do.

That capital causes wealth to condense at high levels in this way is in fact a simple statement of the obvious. To the man on the street it is clear that the more money you have, the easier it is to make more, and the question of whether money that is gained by investment is 'earned' or justified remains open to debate.

The fact that paying interest unfairly benefits the rich has of course been noted by Proudhon, Marx, Gesell and others. For the same reasons usury was also condemned by the writers of Exodus, Leviticus and Deuteronomy. In these circumstances, the failure of mainstream economists to notice this basic problem with capitalism is puzzling.

The actual details of how the wealth is shared out is a consequence of entropy.

An understanding of entropy provides standard methodologies of counting possible states that a multi-body system can occupy. In the case of the GLV, this appears to be a consequence of 'path entropy' the number of different routes through a system that can be taken.

One of the profound things about entropy, and one of the reasons why it can be so useful, is that the statistical power of entropy can make microscopic interactions irrelevant. So important macroscopic properties of multi-body systems can be calculated without a knowledge of detailed microscopic interactions.

It is not proposed to discuss this in detail here; part B of the full paper discusses the concept and consequences of entropy in much more detail.

The essential point that needs to be understood at this point is that the GLV distribution is the only possible output distribution in this model because of simple statistical mechanical counting. No other output distribution is possible given the restraints on the system.

The invisible hand in this system is the hand of entropy.

As has been repeatedly noted, a GLV, complete with power tail, and gross inequality, can be produced from model 1A which uses absolutely identical agents.

In this regard, it is worth noting; and this is extremely important, some of the many things which are not needed to produce a wealth distribution model that closely models real life.

It is clear that to produce such a model, you don't need any of the following:

- Different initial endowments
- Different saving/consumption rates
- Savings rates that change with wealth
- Different earning potentials
- Economic growth
- Expectations (rational or otherwise)
- Behaviouralism
- Marginality
- Utility functions
- Production functions

In this equilibrium, utility theory is utterly irrelevant. The GLV distribution is a direct consequence of the power of entropy combined with the simple concept of a rate of return on capital. It is a full equilibrium solution, a dynamic equilibrium, but an equilibrium nonetheless.

In economic systems utility is not maximised. In fact it appears that there is an alternate maximisation process controlling economics, the maximisation of entropy production, and that this is of profound importance, this is discussed in 7.3 of the full paper.

The non-maximisation of utility of course has important consequences; the distributions of wealth and income dictated by the GLV are neither efficient or rational, never mind fair.

In real life human beings are not rewarded proportionally for their abilities or efforts.

It should be noted that, though constructed very differently, Wright's models produce similar results to my own, see full paper for more discussion.

### 1.6 Enter Sir Bowley - Labour and Capital

All the income models above were carried out using a 50%/50% split in the earnings accrued from capital and labour. So in all the previous models the profit ratio ρ and the Bowley ratio β are both equal to 0.5. In this section the effects of changing these ratios is investigated.

It was noted in model 1B that the input wage distribution, of itself, has no effect on the output distribution. That is to say; the input wage distribution is copied through to the output distribution. It is the consumption/savings ratios that generate the power tails and make things interesting. To keep things clearer, model 1C was therefore chosen, as this has a uniform wage distribution. This is less realistic, but makes analysis of what is happening in the model easier.

Reruns of the simulations were carried out for model 1C with varying proportions of returns to capital and labour. The profit ratio ρ; the ratio of returns to capital over total returns, was varied from 0 to 1, ie from all returns to labour to all returns to capital.

From the resulting distributions it was possible to calculate the Gini coefficients and the ratio of wealth/income between the top 10% and the bottom 10%.
The poverty ratio, the proportion of people below half the average wealth/income is also shown.

The data for this model is included in figure 1.6.1. The variation of Gini coefficients and poverty ratios with profit ratio are shown in figure 1.6.2. Figure 1.6.3 shows how the ratio of the top 10% to the bottom 10% changes with profit ratio.
The results are dramatic.

| Figure 1.6.1 | | | | | | | | | | | |
|---|---|---|---|---|---|---|---|---|---|---|---|
| **Profit Ratio** | **0.00** | **0.10** | **0.20** | **0.30** | **0.40** | **0.50** | **0.60** | **0.70** | **0.80** | **0.90** | **1.00** |
| Bowley Ratio | 1.00 | 0.90 | 0.80 | 0.70 | 0.60 | 0.50 | 0.40 | 0.30 | 0.20 | 0.10 | 0.00 |
| Gini coefficient wealth | 0.06 | 0.06 | 0.07 | 0.08 | 0.10 | 0.12 | 0.15 | 0.37 | 0.63 | 0.84 | 1.00 |
| Gini coefficient total income | 0.00 | 0.01 | 0.01 | 0.02 | 0.04 | 0.06 | 0.09 | 0.26 | 0.50 | 0.75 | 1.00 |
| decile ratio wealth | 1.43 | 1.49 | 1.57 | 1.68 | 1.84 | 2.09 | 2.58 | 7.81 | 22.68 | 67.31 | Inf |
| decile ratio income | 1.00 | 1.04 | 1.10 | 1.17 | 1.28 | 1.45 | 1.78 | 4.60 | 12.46 | 36.04 | Inf |
| poverty ratio wealth | 0.00 | 0.00 | 0.00 | 0.00 | 0.00 | 0.00 | 0.00 | 0.07 | 0.76 | 0.99 | 1.00 |
| poverty ratio income | 0.00 | 0.00 | 0.00 | 0.00 | 0.00 | 0.00 | 0.00 | 0.00 | 0.37 | 0.99 | 1.00 |

Figure 1.6.2 here

Figure 1.6.3 here

The model used is model 1C In which the earnings potential is a uniform distribution and so is equivalent for all individuals, that is all the agents have equal skills. However in model 1C savings rates are different for different agents. Clearly when all earnings are returned as wages $\rho = 0$, $\beta = 1$, and the Gini index is zero. In contrast, when all earnings are returned as capital, one individual, the one with the highest saving propensity, becomes the owner of all the wealth, and the Gini index goes to 1.

(From a profit ratio of 0.65 upwards, the Gini coefficient for wealth appears to vary linearly with the profit ratio, though the mathematics of this were not investigated.)

Figures 1.6.4 and 1.6.5 show the variation of the power exponent (which describes the power tail of the distribution) with the profit ratio.

| **Figure 1.6.4** | | | | | | | |
|---|---|---|---|---|---|---|---|
| Bowley Ratio | 1.00 | 0.90 | 0.80 | 0.70 | 0.60 | 0.50 | 0.40 |
| **Profit Ratio** | **0.00** | **0.10** | **0.20** | **0.30** | **0.40** | **0.50** | **0.60** |
| Power Tail Slope Wealth | na | -17.42 | -14.81 | -12.20 | -9.59 | -6.97 | -4.23 |

Figure 1.6.5 here

For very low and very high values of the profit ratio the power tail is not well defined, but for a range of values in the middle the results are mathematically very interesting.

For model 1C The relationship between alpha and the profit ratio $\rho$ is strikingly linear. If the plot is limited to the thirteen data points between 0.05 and 0.65 the $R^2$ value is 0.9979. If the plot is further restricted to the eleven points between 0.1 and 0.6 the $R^2$ value rises to 0.9999.

It appears that in this case there is a direct mathematical relationship between the Bowley Ratio and the $\alpha$ that defines the power tail in the GLV equation.

This relationship was investigated further by rerunning the model and varying the various parameters in the model systematically. The value of $\alpha$ was calculated in the model using the top 400 data points and the formula:

$$\alpha = 1 + n\left[\sum \ln(x_i/x_{min})\right]^{-1} \qquad (1.6a)$$

where n is 400, and the sum is from 1 to n.

The parameters available to change are as follows. Firstly the ratio of total income to total capital; that is the total income to both labour and capital (wages plus dividends) as a proportion of total capital, this was defined as the income rate, $\Gamma$, in equation (1.3s).

Secondly relative returns to labour and capital; that is either the profit ratio $\rho$ or the Bowley ratio $\beta$. Either can be used as they sum to unity.

Thirdly the average value of the consumption rate $\Omega$, and fourthly the variance of this consumption rate.

The first interesting thing to come out of this analysis was that the income rate, the ratio of total returns to total capital $\Gamma$ had no effect on $\alpha$ whatsoever.

The second attribute to drop out of the model was that seen in figure 1.6.5 above; for fixed values of the other parameters there was a substantial central section of the profit ratio $\rho$ for which (absolute) $\alpha$ declined linearly with increasing $\rho$.

Like the total returns, varying the absolute value of the consumption rate $\Omega$ had no effect whatsoever on the value of $\alpha$.

Although the absolute value of Ω had no effect on α, changing the variance of Ω had a significant effect. In this model Ω is distributed normally, and v is used to denote the matlab variance ($\sigma^2$) parameter compared to the total value of Ω.

In this model the value of α appears to vary as a power law of v. It should be noted that the value of v could only be increased from 0 to roughly 0.25. Around this value of 0.25 the outliers in the distribution of Ω become similar to the average size of Ω. This creates negative values of Ω for some individuals which results in no consumption, and so hyper-saving for these individuals. This is both unrealistic and results in an unstable model. (a better model would treat this as a new boundary condition.)

A first attempt at fitting of the data gave very good fits across the range of ρ and v using the following equation for (absolute) α:

$$\alpha = \frac{1.5}{v^{1.30}} - \frac{1.9\rho}{v^{1.07}} \qquad (1.6b)$$

The presence of power laws for v under both terms, with similar powers, was too tempting. So a second fit was attempted using a common denominator. This gave the equation below which gave a fit to the data almost as good as equation (1.6b):

$$\alpha = \frac{(1.37 - 1.44\rho)}{v^{1.15}} \qquad (1.6c)$$

now the two constants had moved suspiciously close together, so a further fit was carried out using a common constant, again this gave a data fit almost as good as (1.6b) and (1.6c):

$$\alpha = \frac{1.36(1-\rho)}{v^{1.15}} \qquad (1.6d)$$

Of course (1.6d) can more simply be written as:

$$\alpha = \frac{1.36\beta}{v^{1.15}} \qquad (1.6e)$$

Where β is of course the Bowley ratio.

Equations (1.6d) and (1.6e) are deceptively simple and appealing, and their meaning is discussed below in more detail.

Before this is done, it is worth stressing some caveats.

Firstly the two equations (1.6d) and (1.6e) have been extracted empirically from a model. They have not been derived mathematically. Neither have they been extracted from real data. Although it is the belief of the author that the equations are important and are sound reflections of economic reality, this remains solely a belief until either the equations are derived exactly or supporting evidence is found from actual economic data; or, ideally, both.

Secondly the nature of the two variables β and ν are different. The Bowley ratio is well known in economics and is an easily observed variable in national accounts. In contrast ν is the variance in an assumed underlying distribution of consumption saving propensity. In real economics the shape of such a distribution is highly controversial and is certainly not settled.

Thirdly, the two equations are limited by the parameters included in a highly simplified model. In real economies it is likely that other parameters will also effect α.

Finally, the two equations are for wealth, and do not fit the income data. A similar investigation was carried out to look at the variation of the α for the income distribution power tails. The results were much more complex, and beyond this authors mathematical abilities to reduce to a single equation. As with the wealth distributions, neither the total returns or the average value of the consumption ratio Ω had any effect on the value of α for income.

For any fixed value of ν, the absolute value of α declined with increasing ρ, however the decline appeared to be exponential rather than linear. Similarly for any fixed value of ρ the value of α appeared to decline exponentially with ν. Attempts to combine these facts together necessitated introductions of increasing numbers of terms and proved fruitless. Hopefully somebody with greater mathematical skills than myself should be able to illuminate this.

Despite this failure to extract a meaningful formulation, it is clear that increasing the value of the profit ratio ρ, or reducing the Bowley ratio β has a direct causal relationship on α resulting in reducing the absolute value of α for income, just as it does for the α for wealth.

This is of the utmost importance for the welfare of human beings in the real world.

It is of course trivially obvious that decreasing the Bowley ratio and increasing the profit ratio is bad for wealth and income distribution. If more income is moved to the small numbers of capital holders, at the expense of the much larger number of wage earners, then income distribution as a whole is going to get worse.

But equation (1.6d) shows that it is in fact much worse than that.

The α of the GLV defines the log law of differences in wealth for people in the power tail. As the absolute value of α decreases, inequality increases. Because α is the 'slope' of an inverse law curve (rather than say the slope of a straight line), small changes in α produce very large changes in distribution of wealth. Also by moving wealth around in the main body of the GLV, the α has a profound effect on the wealth and income of all people, not just the rich. The clear link between the Bowley ratio and the α's of the wealth and income distributions means that the changing value of the Bowley ratio has profound effects on the Gini index, relative poverty levels etc. Increasing returns to capital, at the expense of labour produces substantial feedback loops that increases poverty dramatically.

All of this of course begs the question of what exactly controls the values of the profit ratio ρ, the Bowley ratio β and the shape of the consumption rate distribution, so giving ν. The source of the Bowley ratio is discussed in detail in sections 4.5 to 4.8 of the full paper with what appears to be a straightforward derivation.

My answer to the source of ν is more tentative and more subjective, this will be introduced briefly below, but is discussed in more depth in section 7.3 of the full paper.

Before discussing the source of the consumption rate distribution, I would first like to return to equations (1.6d) and (1.6e):

$$\alpha = \frac{1.36(1 - \rho)}{v^{1.15}} \quad (1.6d)$$

$$\alpha = \frac{1.36\beta}{v^{1.15}} \quad (1.6e)$$

Although equation (1.6e) is simpler, equation (1.6d) is the key equation here. Indeed the more diligent readers; those who boned up on their power law background material, may have noted the strong resemblance of equation (1.6d) with the exponent produced from equation (45) in Newman [Newman 2005], which gives a general formula for $\alpha$ as:

$$\alpha = 1 - a/b \quad (1.6f)$$

Where a and b are two different exponential growth rate constants.

This is of course exactly what we have in equation (1.6d) where $\rho$ is the ratio of two different growth constants, r and $\Gamma$.

Going all the way back to equations (1.3h, 1.3p, 1.3v, 1.3s and 1.3w) $\rho$ is the ratio of the different components of Y, which are e and $\pi$.

The total income produced by capital, the amount of value created in each cycle, is given by the sum of wages and profits:

$$\text{Total Income} \quad \sum Y = \sum e + \sum \pi \quad (1.3p)$$

$$\text{Income rate} \quad \Gamma = \frac{\sum Y}{\sum w} \quad (1.3s)$$

The direct returns to capital; that is the returns to the owners of the capital, is given by the profit rate:

$$\text{Profit rate} \quad r = \frac{\sum \pi}{\sum w} \quad (1.3r)$$

but $\rho$ is defined by:

$$\text{Profit ratio} \quad \rho = \frac{\text{direct returns to capital}}{\text{total income from capital}}$$

$$\text{Profit rate} \quad r = \frac{\sum \pi / \sum w}{\sum Y / \sum w} \quad \text{so:}$$

$$\text{Profit ratio} \quad \rho = \frac{r}{\Gamma} \qquad (1.3w)$$

The value of ρ is simply the growth rate that capitalists get on capital, divided by the growth rate that everybody (capitalists and workers) gets on capital.

It is the combination of these two growth rates that creates and defines the power law tail of the wealth and income distributions. This is the first, and simplest class of ways to generate power laws discussed in Newman [Newman 2005].

And a curious thing has happened here.

There are many different ways to produce power laws, but most of them fall into three more fundamental classes; double exponential growth, branching/multiplicative models, and self-organised criticality.

The models in this paper were firmly built on the second group. The GLV of Levy and Solomon is a multiplicative model built along the tradition of random branching models that go back to Champernowne in economics and ultimately to Yule and Simon [Simkin & Roychowdhury 2006].

Despite these origins we have ended up with a model that is firmly in the first class of power law production, the double exponential model.

It is the belief of the author that this is because the first two classes are inherently analogous, and are simply different ways of looking at similar systems.

Much more tentatively, it is also the belief of the author that both the first two classes are incomplete descriptions of equilibrium states, and further input is need for most real systems to bring them to the states described by the third class; that of self organised criticality (SOC).

Going back to the wealth and income distributions, equation (1.6d) can define many different possible outcomes for $\alpha$. Even with a fixed Bowley ratio of say 0.7, it is possible to have many different values for $\alpha$ depending, in this case, on the value of v.

It is worth noticing that there is a mismatch between the values for $\alpha$ given by the models and economic reality. The models give values of $\alpha$ of 4 and upwards for both wealth and income. In real economies the value of alpha can vary in extreme cases can between 1 and 8, but is typically close to a value of 2 see for example Ferrero [Ferrero 2010]. While the model clearly needs work to be calibrated against real data, it is the belief of the author that the relationship between $\alpha$ and ρ or β is valid and important.

It is the belief of the author that in a dynamic equilibrium, the value of $\alpha$ naturally tends to move to a minimum absolute value, in this case by maximising v to the point where the model reaches the edge of instability. At this point, with the minimum possible value of $\alpha$ (for any given value of ρ or β) there is the most extreme possible wealth/income distribution, which, it is the belief of the author is a maximum entropy, or more exactly a maximum entropy production, equilibrium. This belief; that self-organised criticality is an equilibrium produced by maximum entropy production, is discussed in more detail in section 7.3 below.

It is the suspicion of the author that the unrealistic distribution for $\Omega$ used in the modelling approach above results in a point of SOC, that is artificially higher than that in real economies. Indeed, it is a suspicion that movement towards SOC may of itself help to define underlying distributions of earnings and consumption. This is returned to in section 7.4.

### 1.7 Modifying Wealth and Income Distributions

The modelling above shows that grossly unequal distributions of wealth and income are produced as a natural output of statistical mechanics and entropy in a free market society.

In particular, the ownership of capital and the function of saving are key to the formation of inequality in wealth and income distributions.

In the following two sections alternative approaches look at how wealth and income distributions might be modified, given the knowledge that these distributions are formed in a statistical mechanical manner. The first approach looks at imposing boundary conditions on a model of society, the second looks at modifying the saving mechanism feedback loop.

#### 1.7.1 Maximum Wealth

The use of a maximum wealth barrier was found to be ineffective for poverty reduction. For more detail see full paper.

#### 1.7.2 Compulsory Saving

The second approach for changing income distributions focuses on the crucial role of saving in the GLV equation. From models 1B and 1C it appears that rates of consumption and saving are critical to the formation of the power tail and so large wealth inequalities. If saving is the problem, it seems sensible to use saving as the solution.

Again model 1D was used as the base model.

In this model a simple rule was introduced. If any agent's current wealth was less than 90% of the average wealth, that agent was obliged to decrease their consumption rate by 20 percent. This could be thought of an extra tax on these individuals, which is automatically paid into their own personal savings plan. It should be noted that this increase, though significant, is not enormous, and is comparable say to the rate of VAT/income tax in many European countries.

Figure 1.7.2.1 here

Figure 1.7.2.2 here

Figures 1.7.2.1 and 1.7.2.2 show the log-log and log-linear cumulative distributions for the model, with and without the compulsory saving rule.

It can be clearly seen in figures 1.7.2.1 and 1.7.2.2 that the number of poor people is much smaller with compulsory saving. For the bottom half of the agents (the top half of figure 1.7.2.2), the distribution is very equal, though it retains a continual small gradient of wealth difference.

The top half of society retains a very pronounced power-law distribution, with approximately the same slope. Each individual in the top half is less wealthy by an amount that varies from roughly 5% for those in the middle to roughly 10% for those at the top. Despite this they remain far richer than the average. This drop in wealth seems a very slight price to pay for the elimination of poverty, and the likely associated dramatic reduction in crime and other social problems. The power tail structure would leave in place the opportunity for the gifted and entrepreneurial to significantly better themselves.

Figure 1.7.2.3 shows various measures of equality with and without the saving rule.

| Figure 1.7.2.3 | No Compulsory Saving | Compulsory Saving |
| --- | --- | --- |
| Gini Earnings | 0.056 | 0.056 |
| Gini Wealth | 0.131 | 0.077 |
| Gini Income | 0.082 | 0.058 |
| Earnings Deciles Ratio | 1.429 | 1.429 |
| Wealth deciles ratio | 2.268 | 1.617 |
| Income deciles ratio | 1.686 | 1.451 |

The results are dramatic and also very positive.

Without compulsory saving the input earnings distribution was magnified through saving in the GLV into a more unequal distribution for wealth and income. This can be seen in both the Gini indices and also the ratio of the wealth or income of the top 10% to the bottom 10%.

With compulsory saving the output distribution for income has almost the same inequality values as the original earnings distribution for both the Gini index and deciles ratio. Wealth is more unequal, but much less so than in the model without compulsory saving.

In fact the shapes of this output income distribution (in figures 1.7.2.1 & 2 above) is significantly different in shape to the input earnings distribution, which in this case is a normal distribution. But by smoothing out the rough edges of the GLV, compulsory saving provides an output that is similar in fairness to the skill levels of the inputs. This is probably a distribution that society could live with.

In practice poverty has been eliminated for all except those that combine a very poor earnings ability with a very poor savings rate – individuals who in real life would be necessarily be candidates for intervention by the social services.

It is also worth noting the form in which this transfer of wealth takes place.

In this model the rich are not taxed.

In this model the poor are compelled to save.

The rich would only notice this form of financial redistribution in the form of increased competition for the purchase of financial assets.

In practice a compulsory saving scheme would be highly effective once the new, more equal, distribution was in place. However expecting people who are currently very poor to save their way out of poverty is not reasonably realistic.

In section 1.8 of the full paper more detailed proposals are made for modifying wealth and income distributions; based on the outcomes of the models above. It is hoped that these proposals will provide solutions that are more practical, effective and far less costly than current mechanisms such as welfare and subsidised housing.

## 11. The Logic of Science

In their abstract to 'Worrying trends in econophysics' Mauro Gallegati, Steve Keen, Thomas Lux and Paul Ormerod wrote:

*'Our concerns are fourfold. First, a lack of awareness of work that has been done within economics itself. Second, resistance to more rigorous and robust statistical methodology. Third, the belief that universal empirical regularities can be found in many areas of economic activity. Fourth, the theoretical models which are being used to explain empirical*

*phenomena. The latter point is of particular concern. Essentially, the models are based upon models of statistical physics in which energy is conserved in exchange processes. There are examples in economics where the principle of conservation may be a reasonable approximation to reality, such as primitive hunter–gatherer societies. But in the industrialised capitalist economies, income is most definitely not conserved. The process of production and not exchange is responsible for this. Models which focus purely on exchange and not on production cannot by definition offer a realistic description of the generation of income in the capitalist, industrialised economies.'* [Gallegati et al 2006]

I am slightly embarrassed to admit that, due to both time constraints and limited experience in econometrics, the present paper remains significantly remiss with regard to the second criticism above.

But, then again, to rephrase Ernest Rutherford; if you need to use statistics to prove your theory, you ought to have thought of a better theory.

In the event of some party choosing to award me remuneration for my ongoing research I would hope to remedy these shortcomings in future papers.

However I believe the present paper has come a long way in answering the other criticisms.

In particular, I believe criticisms one and four have been fully addressed in this paper.

I believe however that the authors' third criticism is fundamentally flawed.

It is the nature of science that a field can appear complex and difficult to make any sense of until a significant insight can bring sudden clarity. It has taken time for physicists to bring this clarity to economics, but to physicists, the multi-body nature of economic and financial systems meant that the belief that universal empirical regularities would be explained was only a matter of time. It is this insight that drove Champernowne half a century ago. It is this insight that resulted in Wright, Souma & Nirei and myself independently producing similar models near simultaneously.

Indeed the '*universal empirical regularities'* pooh-poohed by Gallegati et al where always there.

Economics has systematically treated such persistent 'anomalies' as anomalies, ignoring raw data while retreating into the comforts of intellectual hypothesis, whether this be neoclassical, Keynesian, Marxian, behavioural or other.

This behaviour is the behaviour that has kept economics as a branch, to be generous, of political philosophy. It is this behaviour, understood intuitively by the general public, and explicitly by natural scientists, that is responsible for the very low regard that both have for economics as a science.

It is precisely by investigating 'anomalous' but persistent data outputs that the natural sciences have progressed. By definition, if data output is persistent, it is not 'anomalous'. If data output is persistent, it is normal. It may be 'anomalous' with regard to current theory. But that simply makes the current theory by definition 'anomalous', not the data. In these circumstances the theory must be abandoned, not the data.

Einstein, for example, is usually characterised as a theoretical physicist. But his biggest single insight (amongst many) was to treat the experimental fact of the constancy of the speed of light as a given. From this he abandoned 'common sense' and simply worked out the mathematical consequences of this fact. Thus was relativity born.

Science can not be built simply on common sense, intuition and intellectual rigour. Science must start with the observed facts if it is to make progress. This, at a much deeper level than that intended by Jaynes, is the logic of science.

For any multi-body system, entropy has to be the guiding force, it has taken time for physicists and mathematicians to get to the root of this, mainly because the entropy was dynamic path entropy rather than static state entropy, but the driving power of entropy in economics is immediately obvious to anybody who has a passing understanding of entropy.

## 11.1 Afterword

As noted in the introduction, this paper is a condensed extract of the full paper 'Why Money Trickles Up'. The full paper applies the same basic model to explain the power tail seen in company size distributions, it also provides models for booms and busts in commodity prices and macroeconomic business cycles. The full paper explains the Bowley ratio; the ratio of returns to labour and capital. The full paper also contains extensive background material on chaos, statistical mechanics, entropy and heterodox economics and finance. The abstract and paper structure for the full paper are given below in section 11.2. The full paper is available at econodynamics.org.

## 11.2 Abstract & Structure of full paper 'Why Money Trickles Up'

**Abstract**

This paper combines ideas from classical economics and modern finance with Lotka-Volterra models, and also the general Lotka-Volterra models of Levy & Solomon to provide straightforward explanations of a number of economic phenomena.

Using a simple and realistic economic formulation, the distributions of both wealth and income are fully explained. Both the power tail and the log-normal like body are fully captured. It is of note that the full distribution, including the power law tail, is created via the use of absolutely identical agents.

It is further demonstrated that a simple scheme of compulsory saving could eliminate poverty at little cost to the taxpayer. Such a scheme is discussed in detail and shown to be practical.

Using similar simple techniques, a second model of corporate earnings is constructed that produces a power law distribution of company size by capitalisation.

A third model is produced to model the prices of commodities such as copper. Including a delay to capital installation; normal for capital intensive industries, produces the typical cycle of short-term spikes and collapses seen in commodity prices.

The fourth model combines ideas from the first three models to produce a simple Lotka-Volterra macroeconomic model. This basic model generates endogenous boom and bust business cycles of the sort described by Minsky and Austrian economists.

From this model an exact formula for the Bowley ratio; the ratio of returns to labour to total returns, is derived. This formula is also derived trivially algebraically.

This derivation is extended to a model including debt, and it suggests that excessive debt can be economically dangerous and also directly increases income inequality.

Other models are proposed with financial and non-financial sectors and also two economies trading with each other. There is a brief discussion of the role of the state and monetary systems in such economies.

The second part of the paper discusses the various background theoretical ideas on which the models are built.

This includes a discussion of the mathematics of chaotic systems, statistical mechanical systems, and systems in a dynamic equilibrium of maximum entropy production.

There is discussion of the concept of intrinsic value, and why it holds despite the apparent substantial changes of prices in real life economies. In particular there are discussions of the roles of liquidity and parallels in the fields of market-microstructure and post-Keynesian pricing theory.

Structure of the Paper

Part A of this paper discusses a number of economic models in detail, Part A.I discusses a number of straightforward models giving results that easily accord with the real world and also with the models of Ian Wright. Part A.II discusses models that are more speculative.

Part B discusses the background mathematics, physics and economics underlying the models in Part A. The mathematics and physics is discussed in Part B.I, the economics in part B.II, the conclusions are in part B.III. Finally, Part C gives appendices.

Within Part A; section 1 discusses income and wealth distributions; section 1.1 gives a brief review of empirical information known about wealth and income distributions while section 1.2 gives background information on the Lotka-Volterra and General Lotka-Volterra models. Sections 1.3 to 1.5 gives details of the models, their outputs and a discussion of these outputs.

Section 1.6 discusses the effects that changing the ratio of waged income to earnings from capital has on wealth and income distributions.

Sections 1.7 and 1.8 discuss effective, low-cost options for modifying wealth and income distributions and so eliminating poverty.

Finally, section 1.9 looks at some unexplained but potentially important issues within wealth and income distribution.

Sections 2.1 to 2.4 go through the background, creation and discussion of a model that creates power law distributions in company sizes.

Sections 3.1 to 3.4 use ideas from section 2, and also the consequences of the delays inherent in installing physical capital, to generate the cyclical spiking behaviour typical of commodity prices.

Sections 4.1 to 4.4 combine the ideas from sections 1, 2 and 3 to provide a basic macroeconomic model of a full, isolated economy. It is demonstrated that even a very basic model can endogenously generate cyclical boom and bust business cycles of the sort described by Minsky and Austrian economists.

In section 4.5 it is demonstrated that an exact formulation for the Bowley ratio; the ratio of returns to labour to total returns, can easily be derived from the basic macroeconomic model above, or indeed from first principles in a few lines of basic algebra.

In section 4.6 and 4.7 the above modelling is extended into an economy with debt. From this a more complex, though still simple, formulation for the Bowley ratio is derived. This formulation suggests that excessive debt can be economically dangerous and also directly increases income inequality. The more general consequences of the Bowley ratio for society are discussed in more depth in section 4.8.

In section 4.9 two macroeconomic models are arranged in tandem to discuss an isolated economy with a financial sector in addition to an ordinary non-financial sector. In section 4.10 two macroeconomic models are discussed in parallel as a model of two national economies trading with each other.

To conclude Part A, section 4.11 introduces the role of the state and monetary economics, while section 4.12 briefly reviews the salient outcomes of the modelling for social equity.

In Part B, section 6.1 discusses the differences between static and dynamic systems, while section 6.2 looks at the chaotic mathematics of differential equation systems. Examples of how this knowledge could be applied to housing markets is discussed in section 6.3, while applications to share markets are discussed in section 6.4. A general overview of the control of chaotic systems is given in section 6.5.

Section 7.1 discusses the theory; 'statistical mechanics', which is necessary for applying to situations with many independent bodies; while section 7.2 discusses how this leads to the concept of entropy.

Section 7.3 discusses how systems normally considered to be out of equilibrium can in fact be considered to be in a dynamic equilibrium that is characterised as being in a state of maximum entropy production. Section 7.4 discusses possible ways that the statistical mechanics of maximum entropy production systems might be tackled.

Moving back to economics; in section 8.1 it is discussed how an intrinsic measure of value can be related to the entropy discussed in section 7 via the concept of 'humanly useful negentropy'.

Section 8.2 discusses the many serious criticisms of a concept of intrinsic value in general, with a discussion of the role of liquidity in particular.

Section 9.1 looks at theories of supply and pricing, the non-existence of diminishing returns in production, and the similarities between the market-microstructure analysis and post-Keynesian pricing theory. Section 9.3 looks for, and fails to find, sources of scarcity, while section 9.4 discusses the characteristics of demand.

In section 10 both the theory and modelling is reviewed and arranged together as a coherent whole, this is followed by brief conclusions in section 11.

Sections 12 to 16 are appendices in Part C.

Section 12 gives a history of the gestation of this paper and an opportunity to thank those that have assisted in its formation.

Section 13 gives a reading list for those interested in learning more about the background maths and economics in the paper.

Section 14 gives details of the Matlab and Excel programmes used to generate the models in Part A of the paper.

Sections 15 and 16 give the references and figures respectively.

## 12. History and Acknowledgements

Between 1980 and 1982 I was taught A-level physics by Malcolm Ruckledge using the innovative Nuffield Foundation Physics course. This was a powerful combination of an outstanding teacher with outstanding material. The section on statistical mechanics was particularly well written and taught, and gave me an early and profound intuitive insight into the power and simplicity of entropy. I suspect this paper would not have been written without this insight.

Sometime in my first year studying physics at the University of Manchester, in 1992/3, while looking at a picture of the Maxwell-Boltzmann distribution of molecular velocities on a blackboard, it occurred to me that wealth in a society was shared out in a similar manner; a lot of people with a little wealth and a few with a lot of wealth. It further occurred to me that the underlying systems, involving a lot of freely interacting particles/individuals, where fundamentally similar. At the time I imagined this was a unique and very clever insight, however it turned out that a lot of other physicists and mathematicians have had similar insights, some preceding mine by many decades.

After this, nothing very much happened for a decade or so, though the idea refused to go away, and being by nature an engineer at heart, I thought a lot about how income and wealth inequality might be tackled as well as to why it exists.

In 2003 I had a letter published in the New Scientist. This encouraged me to take my ideas more seriously, and while working abroad in 1995 I had the opportunity to write down my ideas at that stage into a fairly amateurish paper.

On returning to the UK I circulated the paper around various individuals I thought might be interested. The paper was greeted on a spectrum that largely went from disinterest through to derision.

One exception was Michael Stutzer, who suggested I forward it to Duncan Foley, with whom I had a brief but very rewarding correspondence. I remain very thankful to both these individuals and especially to Duncan Foley for encouraging my work even when it was at this very early and amateurish stage.

After this nothing very much happened again for some years, as I lacked the skills, in both economics and mathematics to take the work forward. I did however read a paper by Ayres & Nair 'Thermodynamics and economics' which I found very useful in linking the concept of entropy to the economic concept of value.

This changed in August 2000 when, via the New Scientist, I discovered the work of Bouchaud & Mézard and other researchers, primarily physicists but also some heterodox economists, working in the new field of econophysics. The majority of the work was in the field of asset pricing in finance, but there was also a parallel stream looking at wealth and income distributions.

Over the next few years I attended a number of econophysics and related conferences where I learned a lot more about both the maths and economics from the other participants.

During this period I was given support and guidance, from Steve Keen, Thomas Lux and others, but most particularly from Juergen Mimkes, for which I would like to give thanks. Thomas Lux gave me some very useful insight into the real meaning of value and wealth that helped to generate the ideas in this paper. Steve Keen gave interesting discussions on economics and also pointed me in the direction of James Galbraith who was also very supportive.

As stated in the introduction I met Wataru Souma at the Econophysics of Wealth Distributions conference at Kolkata in 2005. I almost certainly attended his lecture on his paper 'Universal Structure Of The Personal Income Distribution'. I found Souma & Nirei's model complex and difficult to follow, and did not knowingly use it further.

Judging from the pile of papers that I rediscovered it in; it appears that I read Ian Wright's 'The Social Architecture of Capitalism' some time shortly after the Kolkata conference. I remember reading this paper quite clearly, as the style of the paper was unusual. The paper is very strongly a modelling paper, with very little formal mathematical content. This resulted in my finding it very difficult to make much sense of, and in fact I didn't understand the paper until some years later. I also, at the time, found the Marxian approach very naïve and off-putting, particularly in the insistence on the use of the labour theory of value. This seemed to me plainly wrong; so at this stage I dumped this paper in the 'irrelevant' pile and forgot about it. That was a big mistake.

In 2006 it was suggested to me that the general Lotka-Volterra distribution might make a good fit to some high quality income data I had acquired from the UK Statistical Office. It turned out that the data did fit the GLV exceptionally well; better than alternative distributions.

As a scientist, this dictated that building models along the lines of the GLV would be the most sensible way forward.

By this stage, my knowledge of economics had expanded a little, and I was somewhat dismayed by the naivety and complexity of the approaches taken to economics by most physicists. It seemed to me that power law distributions, and gross inequality, had a universality through geography and more importantly history (cf the paper regarding inequality in ancient Egypt [Abul-Magd 2002]), and that they appeared to be valid in any society where wealth, including land, was traded.

This could be contrasted with, for example, community owned land systems in Africa, which though associated with general poverty appeared to be characterised by low levels of inequality. In my view any model for wealth distributions should be able to accommodate payments to capital in the broadest sense, whether this be via dividends, interest rates, or rent on land and property.

With this in mind I attempted to fit, in the simplest way possible, basic economic concepts to two different generating equations that I was aware were capable of producing GLV distributions. These two systems were the exchange system of Slanina and the GLV system of Levy & Solomon. I wrote a note and circulated it to a number of academics in early 2006, I have reproduced the note in full below in section 12.1.

Unfortunately, none of the academics proved interested in my proposals. Also unfortunately, I did not send the note to Wright or Souma & Nirei, as it had been some time since I read their papers, and I didn't consciously connect them to this present work.

I lacked the mathematical or programming skills to take this forward, so once again, nothing much happened for a few years.

In 2009, in the middle of the post-credit-crunch recession, I took the opportunity to start an MSc in Finance at Aston University. Due to some very unfortunate circumstances I was unable to complete the course.

However in the two terms I attended the course I acquired a lot of useful knowledge regarding basic finance and economics. I would also like to give thanks to Patricia Chelley-Steely for giving me important insights into the role of market-microstructure in general and liquidity in particular.

I was also able to gain invaluable assistance from George Vogiatzis and Maria Chli with regard to producing simulations of my models proposed in 2006. The exchange model proved difficult to construct. However, in March 2010 Maria and George produced the first Matlab model for me based on the GLV model in the second part of my 2006 note. Somewhat to my surprise, this produced a perfect GLV distribution on its first run, though no power law.

It turned out that, to generate the power law, the profit ratio had to be increased substantially from the initial 5% proposed to somewhere near 50%. A little investigation revealed that the returns to capital where indeed on this scale, and so this was realistic.

At this point George and Maria politely, but firmly, suggested that I conquer my technophobia and learn to program in Matlab myself. I followed their advice and discovered that it is a lot easier than other programming languages I had encountered. From the first programme, I produced all the other programmes in this paper in short order, with almost all programming work being done in May 2010. I remain deeply indebted to Maria and George for their initial assistance and support with this work.

The income model followed naturally from the wealth model. The companies model followed naturally from the wealth and income models. The commodity model followed naturally from the companies model.

During the modelling process I was rereading Steve Keen's 'Debunking Economics' and had also read some of the Goodwin modelling work while investigating the ratio of returns to labour and capital. It seemed to me that by combining the wealth, company and commodity models it would be possible to generate a much simpler but effective Goodwin style macroeconomic model. This proved to be the case, with a resultant simple base model that appeared to produce Minskian/Austrian cycles endogenously.

At some point after the modelling was largely complete, while rereading a large volume of papers I had collected over the years, I reread Wright's 'The Social Architecture of Capitalism'. For the second time I found it difficult to follow, and found the labour theory of value difficult to accept. However something in the paper was nagging at me. I reread the paper for a second time, more carefully; and slowly realised that, though coming from a completely different angle, Wright had built a model that was both making the same base assumptions as my own, and producing many of the same outputs. Indeed, in many ways Wright's models produced better results than my own.

Given the very different ways that Wright and myself produced our models, I believe that my approach was not influenced by Wright. My original proposals of 2006 were deliberately, mathematically based on the GLV, and were also focused on a financial sector with returns paid on capital. Wright's models are significantly different to my own, most notably in not involving a financial sector. Also, unlike the present paper Wright takes a 'black box' and 'zero intelligence' approach to modelling which eschews formal fitting of the models to mathematical equations.

Despite this belief, I am obliged to accept that I may have been influenced subconsciously by Wright's work.

Much later in the writing of this paper, close to it's completion, I reread the work of Souma & Nirei. Again I found the complexity of the mathematical approach of Souma & Nirei very difficult to follow, and I believe this complexity is unnecessary, and that my own approach is more useful as a basis for analysing economics. However the parallels between their work and my own are significant. Most notably Souma & Nirei use consumption as a dissipative part of their model in a way that is almost identical to my own models.

They also use capital as a main source of new wealth in their model, which is analogous to my own, though less strongly than with consumption. Souma & Nirei use capital growth as the main form of supplying new wealth to their model. They justify this by using supporting data from the Japanese economy. While this may have seemed sensible at the time, given the collapse of the Japanese stock-market and property prices over the last two decades, this now looks less sensible. Although I believe that capital growth can form a part of wealth generation, on a long-term cyclical basis this is likely to be very small. I believe that my simple model of returns to capital in the form of interest, dividends and rent is a better basis for future economic modelling.

As with Wright, I do not believe I was influenced directly by Souma & Nirei. My first model in 2006 was a simple exchange model, quite different to that of Souma & Nirei, while I generated the second model by simply substituting what to me were the most obvious and simple economic terms into Levy & Solomon's generating equation. Indeed my original model was a little over-complex and significantly different to that of Souma & Nirei.

However, even more so than Wright's work, the parallels between the models of Souma & Nirei and my own are striking. And the possibility that I was subconsciously assisted by their work seems significant.

I would like to state in the strongest terms that I believe that the work of Wright, Souma & Nirei is of considerable importance. These three academics have been able to bridge the gap between the physics and the economics in a way that no other academics have been able to. Also they all carried out this work prior to my own.

Where my own work differs to that of the gentlemen above is that it has a clear mathematical basis, unlike that of Wright, and that the mathematical basis is dramatically simpler than that of Souma & Nirei.

It is my hope that Wright, Souma, Nirei and myself can share the credit for finally bringing an effective mathematical and modelling approach to the understanding of economics.

## 12.1 Proposed Models 2006

Pair exchange process, after Slanina;

$W_{i,t+1} = W_{i,t} + \beta_{ij} - \beta_{ji} - p_i + r * W_{i,t}$

$W_{j,t+1} = W_{j,t} + \beta_{ji} - \beta_{ij} - p_j + r * W_{j,t}$

$\beta_{ij}$ would be a good or service received,
$\beta_{ji}$ would be money exchanged for the good or service,
(or vice versa) you could make this more 'economist friendly' by using:
$\beta_{gs}$ for a good or service received,
$\beta_m$ for money exchanged for the good or service,

typically $\beta_{ij}$ would be a factor smaller than $W_{j,t}$ in size

$\Delta\beta = \beta_{ij} - \beta_{ji}$
is a small random difference in wealth due to the exchange not being exactly equal, typically $\Delta\beta$ would be a few percent of $\beta_{ij}$ (economists would argue that $\Delta\beta$ would be equal to zero at equilibrium, I believe this is not the case, however it is much easier just to argue that there will be small random differences in the wealth exchange, which is a very plausible assumption) I see the $\Delta\beta$'s as the main stochastic driver in this model.

$p_i$

is the profit taken by a third party. If I buy a car directly off you, then $p_i$ equals zero, but if I buy a car off you via e-bay, a small percentage of $β_{ij}$; $p_i$ and/or $p_j$ is taken by e-bay. (In e-bay's case, the seller is charged, so $p_i$ = 0). Ignoring the example of e-bay, I would initially model this by assuming that all $p_i$'s are a fixed small percentage of the exchange. So:

$p_i = β_{gs} * p_{rate}$

r
is the interest rate (factored down to a weekly or daily rate, whatever Δt is) Annual real interest rates (after inflation) are very stable, varying between 0.5 and 4% (annual) over long time periods. I would also initially model this as a small fixed percentage. (To get a working model with equations that balance it may be necessary to have a fixed relationship between and $p_{rate}$ and r ;
$p_{rate}$ = const * r )

I do not see any reason to make the r 's a distribution set. Most peoples investments are stable, poor peoples especially so. Rich people will only hold a portion of investments in riskier, more variable funds. I would only really see a need to introduce a distribution set if it was the only way we could generate the necessary curve.

So in this model the change in wealth comes from a small random element from the exchange, a small element taken in profit, and a small gain of interest which, crucially, is proportional to current wealth.

From a max entropy type approach I would then add the following two conditions:

$Σ W_{i,t} = Σ W_{i,t+1}$

ie, all wealth is conserved (ie. there is no economic growth or recession).

And:

$Σ p_i = Σ r * W_{i,t}$

ie, all profit is recycled as interest on peoples wealth.

In this model the stochastic variability comes from the wealth exchanges; the Δβ's. This combined with the assumption of conservation of wealth would provide a boltzmann type distribution if profits $p_i$ and interest r were equal to zero.

I believe the extra terms of profit and interest will be a circular reinforcing mechanism that should produce the power tail.

If you can solve, this or something similar, hopefully you will get a wealth distribution that is a GLV with alpha = 1.5

GLV type process;

$W_{i,t+1} = W_{i,t} + Inc_i * Δt − p_{Inc} − Con_i * Δt − p_{Con} + r * W_{i,t}$

$Inc_i$
is waged income; income from employment. Realistically I would expect this to be a stable distribution, very much on the lines of Juergen's arguments. (http://arxiv.org/abs/cond-mat/0204234)

$p_{Inc}$

is the small profit taken by the employing organisation. Modelled as previous model.

$Con_i$
is consumption, which includes food, clothes, new cars, petrol, rent#, mortgage payments#, holidays, etc. (# not completely sure about these two). Consumption is the big variable, and is where I would expect the stochastic element to come in strongly.

$p_{Con}$
is the small profit taken by the shop, landlord, building society, etc.

r
As previous model.

Again, from a max entropy type approach, I would then add the following two conditions:

$$\Sigma W_{i,t} = \Sigma W_{i,t+1}$$

again, all wealth is conserved.

And:

$$\Sigma (p_{Inc} + p_{Con}) = \Sigma r * W_{i,t}$$

Again; all profit is recycled as interest.

From this equation you can derive something like:

Total Income = $I_j$

$I_j = [ Inc_i + ( r * W_{i,t} / \Delta t ) ] = [$ wages + interest, etc $]$

$I_j = [ Con_i + ( [ (W_{i,t+1} - W_{i,t}) + (p_{Inc} + p_{Con}) ] / \Delta t ) ]$

If you can solve this or something similar, hopefully you will get an income distribution that is a GLV with alpha = 4 to 5.

## 13. Further Reading

See full paper for details.

## 14. Programmes

See full paper for details.

## 16. Figures

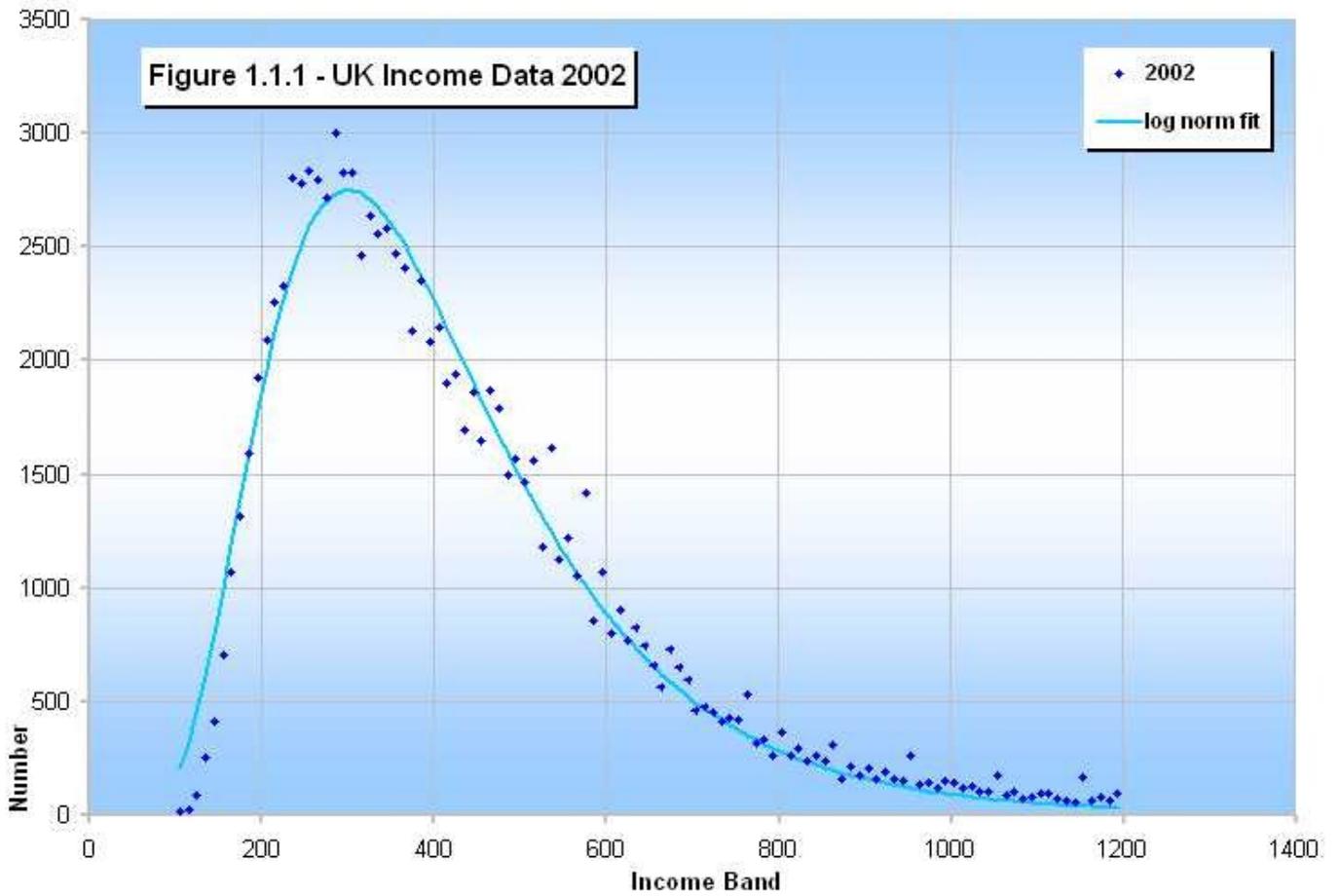

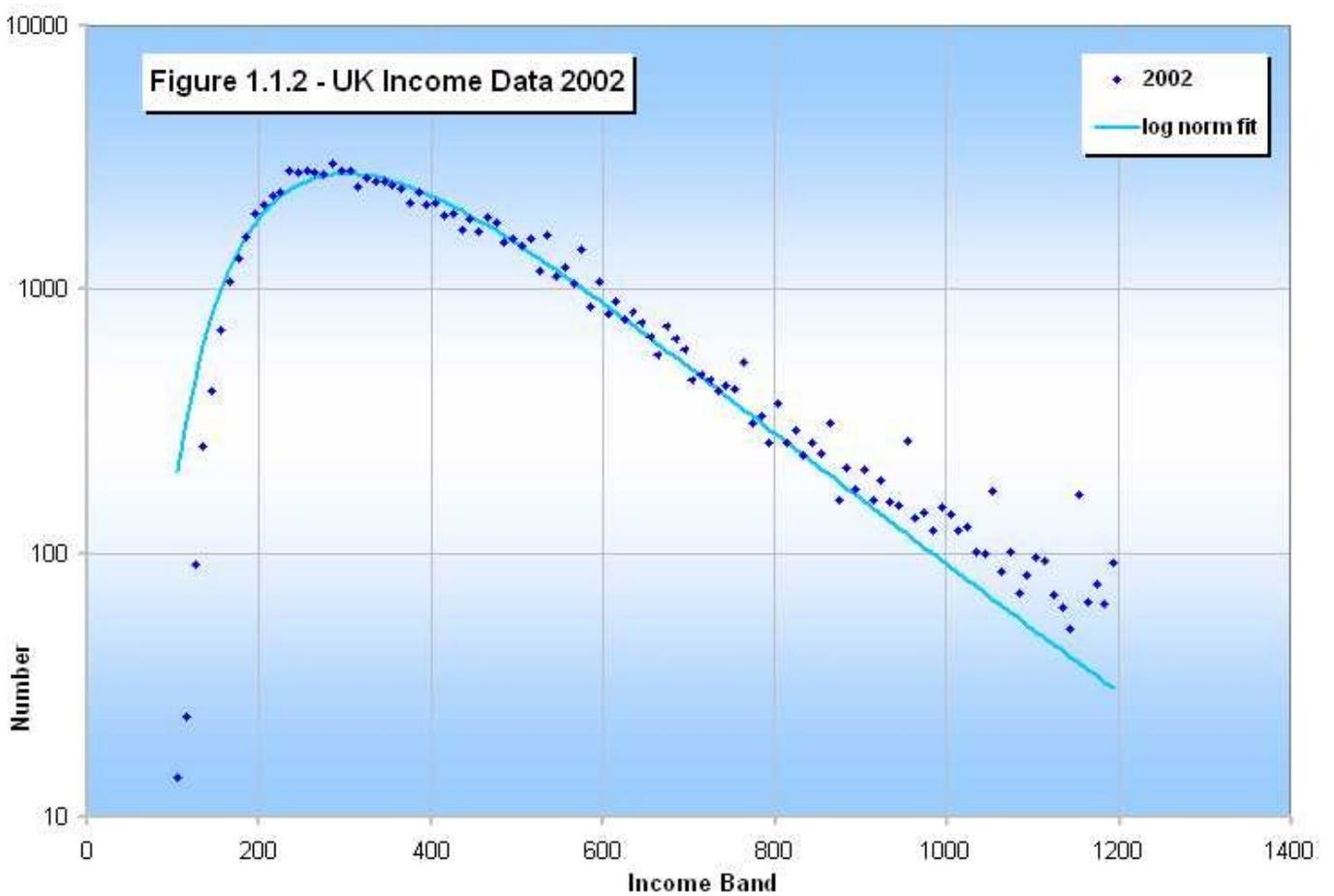

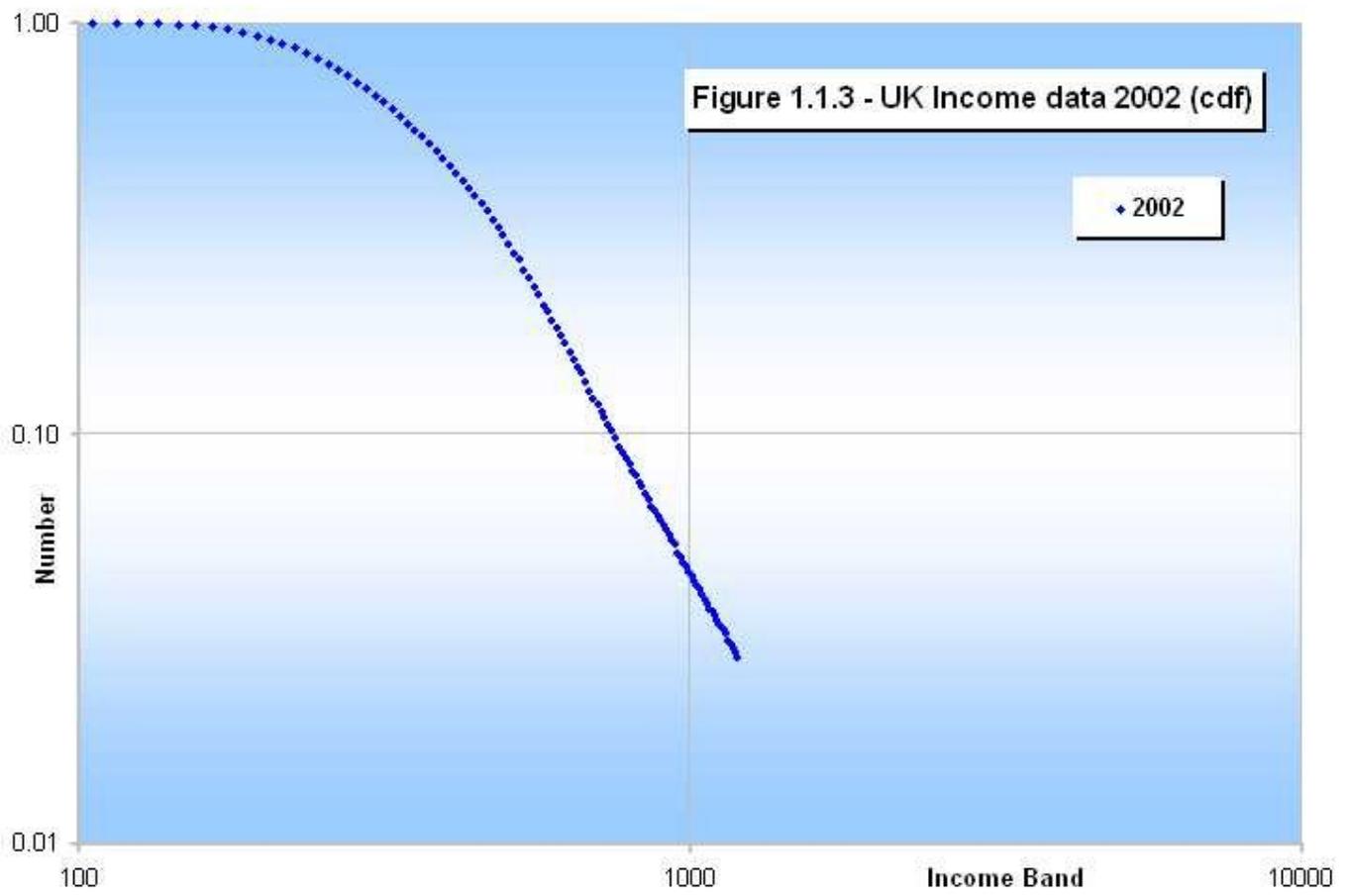

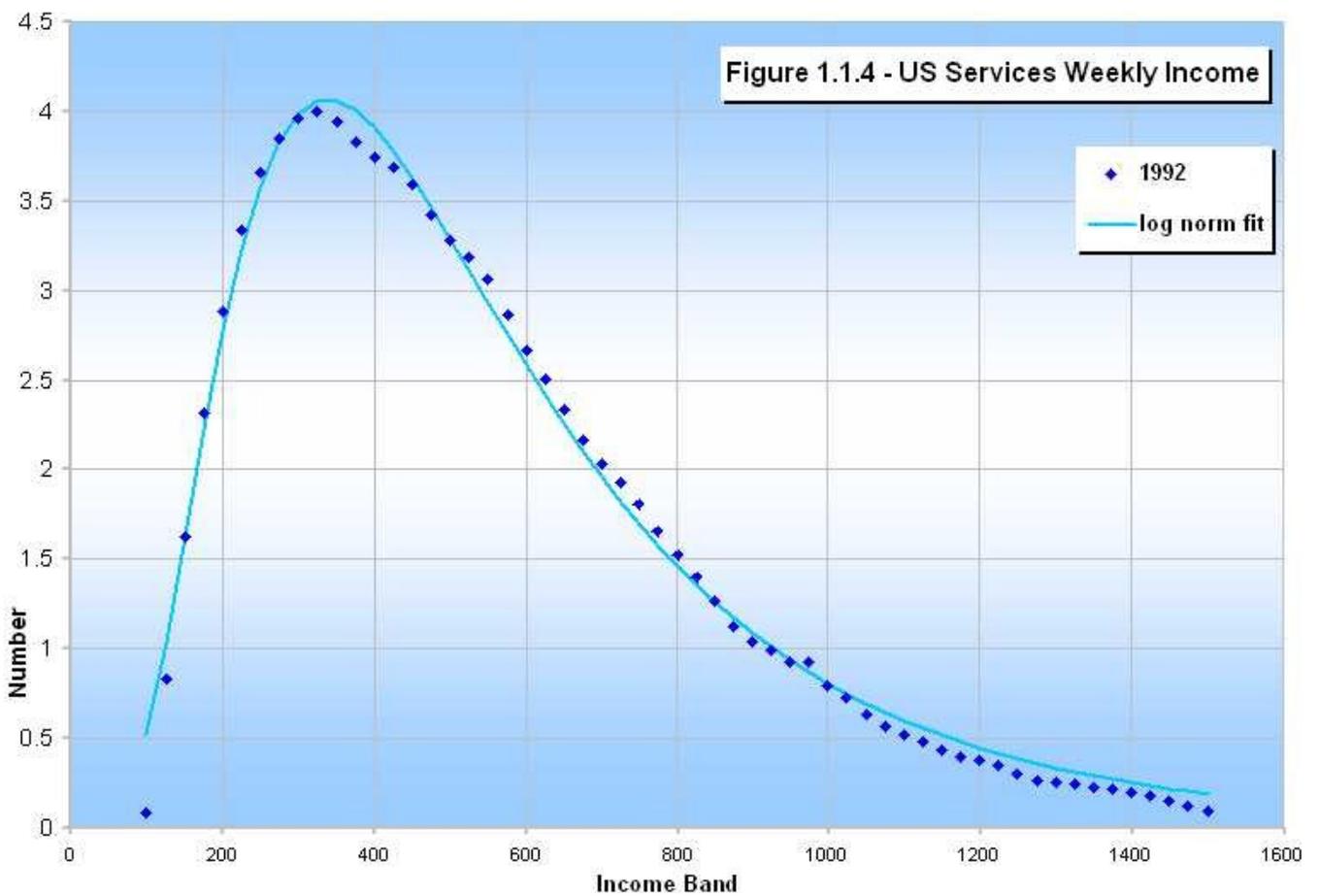

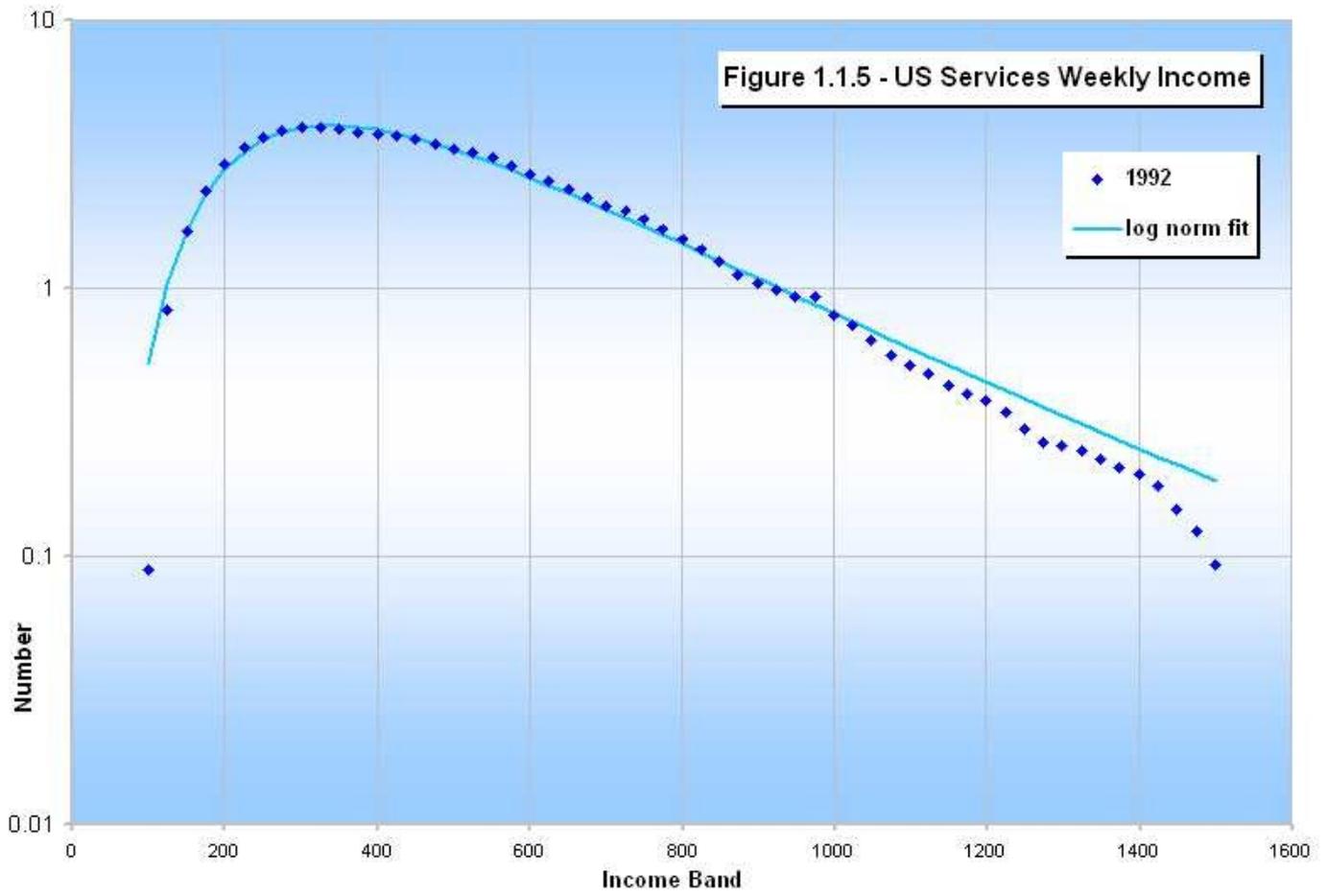

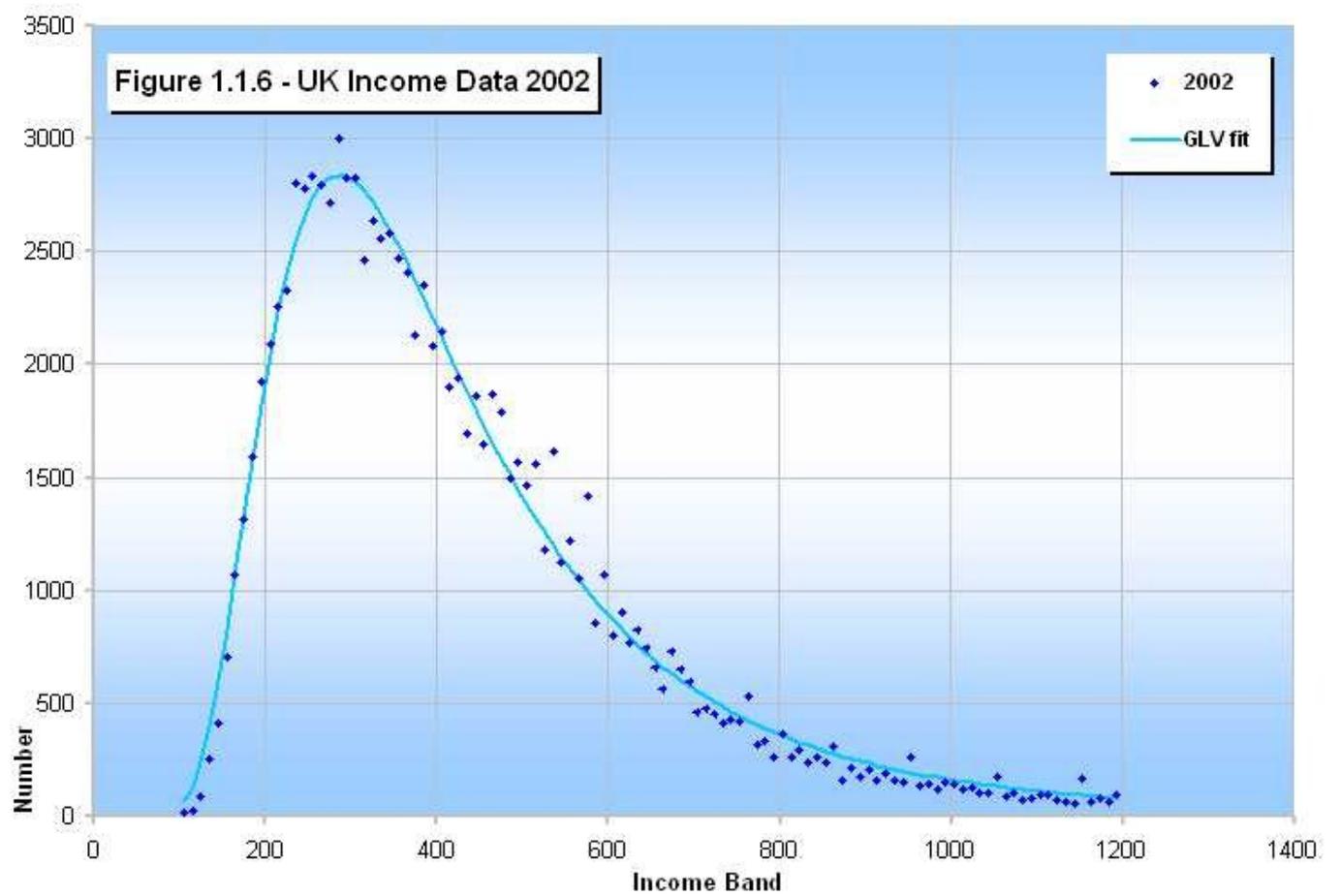

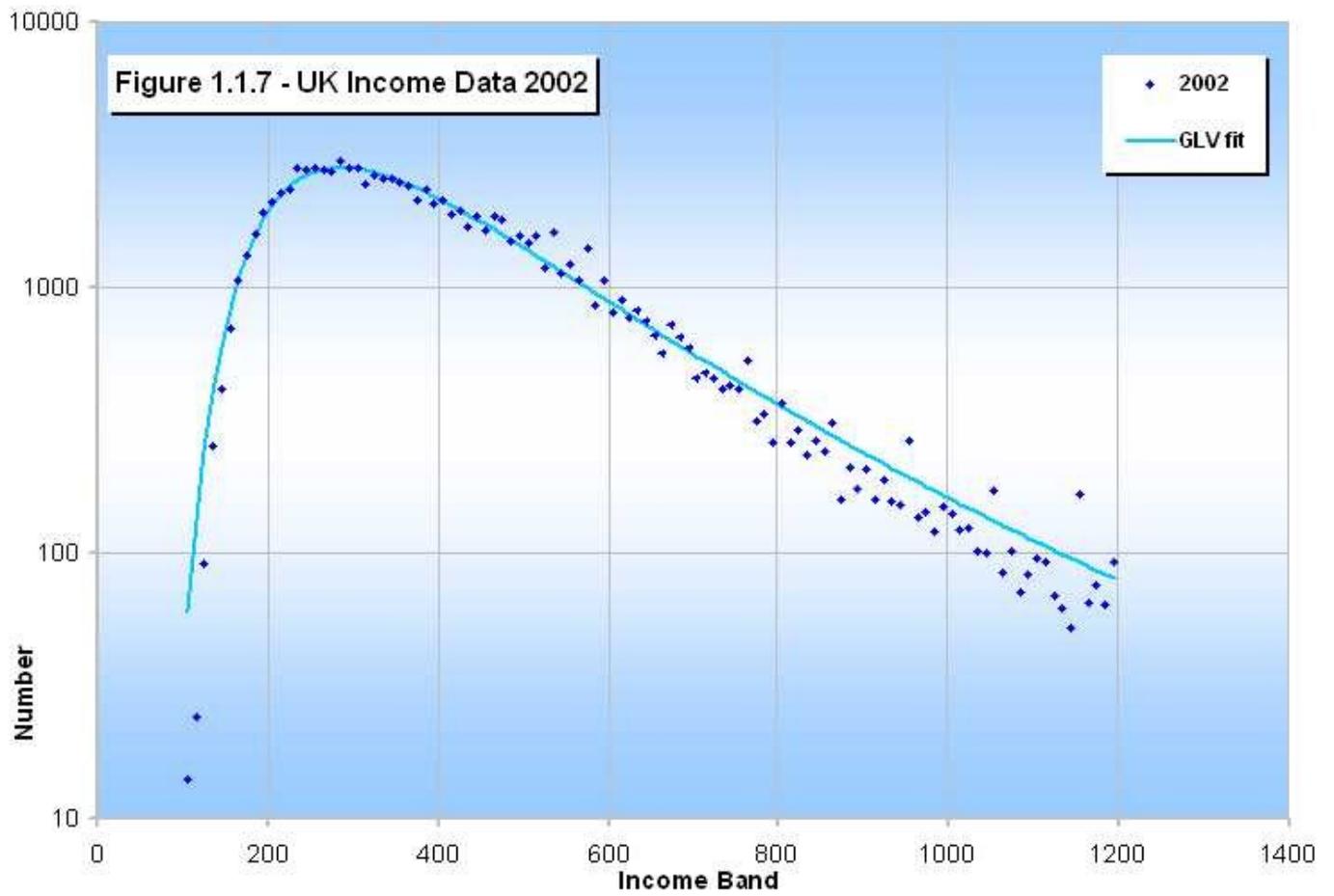

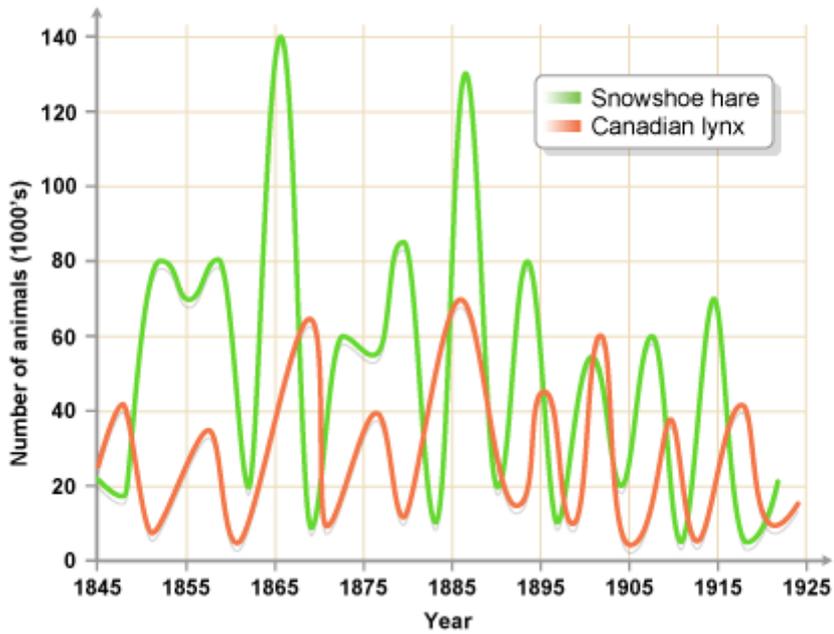

Figure 1.2.1.1

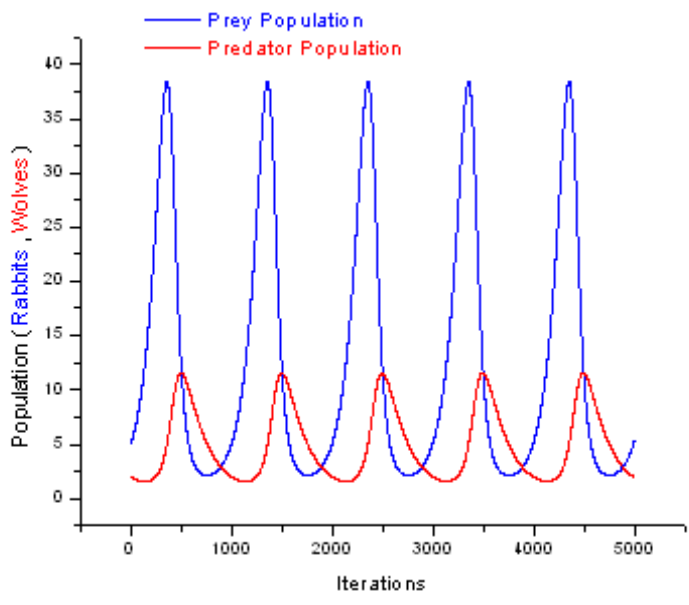

Figure 1.2.1.2

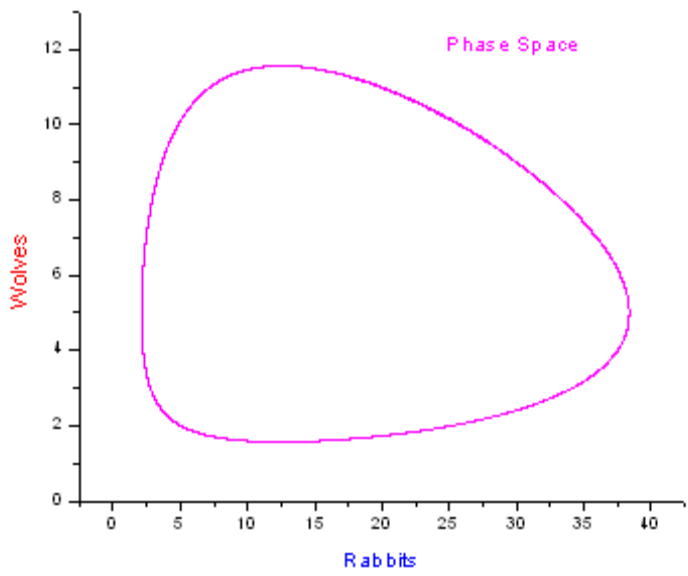

Figure 1.2.1.3

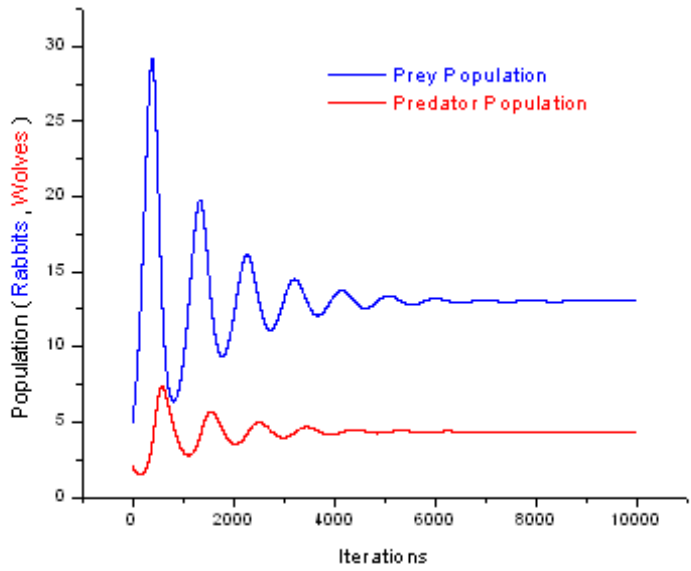

Figure 1.2.1.4

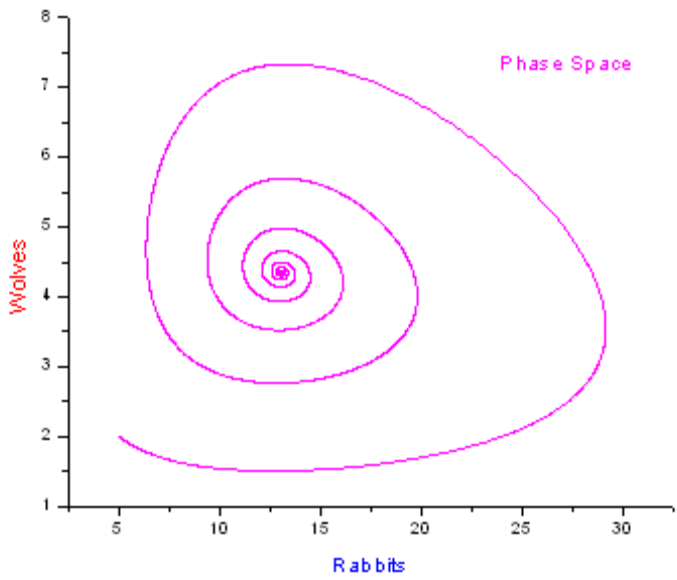

Figure 1.2.1.5

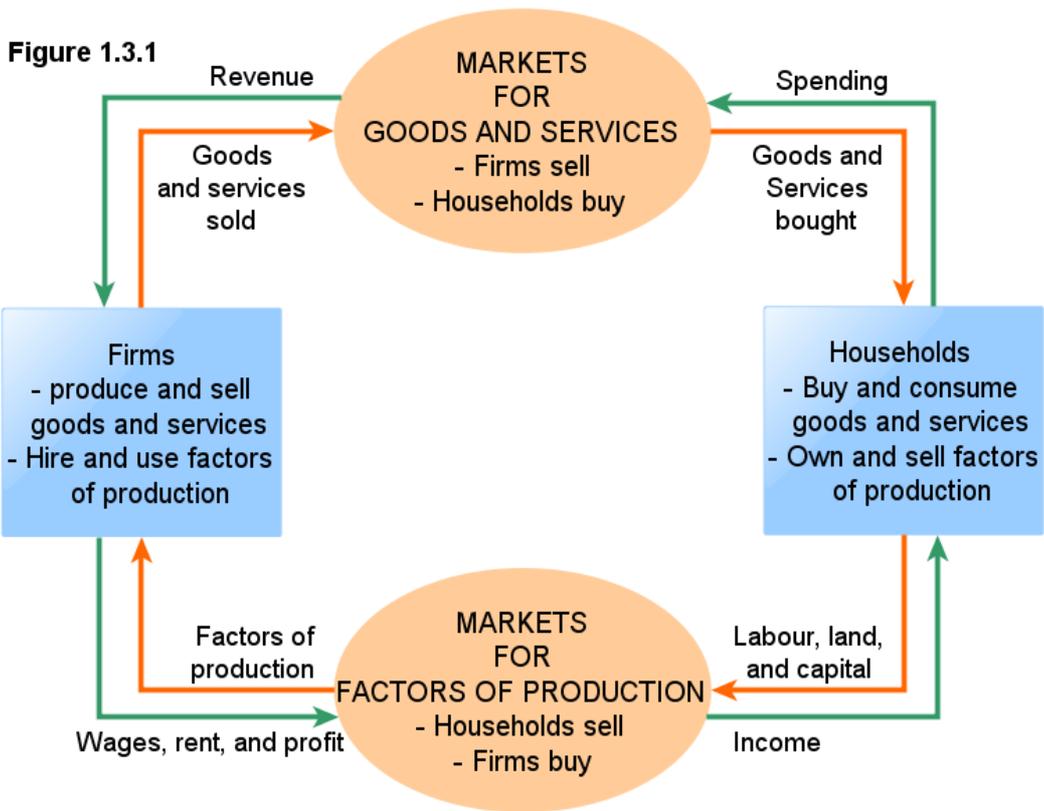

Figure 1.3.1

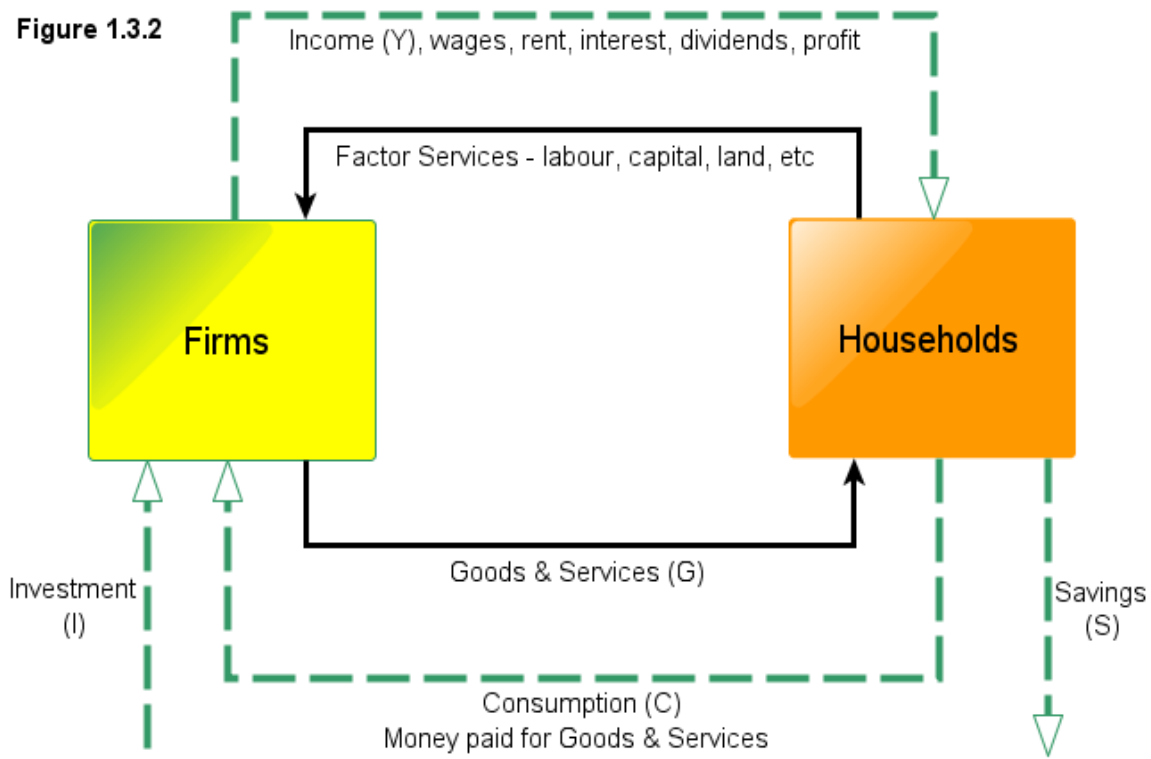

Figure 1.3.2

**Figure 1.3.3**

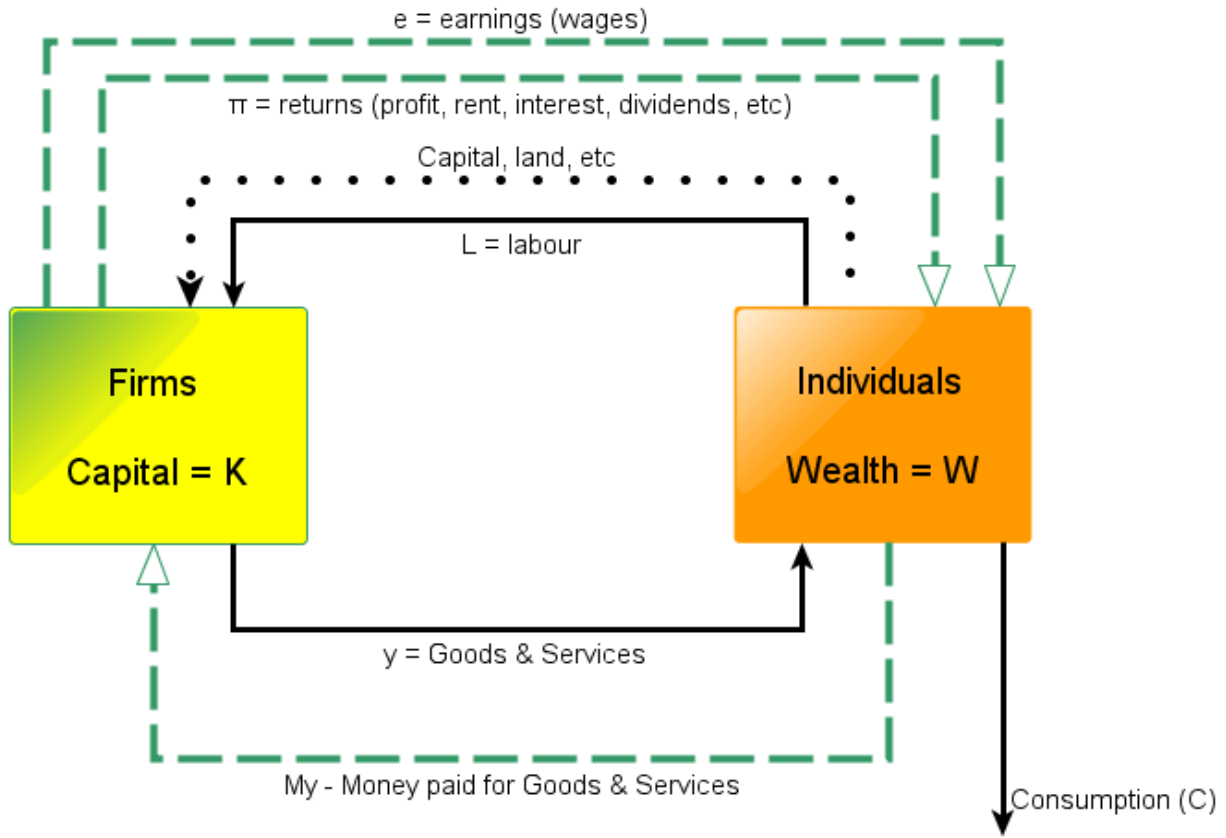

**TABLE 14.1** The Financing of Investment: Flow-of-funds Estimated (%) (1970–1994)

|  | Germany | Japan | UK | USA |
|---|---|---|---|---|
| Internal finance | 78.4 | 69.9 | 95.6 | 94.0 |
| Bank finance | 12.0 | 30.1 | 15.0 | 12.8 |
| Bond finance | −1.0 | 3.4 | 3.8 | 15.3 |
| New equity | −0.02 | 3.4 | −5.3 | −6.1 |
| Other | 10.6 | −6.8 | −9.1 | −16.0 |

*Note*: Internal finance comprises retained earnings and depreciation. The other category includes trade credit and capital transfers. The figures represent weighted averages where the weights for each country are the level of real fixed investment in each year in that country.

*Source*: Corbett and Jenkinson, "How Is Investment Financed?" *The Manchester School* (1996) vol. LXV, pp. 69–94.

Figure 1.3.4

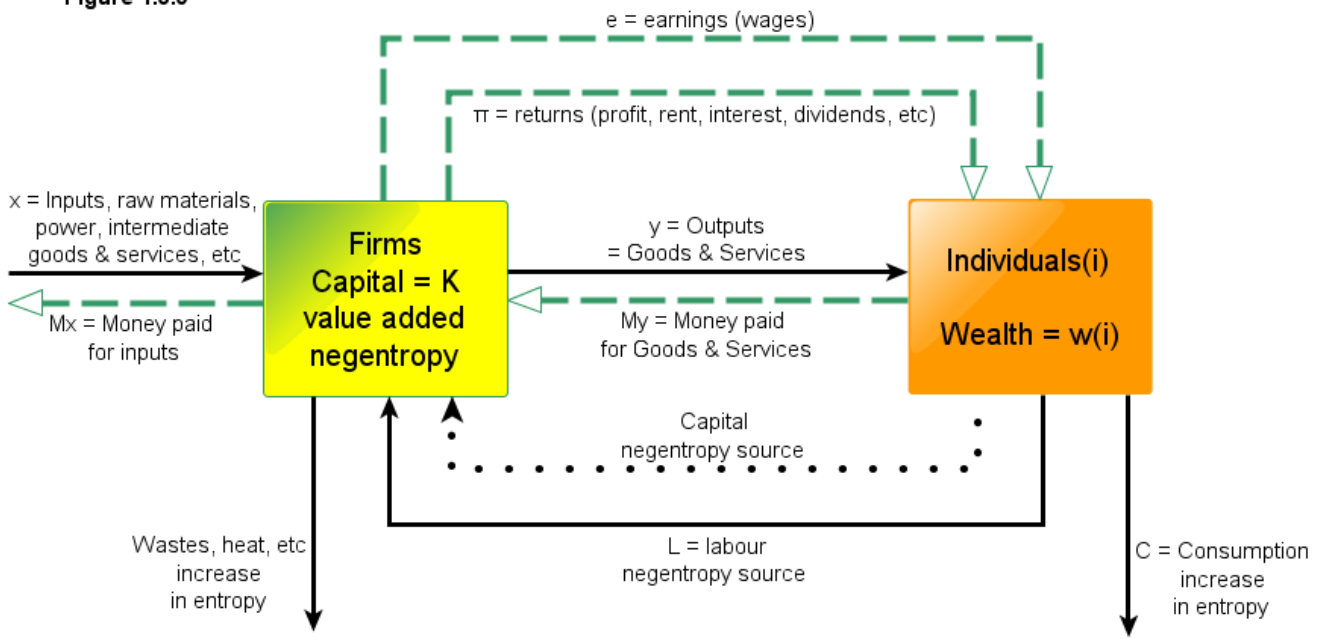

Figure 1.3.5

Figure 1.3.6

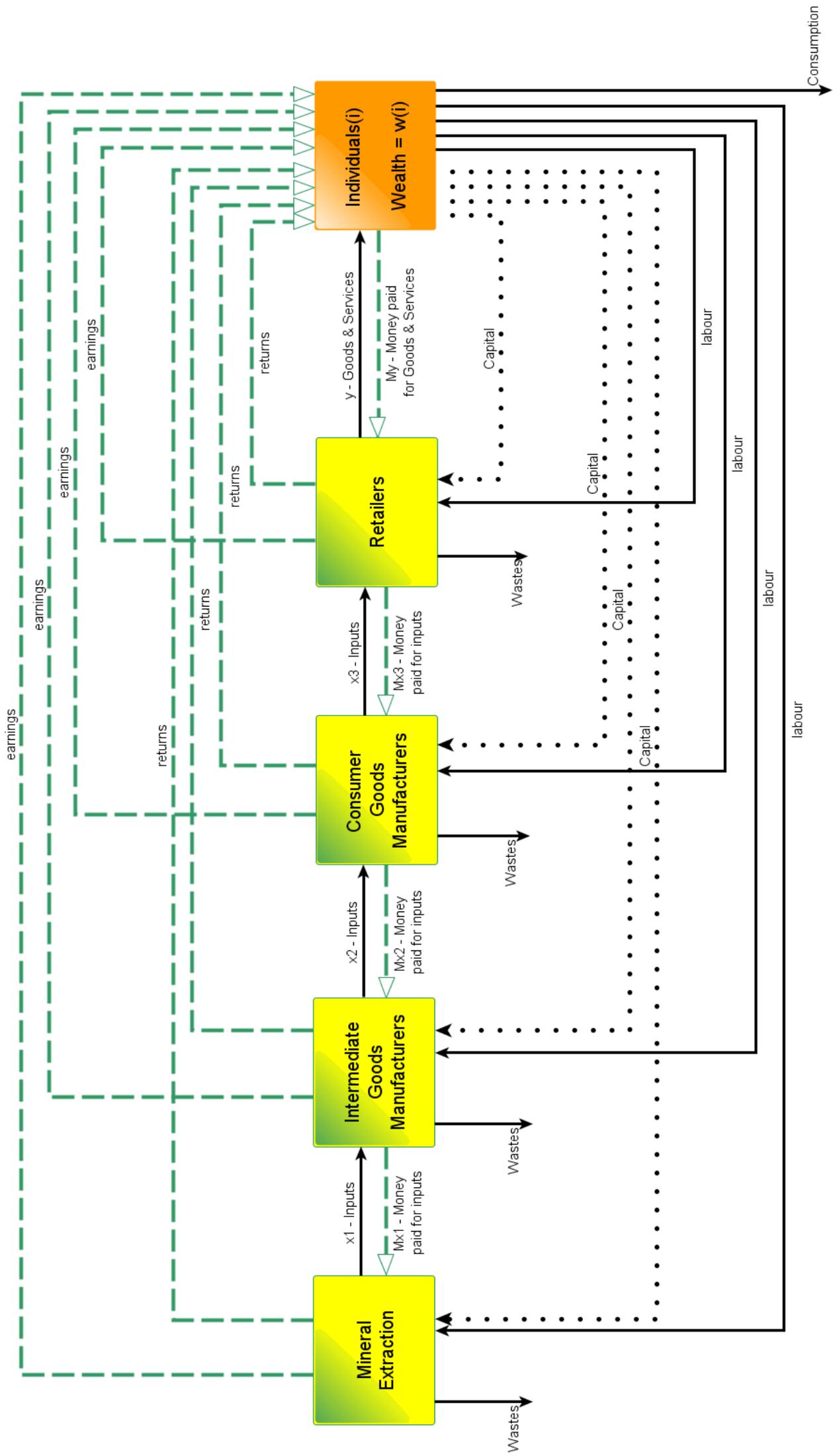

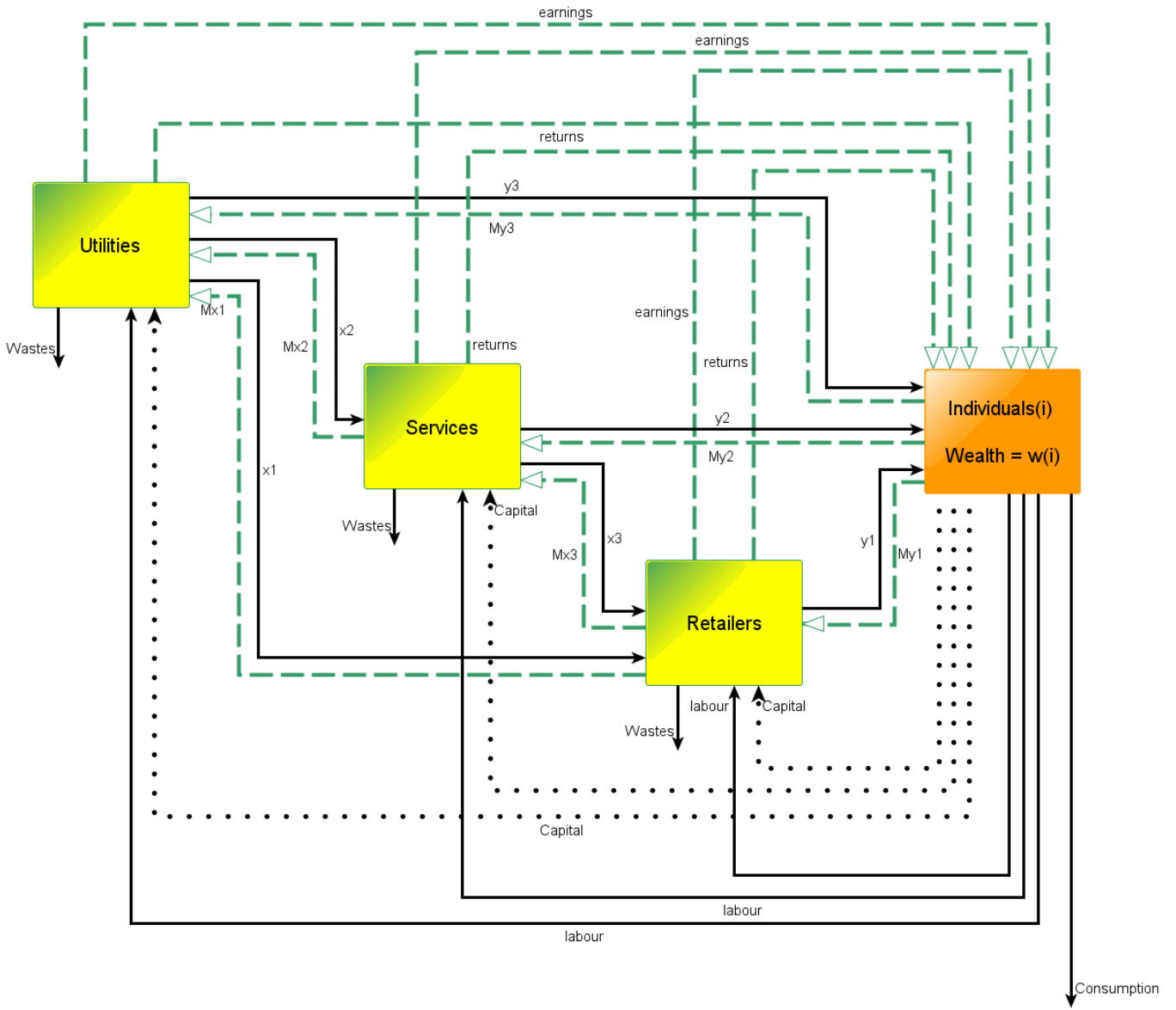

Figure 1.3.7

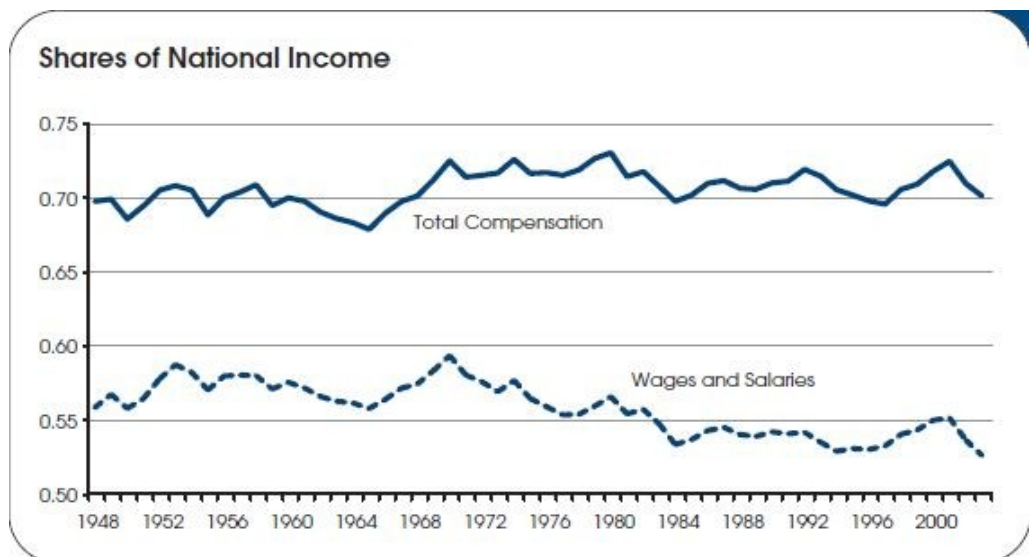

Figure 1.3.8

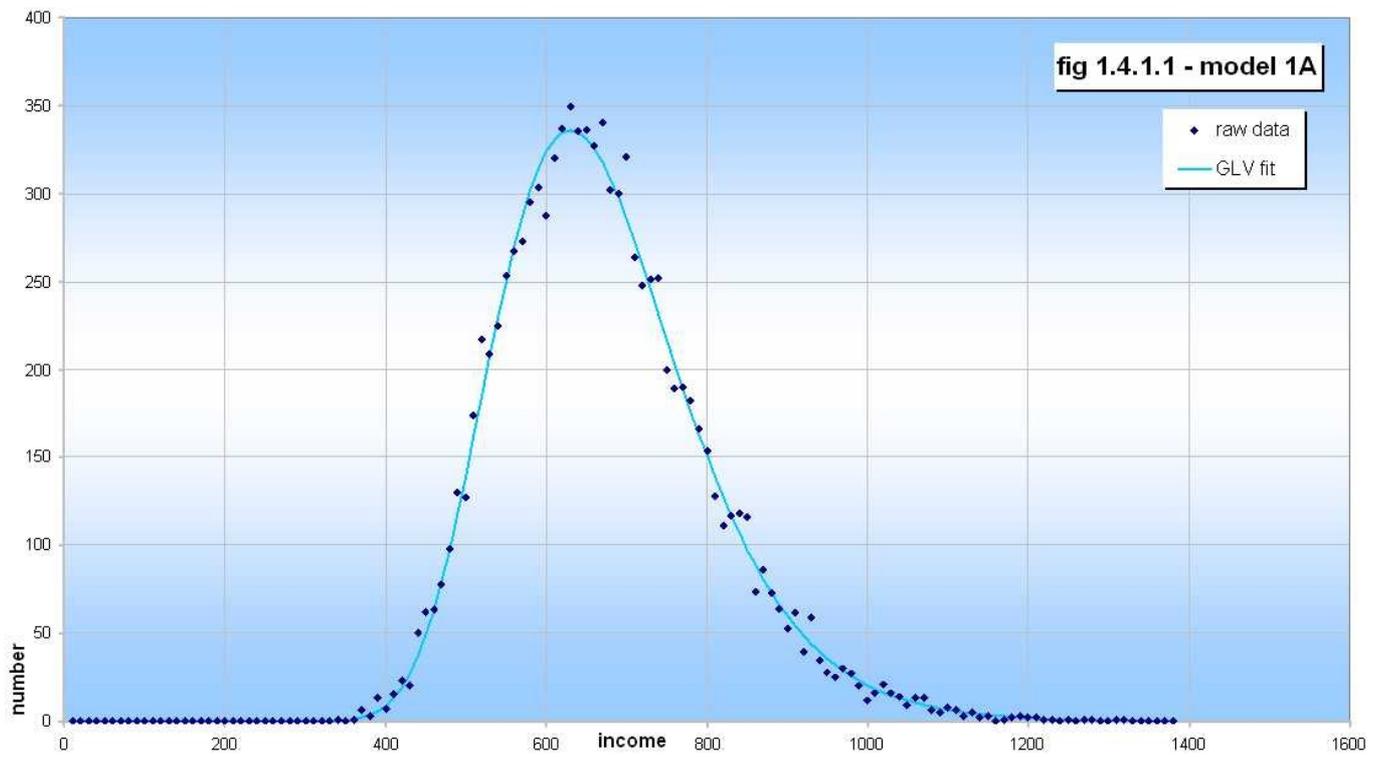

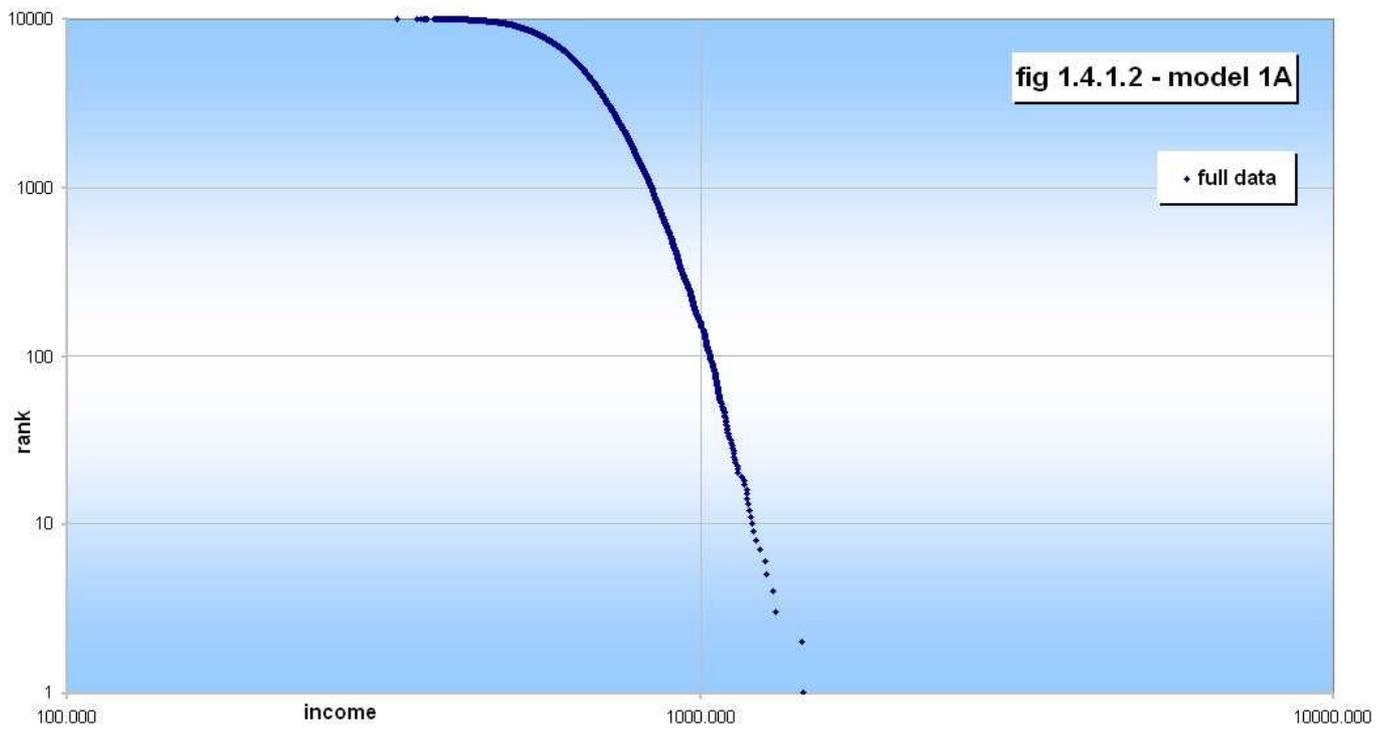

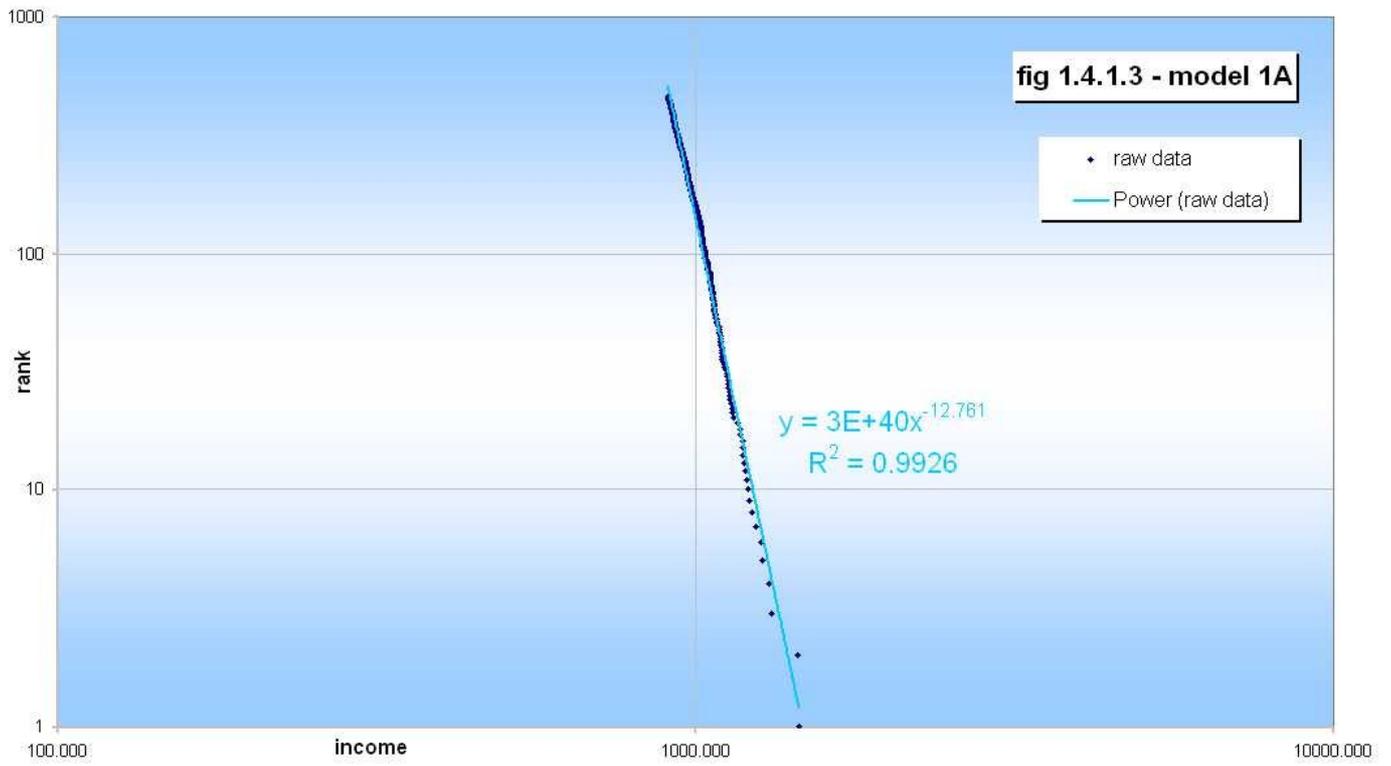

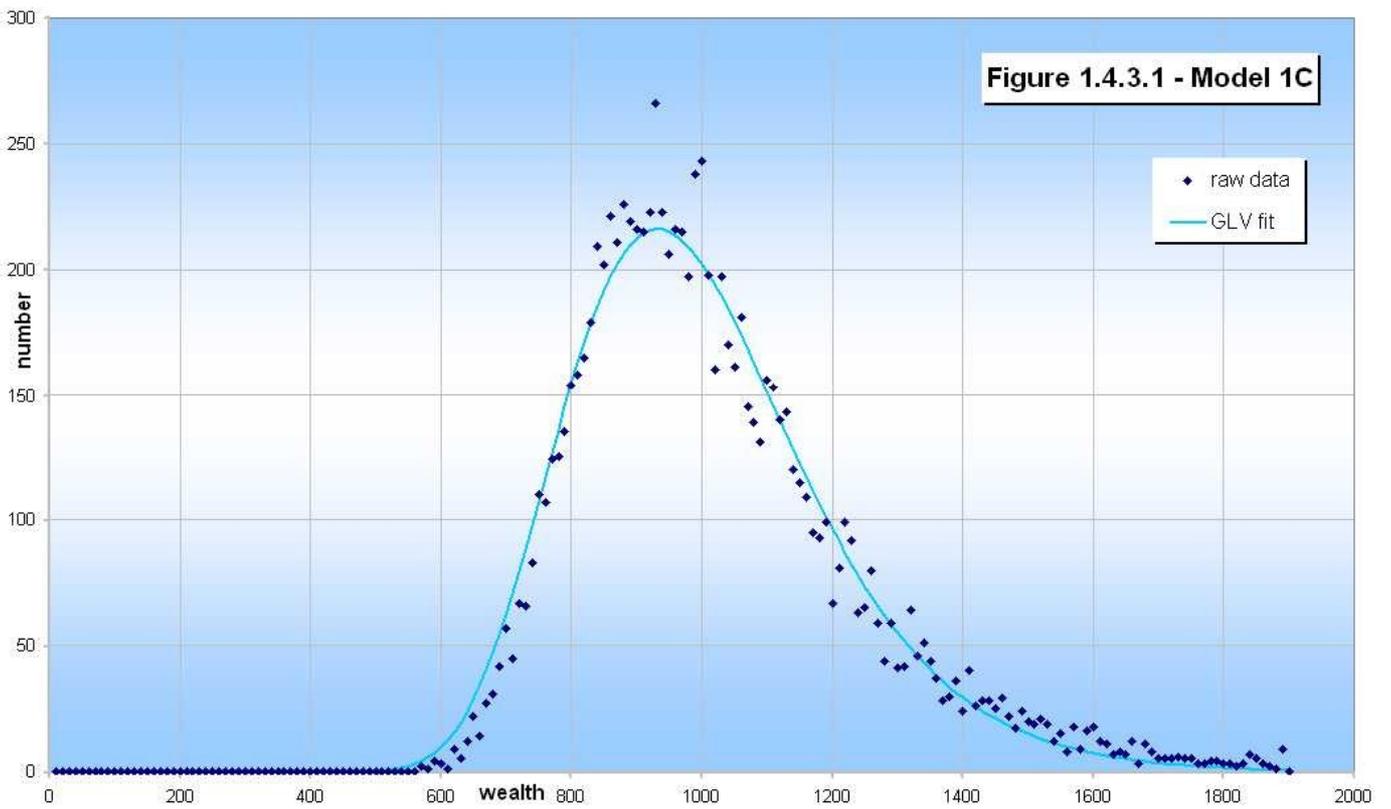

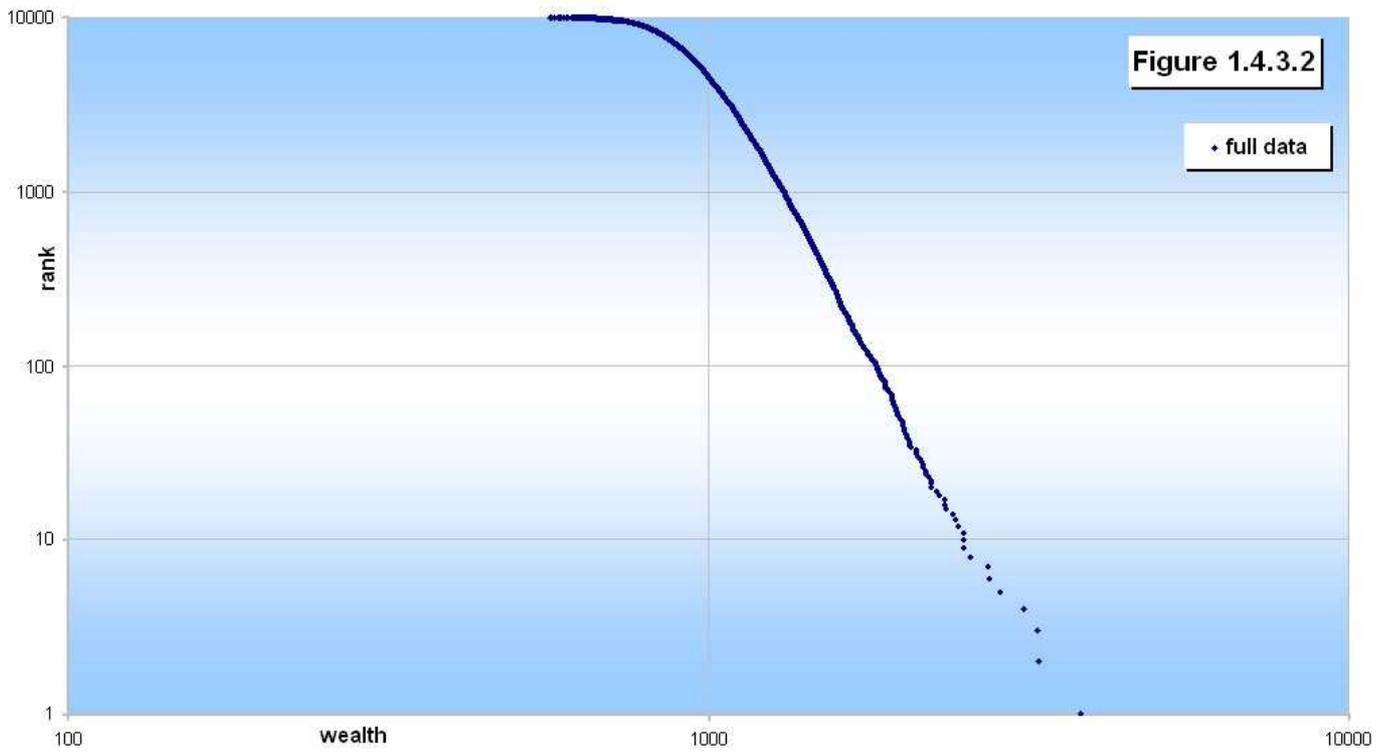

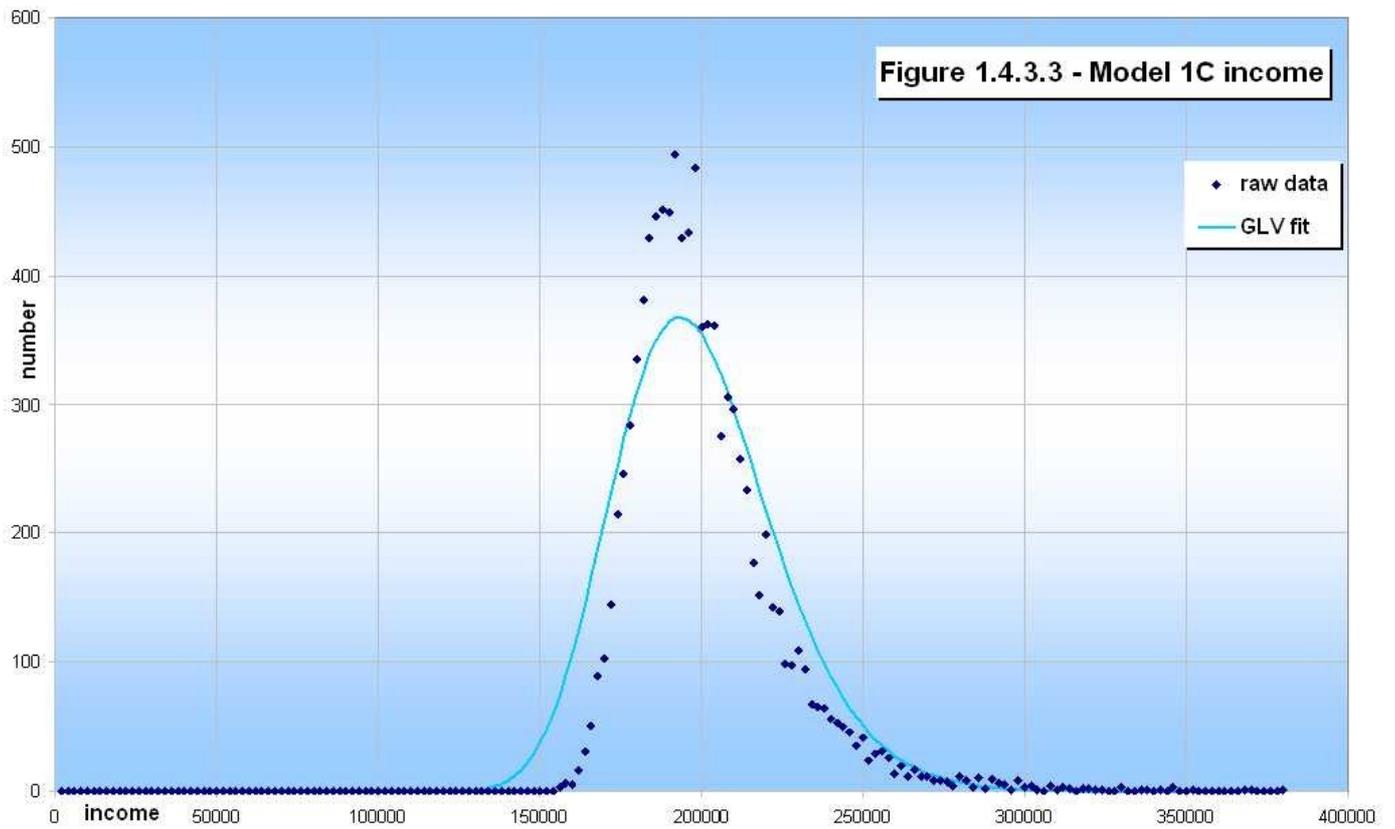

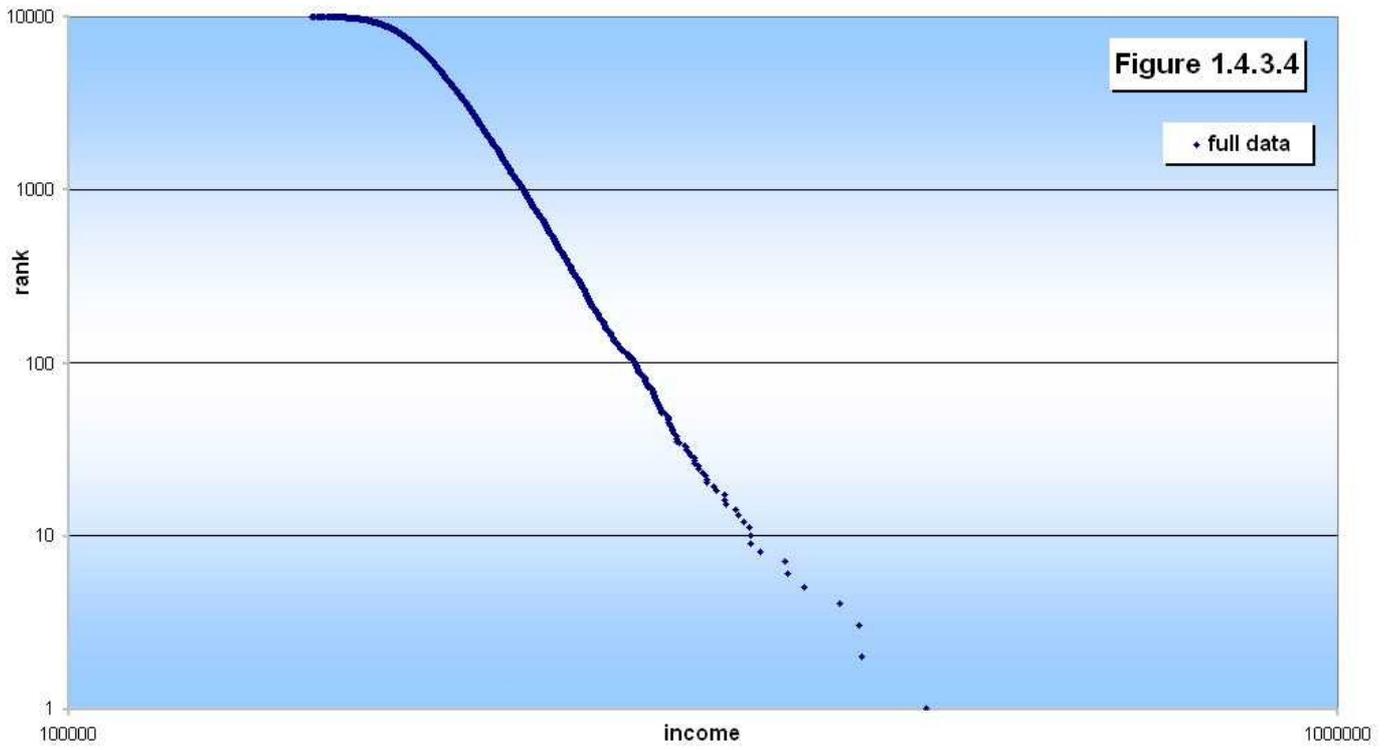

Figure 1.4.3.4

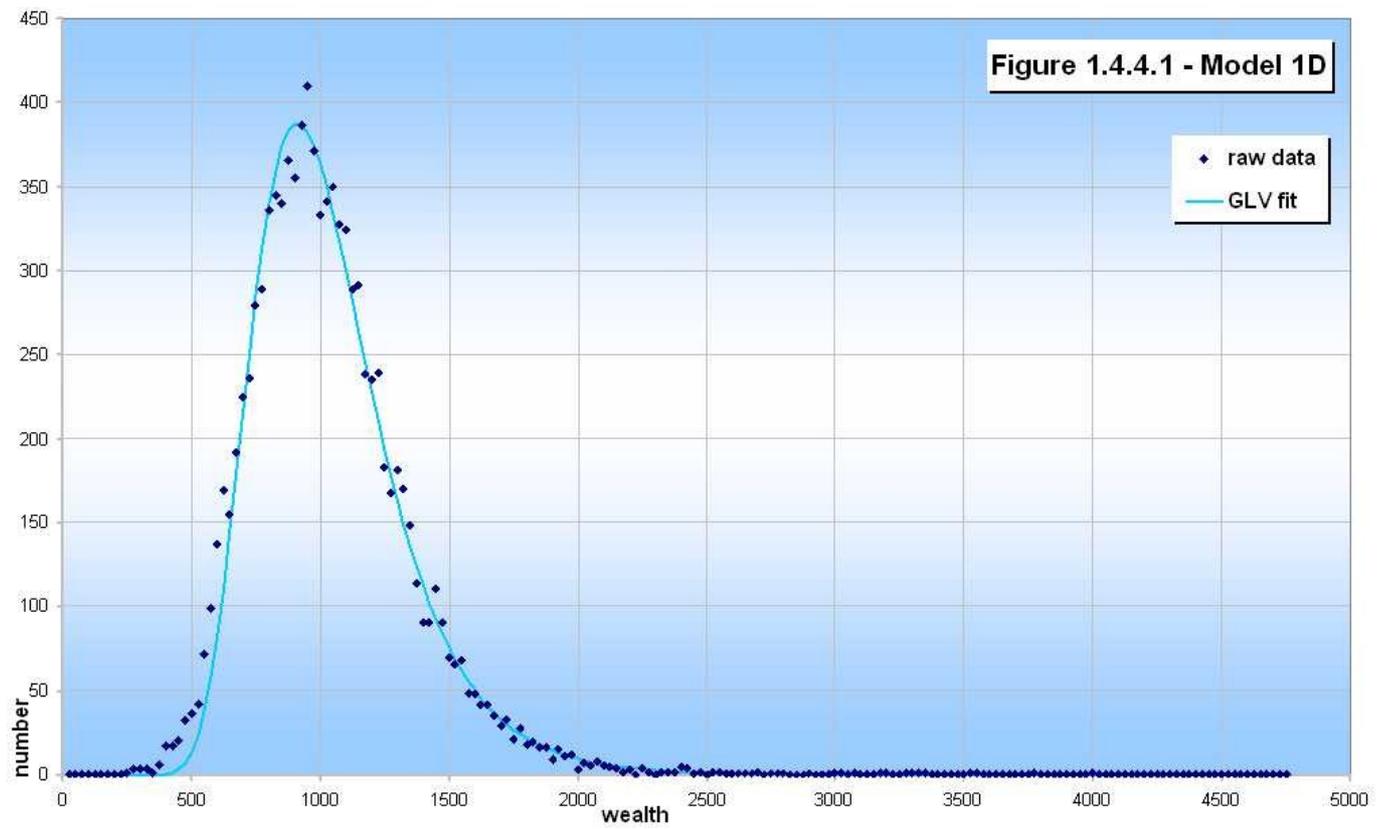

Figure 1.4.4.1 - Model 1D

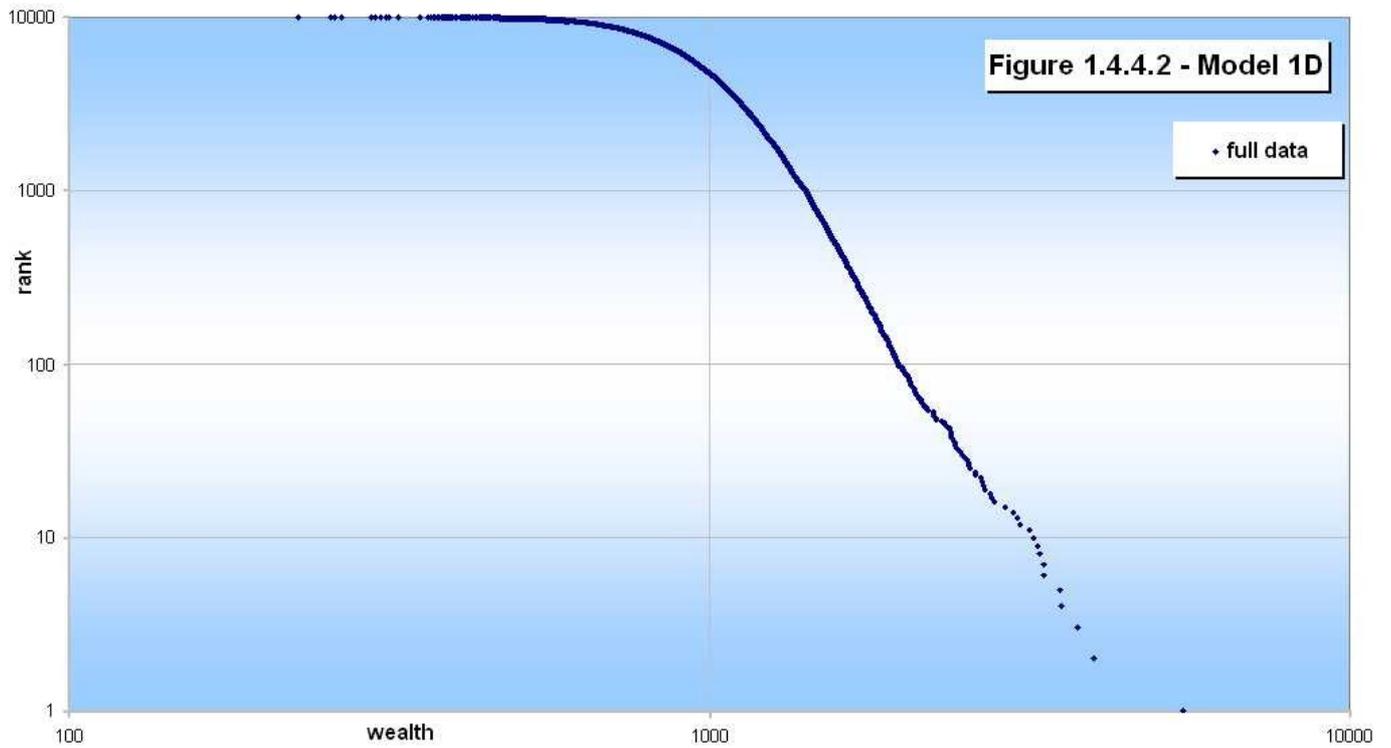

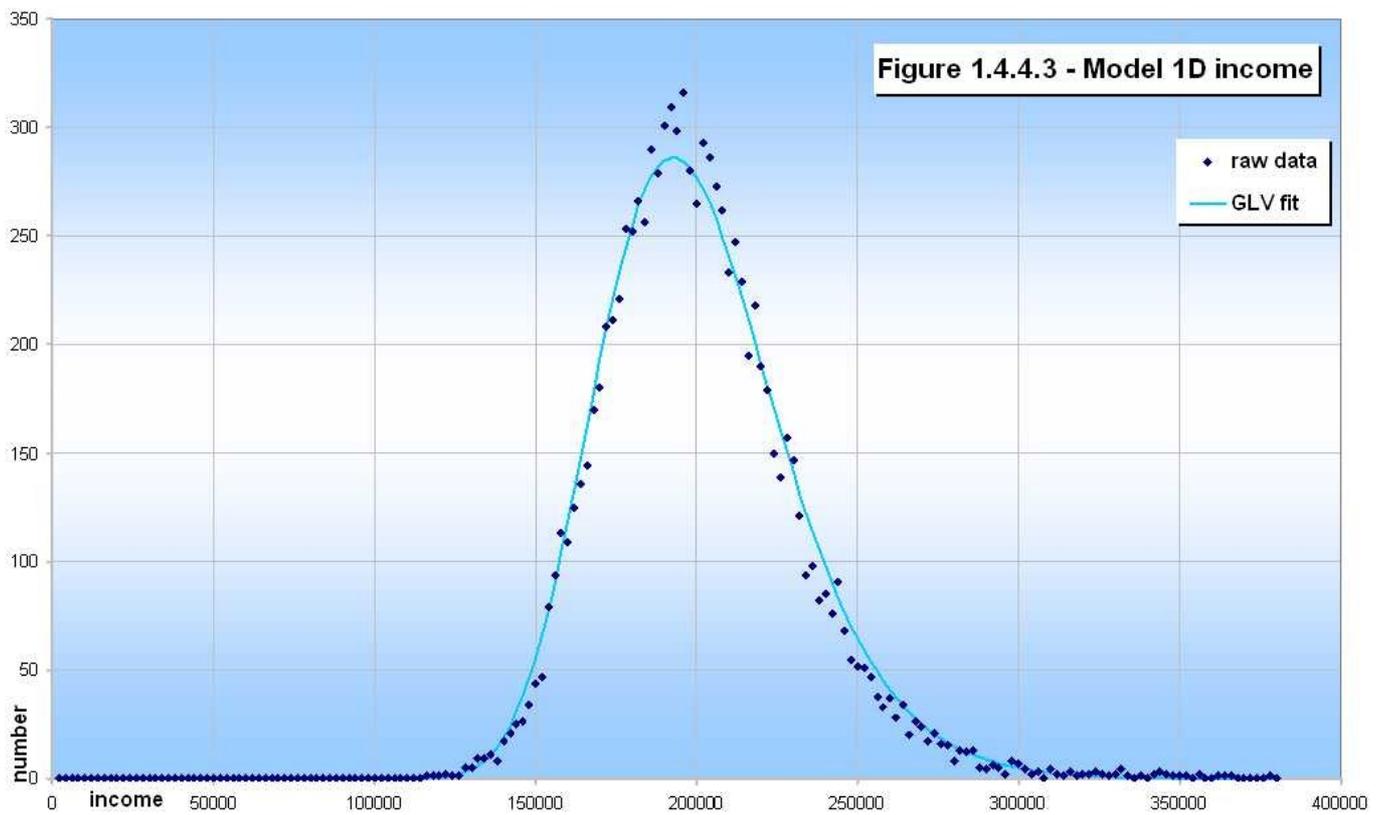

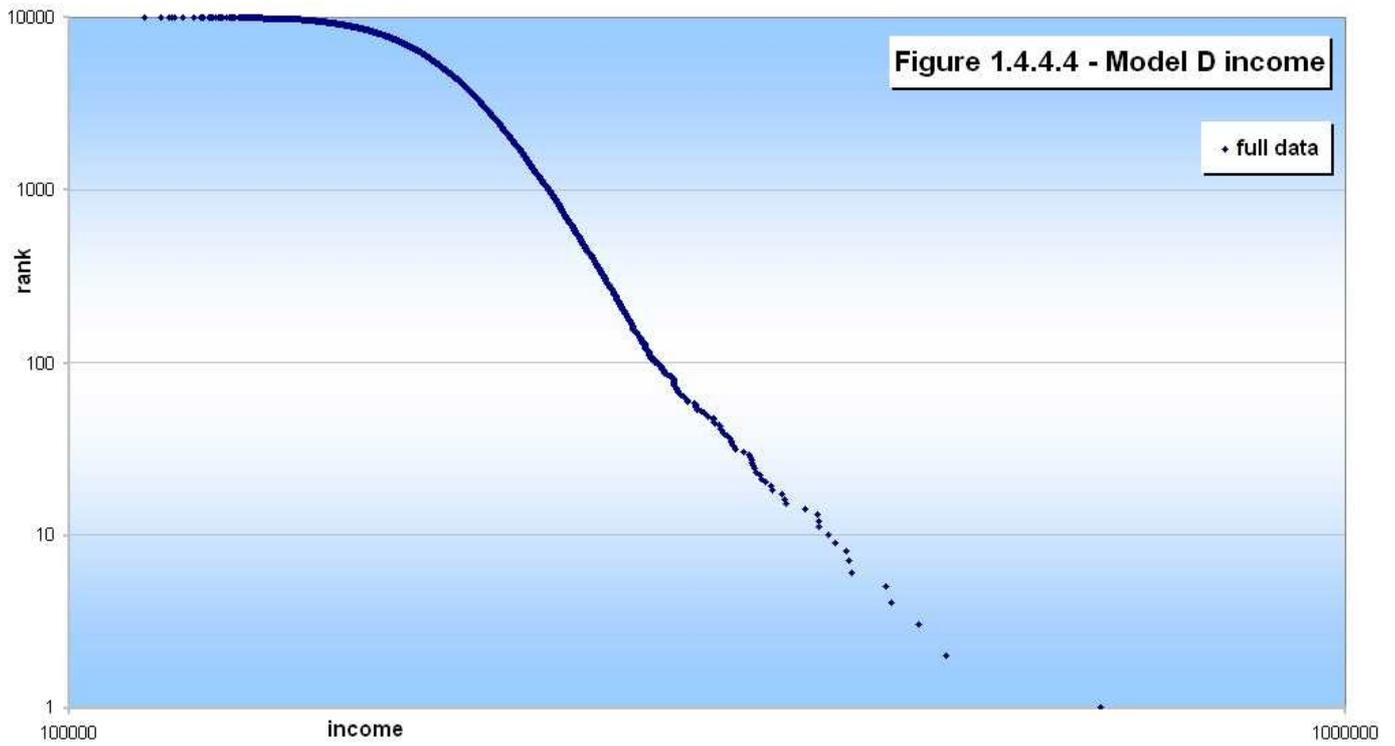

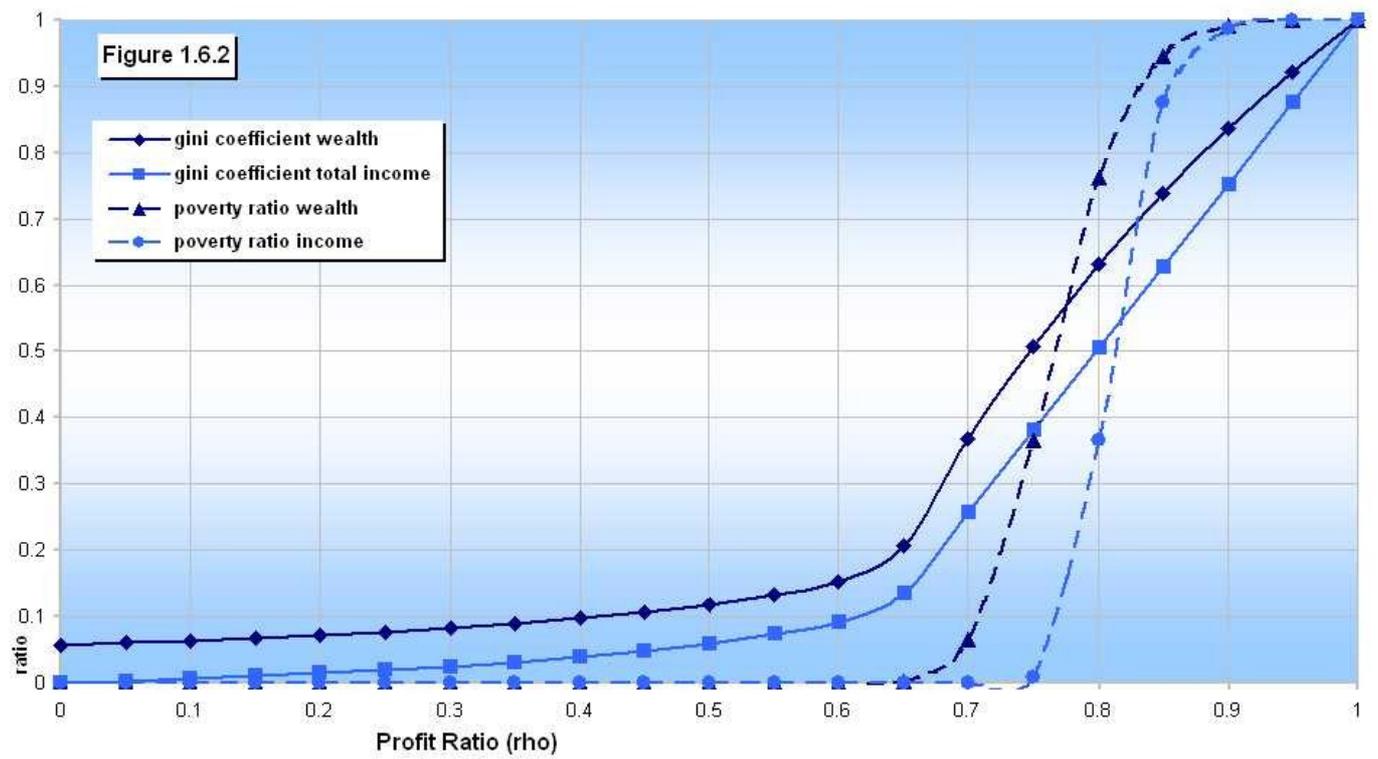

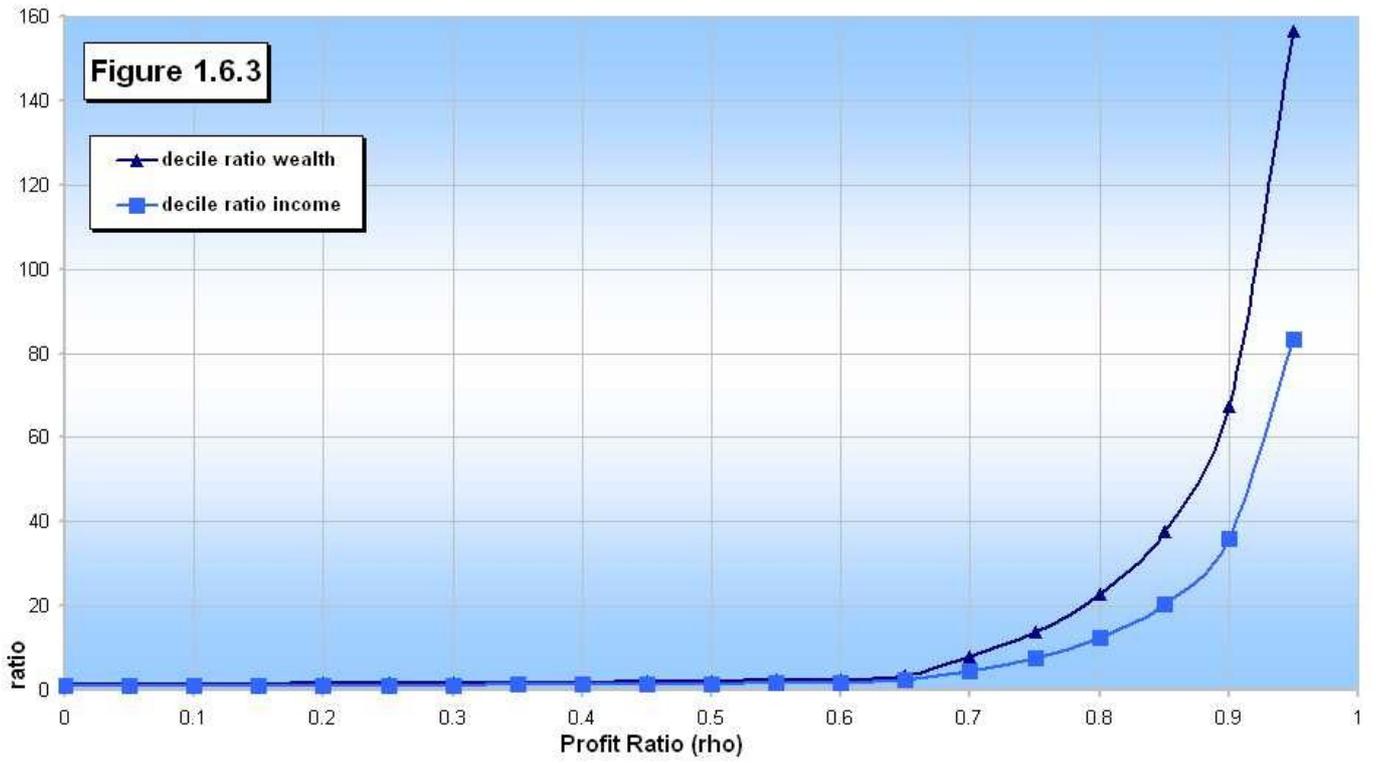

Figure 1.6.3

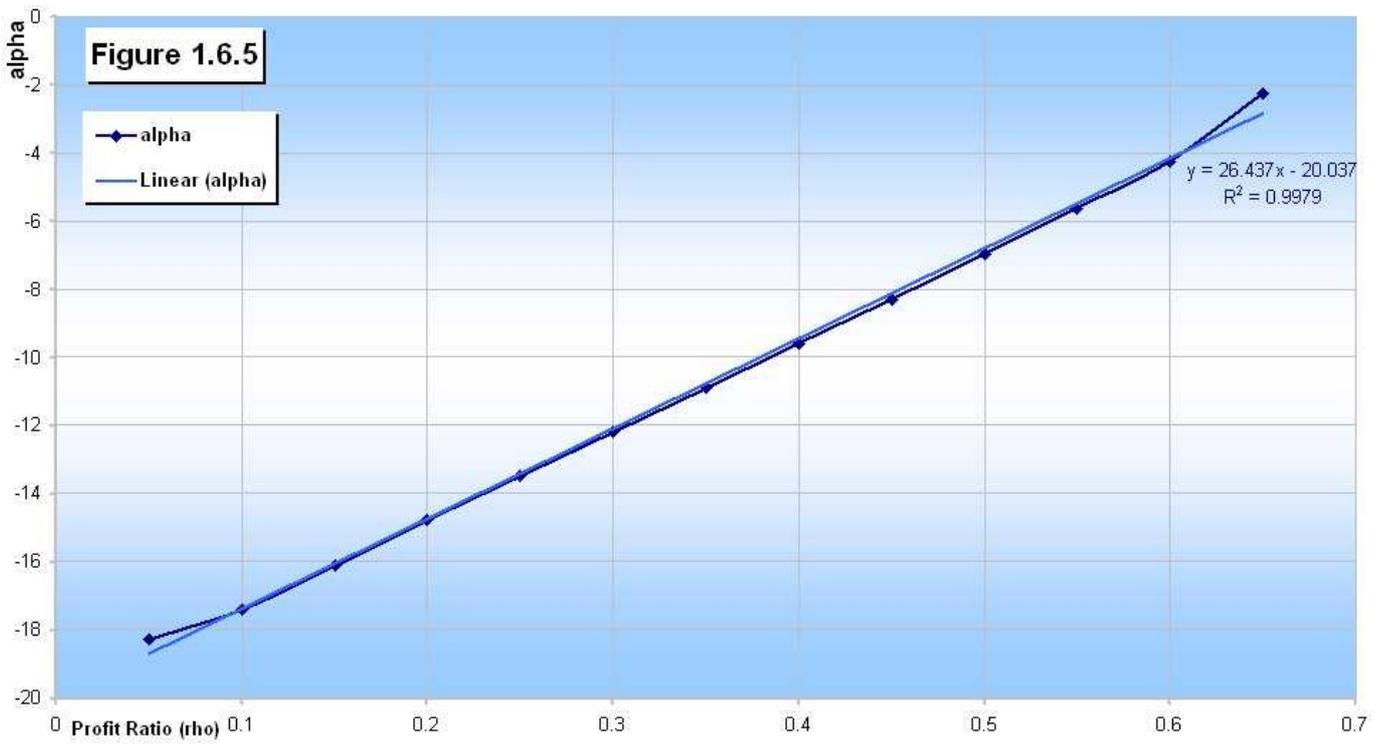

Figure 1.6.5

$y = 26.437x - 20.037$
$R^2 = 0.9979$

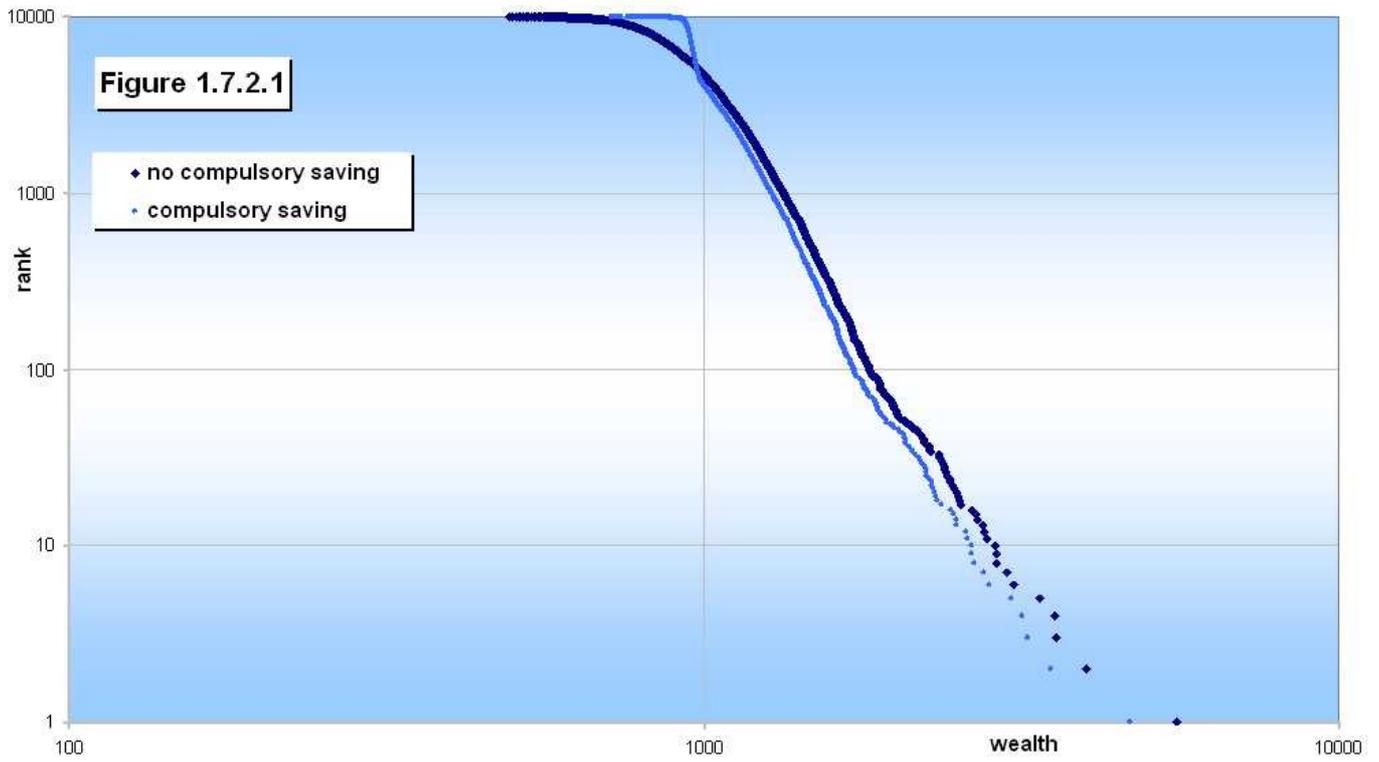

Figure 1.7.2.1

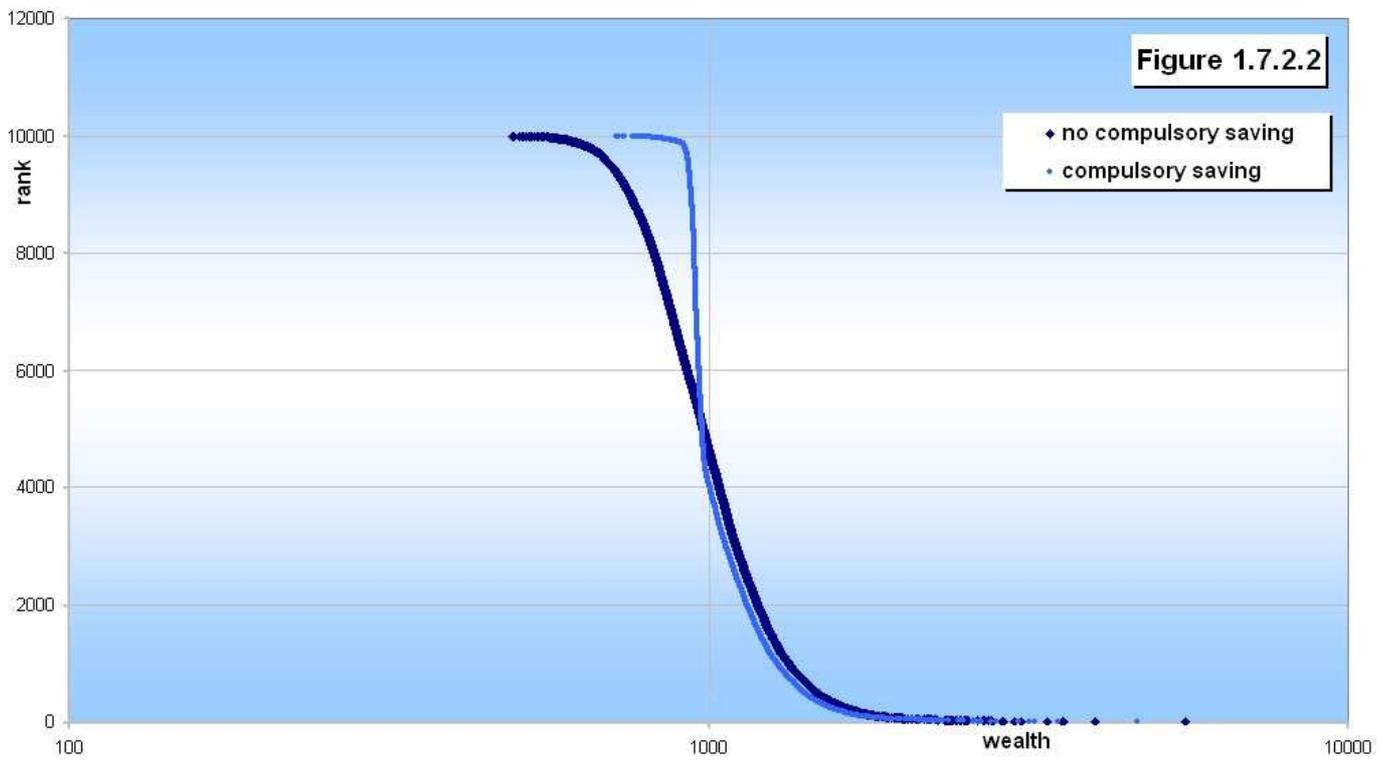

Figure 1.7.2.2